\documentclass[12pt]{iopart}

\usepackage{siunitx}
\usepackage{graphicx}
\usepackage{subcaption}
\usepackage{mathastext}
\usepackage{tikz}
\usetikzlibrary{shapes, arrows, positioning}
\usepackage{pgfplots}
\usepackage{booktabs}
\usepackage{multirow}
\usepackage{xcolor}

\begin{document}

\title[Author guidelines for IOP Publishing journals in  \LaTeXe]{Comparing growth of titania and carbonaceous dusty nanoparticles in weakly magnetised capacitively coupled plasmas}

\author{Bhavesh Ramkorun$^1$\footnote{Author to whom any correspondence should be addressed.}, Gautam Chandrasekhar$^2$, Vijaya Rangari$^2$, Saikat C. Thakur$^1$, Ryan B. Comes$^1$, and Edward Thomas, Jr$^1$.}
\address{$^1$Physics Department, Leach Science Center, 380 Duncan Drive, Auburn, AL, 36832, USA}
\address{$^2$Materials Science and Engineering, John A. Kenney Hall, Old Montgomery Road, Tuskegee, AL, 36088, USA}

\ead{bzr0051@auburn.edu, bhaveshramkorun@gmail.com}
\vspace{10pt}
\begin{indented}
\item[]\today %August 2023
\end{indented}

\begin{abstract}

This study compares the growth cycles and spatial distribution of dust cloud for titania and carbonaceous dusty nanoparticles in capacitively coupled radiofrequency plasmas, with and without the presence of a weak magnetic field  of approximately 500 Gauss. Findings on cycle time, growth rate, and spatial distribution of dust cloud are discussed. The growth of nanoparticles in these plasmas is cyclic, with particles reaching their maximum size and subsequently moving out of the plasma, followed by the generation of a new particle growth cycle. The presence of the magnetic field speeds up the growth cycle in both plasma. The magnetic field also makes the spatial distribution of the two dust cloud different from each other. Langmuir probe measurement of the background plasma parameters such as electron temperature and floating potential reveal radial variations in floating potential but not electron temperature. Furthermore, the magnetic field changes the radial variation of floating potential. These measurements, however, are not sufficient to explain why the two dust clouds appear differently. It is possible that the differences occur due to a gradient in the radial distribution of the magnetic field. 
\end{abstract}

\vspace{2pc}


%%%%%%%%%%%%%%%%%%%%%%%%%%%%%%%%%%%%%%%%%%%

\section{Introduction} \label{sec:introduction}

A dusty plasma is a four component system consisting of electrons, ions, neutrals, and nanometer (nm) to micrometre ($\mu$m) sized charged solids, i.e., dust. All the charged species interact with each other and the background neutrals to form a complex, coupled system \cite{Shukla01, Das14, adamovich20222022, beckers2023physics}. Since the dust particles are significantly more massive than the other species, these systems can exhibit a broad range of phenomena such as self-organization \cite{vladimirov2004dynamic, tsytovich2015self, song2016self, boltnev2018self}, clustering \cite{juan1998observation,ratynskaia2005statistics, schella2011melting, Dharodi2023}, and waves phenomena \cite{nunomura2000transverse, shukla2001dust}. Moreover, the large size of the dust particle enables direct visualization of these phenomena in the laboratory \cite{nakamura2001observation, thomas2007observations, 
kumar2023kelvin}.

Dusty plasmas can be formed in a variety of natural \cite{mendis1994cosmic, horanyi2004dusty, wahlund2009detection, morooka2011dusty, popel2013dusty}, and laboratory environments \cite{ ma1997dust, thomas1999first, ignatov2005basics, chaubey2023controlling}. Many dusty plasma experiments are performed by directly introducing the dust particles into the plasma \cite{jaiswal2017effect, jaiswal2019melting, chaubey2021positive}.  However, if a reactive gas is used to generate the plasma, chemical processes can lead to the spontaneous nucleation and growth of dust particles directly from the gas phase. The background plasma is usually ignited from an inert gas such as argon (Ar). Over the past three decades, different kinds of particles growth have been reported such as silicates \cite{Boudendi94}, carbonaceous \cite{kovavcevic03, Kovacevic09,Kovacevic12}, polyaniline \cite{Pattyn18} and organosilicon \cite{garofano19}. Recently, there has been a shift towards exploring metal-containing nanoparticles such as aluminium \cite{cameron2023capacitively}, titanium \cite{tu2024nonthermal} and titanium dioxide $(TiO_2)$, also known as titania \cite{ramkorun2024introducing}. Studies regarding the formation of dusty plasma have a broad range of applications from understanding the formation of dust in space \cite{kovavcevic05} to semiconductor processing \cite{Boudendi11}. The majority of the visible universe is made up of plasma and dusty plasmas are believed to be prevalent in planetary rings and star formation \cite{Shukla01, Barabash94, horanyi2004dusty}.  

%Silicate dusts has been grown in Ar/silane $(SiH_4)$ plasmas \cite{Boudendi94}. Carbonaceous dusts \cite{Kovacevic12} and carbon nitride \cite{Kovacevic09} have been grown in Ar/acetylene $(C_2H_2)$ and Ar/$C_2H_2$/nitrogen plasmas respectively. Polyaniline, which are conducting polymers, have been grown in Ar/aniline $(C_6H_7N)$ plasmas \cite{Pattyn18}. Organosilicon been grown in Ar/hexamethyldisiloxane (HDMSO) $\left(C_6H_{18}OSi_2\right)$ plasmas \cite{garofano19}; similar precursors have been used to grow low dielectric constant thin films in plasma enhanced chemical vapor deposition (PECVD) \cite{Shirafuji99, Shioya05}. 

%Aluminium nanocrystals have been grown in Ar/hydrogen/aluminium trichloride $\left(AlCl_3\right)$ plasmas \cite{cameron2023capacitively}. Titanium dioxide $(TiO_2)$, also known as titania, nanoparticles have been grown in an Ar/titanium tetraisopropoxide (TTIP) $\left(C_{12}H_{28}O_4Ti\right)$ plasmas \cite{ramkorun2024introducing}. Whereas studies for the past 30 years have mostly focused on the growth of carbonaceous and silicate particles in dusty plasma, there has been a shift towards exploring other particles growth in the past 10 years, as given by the examples above. Therefore, there is a need to study the properties of the new kinds of dusty plasma.

The focus of this paper is to grow titania nanoparticles, using titanium tetraisopropoxide (TTIP) $\left(C_{12}H_{28}O_4Ti\right)$ as the precursor, in a weakly-magnetized ($\sim$ 500 Gauss) dusty plasma environment with an accompanying detailed investigation of the cyclical growth process. This growth will be compared with carbonaceous dust grown using acetylene $(C_2H_2)$. Titania dust, along with other oxides of titanium, have been proposed as candidates for the first condensates in primitive solar nebula \cite{rietmeijer1990, posch2003, simon2017}. In semiconductor processing, TTIP has been commonly used in PECVD to grow $TiO_2$ thin films. In general, it is common in PECVD to take at least 10 minutes to nucleate nm to $\mu$m-sized particles prior to thin film growth \cite{Stoner92, Lee94, Kuhr95, Yang06, Nguyen13, Ramkorun21, lee2022low}.  Dusty plasma particle growth offers the advantage of nucleating nm-sized particles within milliseconds (ms) \cite{ravi09} and growing to $\geq$ 100 nm within tens of seconds (s) \cite{groth2015}.

Dusty plasma particle growth happens in three steps, (i) nucleation, (ii) coagulation, and (iii) agglomeration (accretion) \cite{Boudendi11, Kovacevic12,  Fridman96, samsonov1999instabilities, stefanovic03, Donders22}. 
%A schematic figure is shown in Fig. \ref{growth}. They are depicted in Fig. \ref{nucleation}, \ref{coagulation} and \ref{agglomeration} respectively. 
Energetic electrons from the background plasma constantly bombard the gas molecules to create excited species, ions and radicals. During nucleation, clusters are formed spontaneously from chemical reactions between neutrals, and ions that are trapped in the plasma \cite{hong2002growth}. These clusters have stochastic charge fluctuations \cite{ ravi09, cui1994fluctuations, goree1994charging}. During coagulation, clusters and neutrals collide and react to form bigger particles \cite{girshick2020particle}. As the size of the particles increase, the rate of coagulation reaches a saturation \cite{mankelevich2008dust}. At this stage, the particles have accumulated a negative charge. During agglomeration, also known as accretion, new growth occurs on the surface of the negatively charged particles due to positive ions, radicals and neutrals colliding on its surface \cite{bingham2001new, sodha2010growth, denysenko2020plasma}. The particles' size in the three steps are in the order of 1 nm, 10 nm and 100 nm, respectively. Coagulation and agglomeration are dominated by chemisorption and physisorption respectively \cite{shi2021growth}.

\begin{figure}[ht]
    \centering    \includegraphics[width=\linewidth]{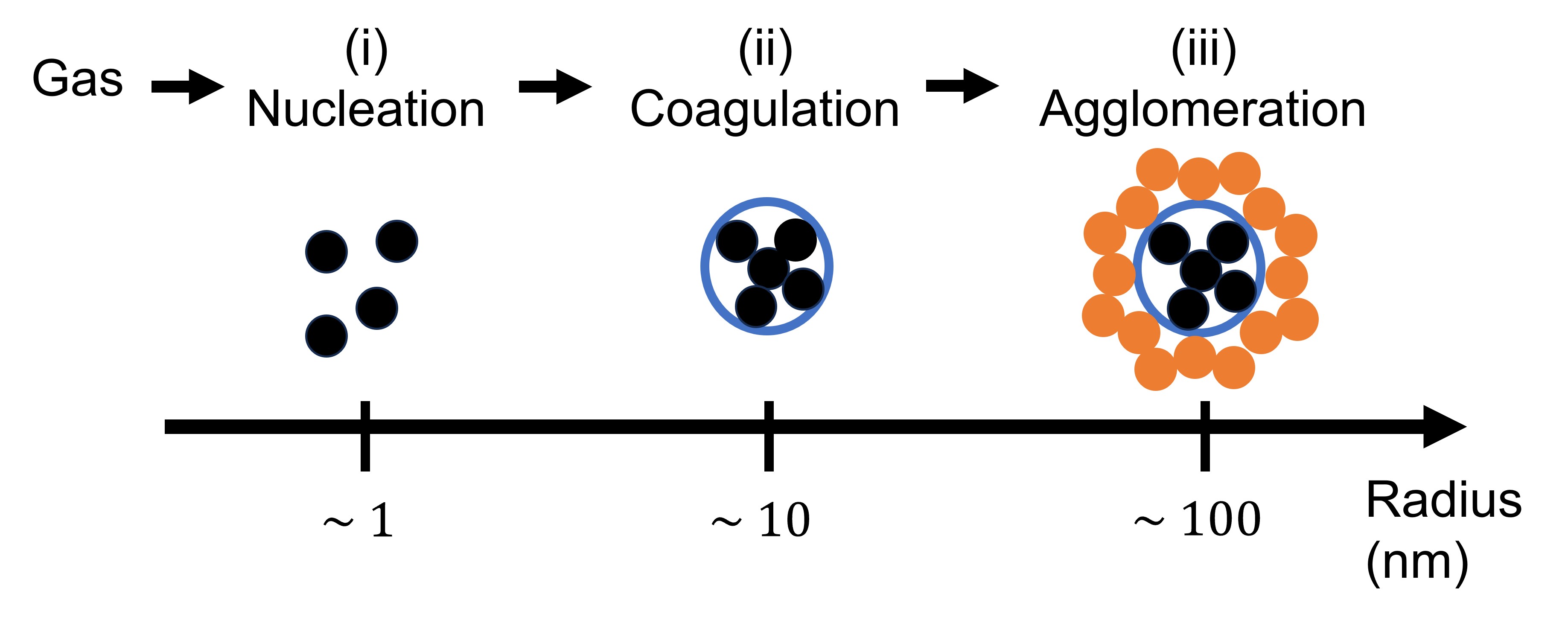}
    \caption{(Schematics not drawn to scale) The particle growth processes from the reactive gas: (i) Nucleation, (ii) Coagulation, and (ii) Agglomeration. After the first generation of growth ends, a new one begins. The particles continuously grow in cycles as long as the plasma is turned on. }
    \label{growth}
\end{figure}

\begin{figure}[ht]
  \centering  
  \begin{subfigure}{0.32\linewidth}
    \includegraphics[width=\linewidth]{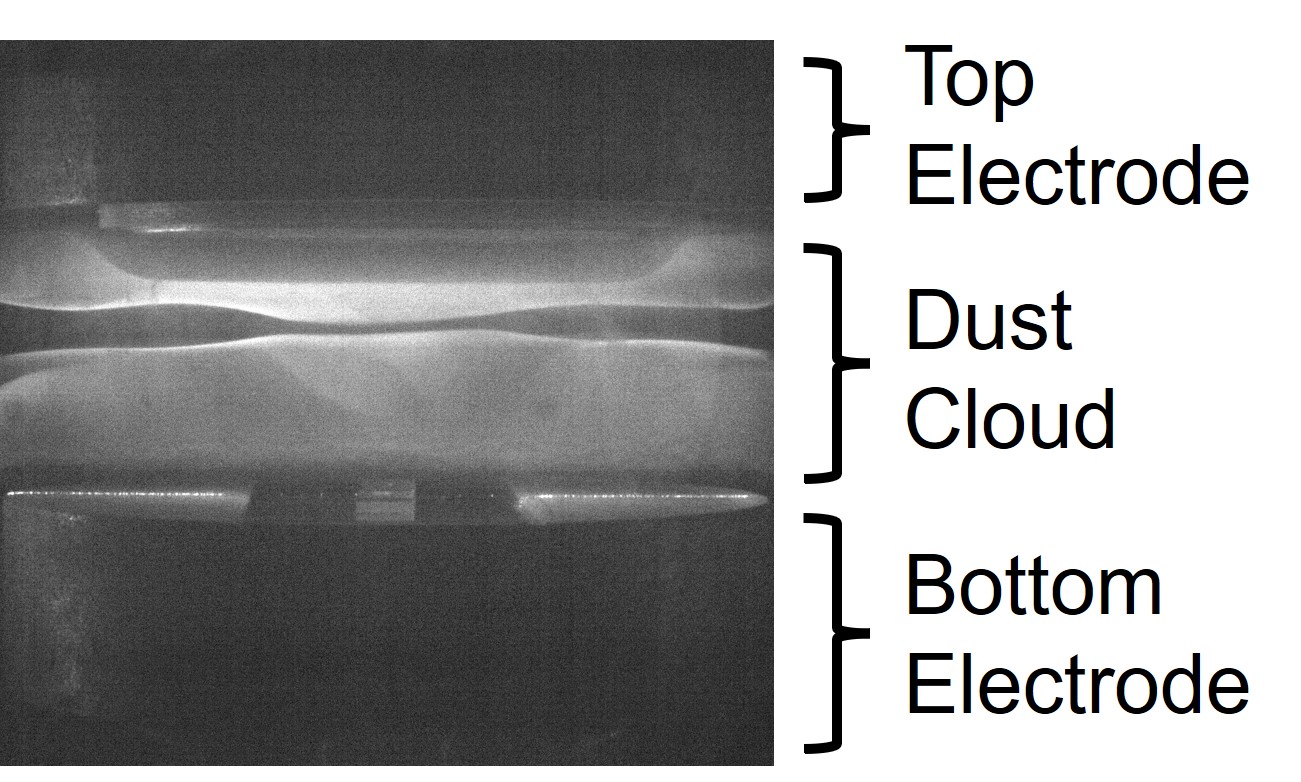}
    \caption{}
    \label{dust_cloud}
  \end{subfigure}
  %\hspace{0.01\linewidth}
  \begin{subfigure}{0.32\linewidth}
    \includegraphics[width=\linewidth]{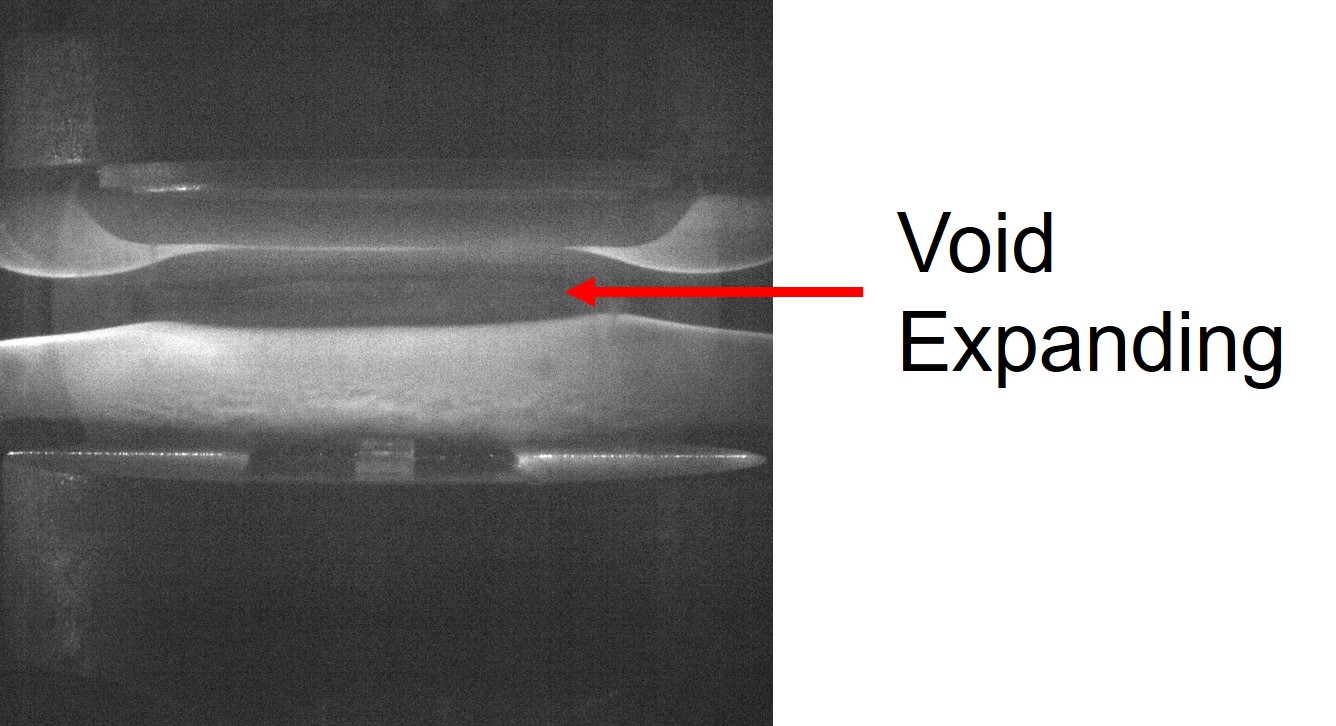}
    \caption{}
    \label{void_expands}
  \end{subfigure}
  %\hspace{0.01\linewidth}
  \begin{subfigure}{0.32\linewidth}
    \includegraphics[width=\linewidth]{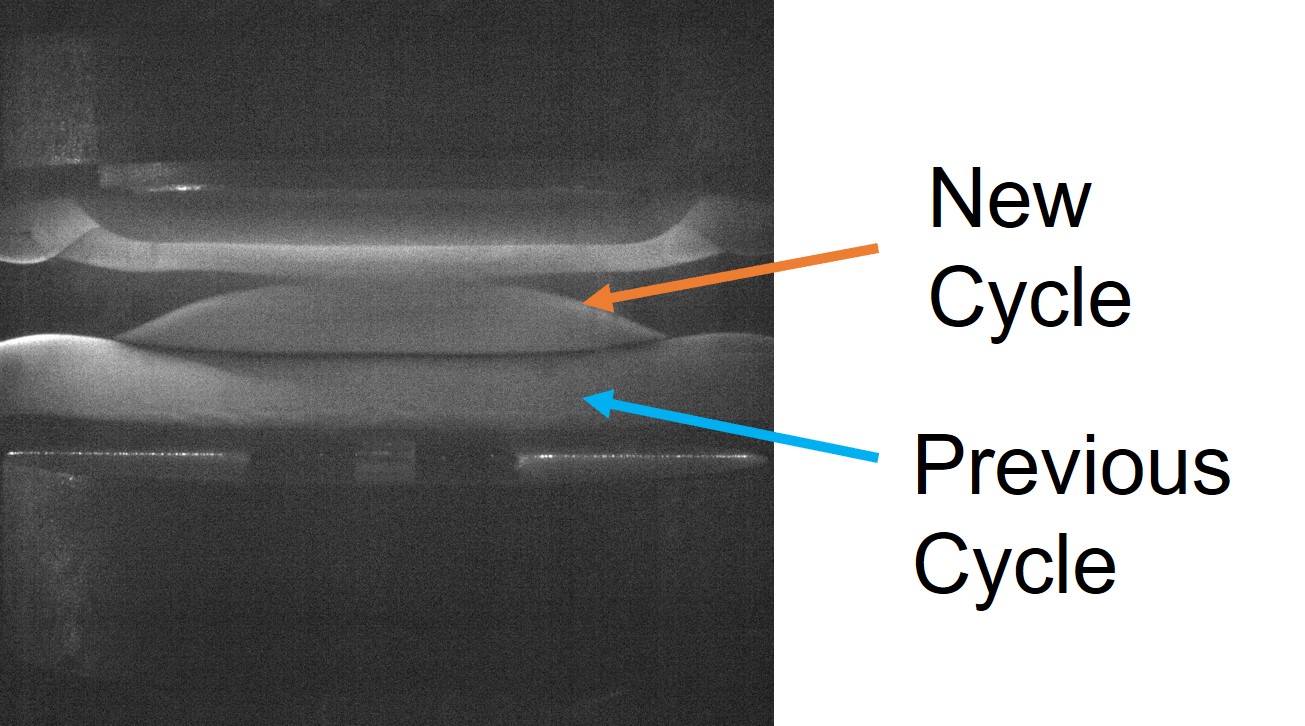}
    \caption{}
    \label{two_cycles}
  \end{subfigure}
  
  \caption{An example of the evolution of the dust cloud during titania particle growth in Ar/TTIP dusty plasma. (a) Dust cloud levitated in the plasma. (b) Void starts to expand, signalling the end of a growth cycle. (c) New growth cycle starts in the void. The previous cycle is gradually moving out of the plasma.}
  \label{evolution}
\end{figure}

During growth, the particles form a dust cloud, as shown in Fig. \ref{dust_cloud}, that is levitated in the plasma due to a balance of several forces such as gravitational, electric, thermophoretic, ion drag, and neutral drag \cite{shukla2009colloquium, merlino06, Beckers11_thesis, van15}. The forces’ magnitudes depend on the radius of the particles, with gravitational being cubic, electric being linear, and the others being quadratic.
Initially, when the particles nucleate and start to grow, gravitational force is negligible and electric and ion drag forces are the dominant forces. As the particles further grow in size ($>$ 100 nm), gravitational force also becomes significant \cite{van15}. Later during the agglomeration, the particles accumulate enough mass and size so that gravitational and ion drag forces cause the particles to move away from the bulk plasma towards the chamber’s wall. Throughout the growth, it is common to have a void, i.e., dust-free region, in the dust cloud. The void expands when the particles move away, as shown in Fig. \ref{void_expands}. The amount of time between nucleation and the void expanding is known as a growth cycle. A new cycle of growth immediately begins in the void, as shown in Fig. \ref{two_cycles}, and similarly reported in other studies \cite{ramkorun2024introducing, feng2004, jaiswal2020}. The cyclic growth of nanoparticles occurs continually when the plasma is on. In capacitively coupled radiofrequency (RF) plasmas, particles have grown from different reactive gases, but the timing of their growth cycles have varied between experiments, likely due to geometrical differences in experimental set ups, RF power, and the concentration and combinations of different reactive gases \cite{winter2009, hundt11}. The cycle time has been measured via different techniques such as microwave cavity resonance spectroscopy \cite{van15}, camera images of the dust cloud \cite{Couedel19}, Rayleigh/Mie scattering of infrared signals \cite{kovavcevic03},  scattered laser beam intensity \cite{winter2009} and optical emission spectroscopy (OES) \cite{despax16, despax2012, garofano2015}. Cycle times as low as 40 seconds \cite{Couedel19} and as high as 35 minutes \cite{kovavcevic03} have been reported.  Nonetheless, a consistent pattern emerges: particles undergo a growth cycle, gradually increasing in size and mass before ultimately losing their force balance and becoming unconfined. 

%{For example, in one Ar/$C_2H_2$ dusty plasma with 30 cm diameter electrodes, 8 cm electrode spacing, and 10 - 20 W of applied RF power, a cycle length of 35 minutes was measured using Rayleigh/Mie scattering of infrared signals \cite{kovavcevic03}. In another Ar/$C_2H_2$ dusty plasma with 75 mm diameter electrodes, 25 mm electrode spacing, and 5 - 30 W of applied RF power, a cycle length of 55 s was measured using pictures taken with a USB camera \cite{Couedel19}. In yet another experiment using Ar/$C_2H_2$ dusty plasma, with 10 cm diameter electrodes, 4 cm electrode spacing, and 20 W of RF power, a 2-minute cycle was measured by analysing the scattered laser beam intensity \cite{winter2009}. Ar/$C_2H_2$ dusty plasma have been well studied in literature. However, as mentioned above, there has been a shift towards exploring different kinds of materials in plasma. Other gases besides Ar/$C_2H_2$ have also demonstrated a growth cycle. For example, in the same experimental set up at the previously mentioned one, when switching to Ar/methane plasma, it required 60 W of RF power to initiate plasma ignition and a 15-minute cycle time was measured \cite{winter2009}. Using techniques such as mass spectrometry, optical emission spectroscopy (OES), and measurements of electron temperature  and electron density, a 150 s cycle was observed in Ar/HDMSO plasma with a top electrode diameter of 10 cm, bottom electrode diameter of 12 cm, electrode spacing ranging from 3.5 to 4 cm, and an RF power of 30 W \cite{despax16, despax2012, garofano2015}}

%%%%%%%%%%%%%%%%%%%%%%%%%%%%%%%%%%%%%%%%%%%

\section{Method} \label{sec:experiment}
\begin{figure*}[ht]
    \centering    
    \begin{subfigure}{0.45\textwidth}
          \centering
        \includegraphics[width=\linewidth]{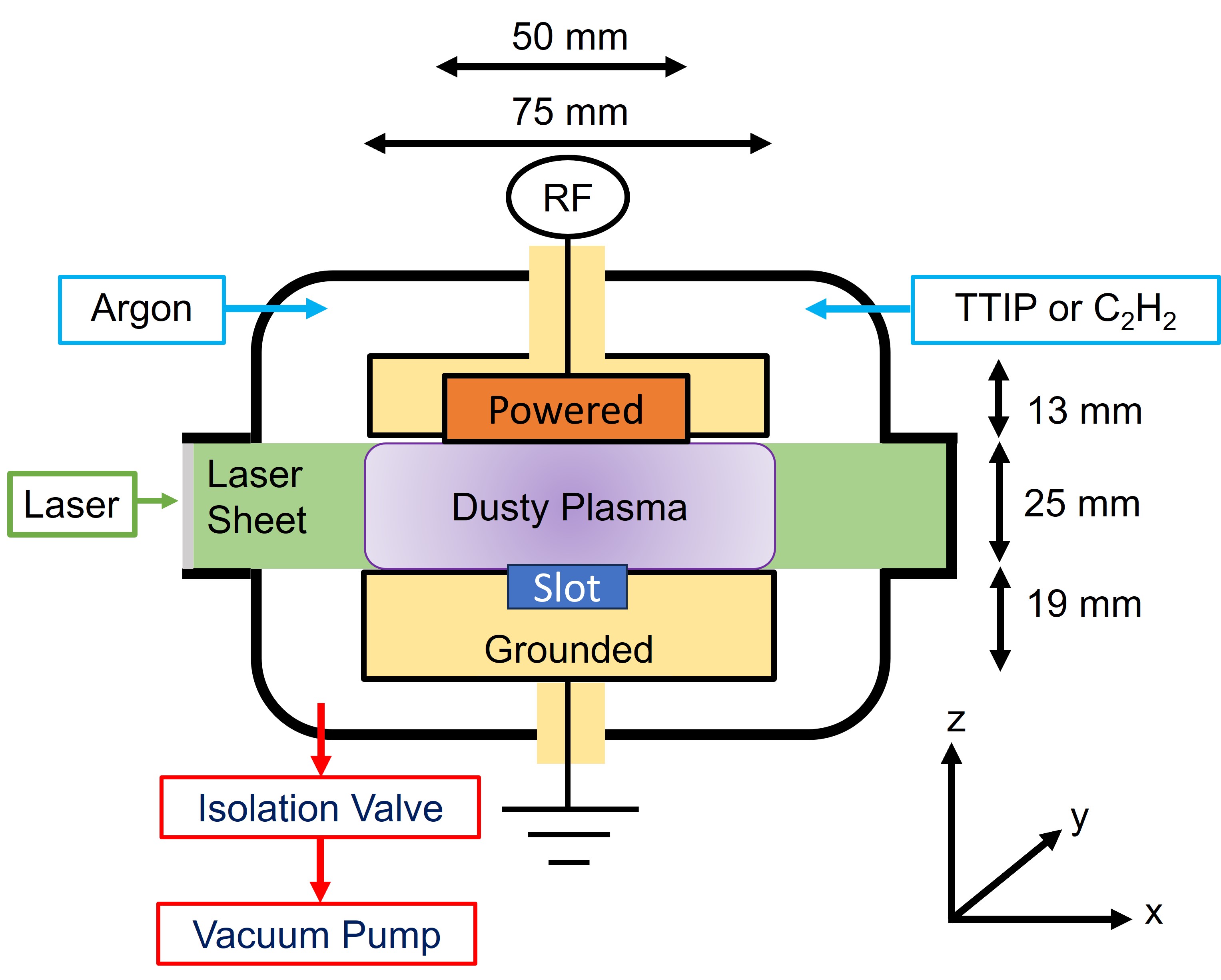}
        \caption{}
         \label{sideview_a}
    \end{subfigure}
    %\hfill
    \begin{subfigure}{0.45\textwidth}
          \centering
        \includegraphics[width=\linewidth]{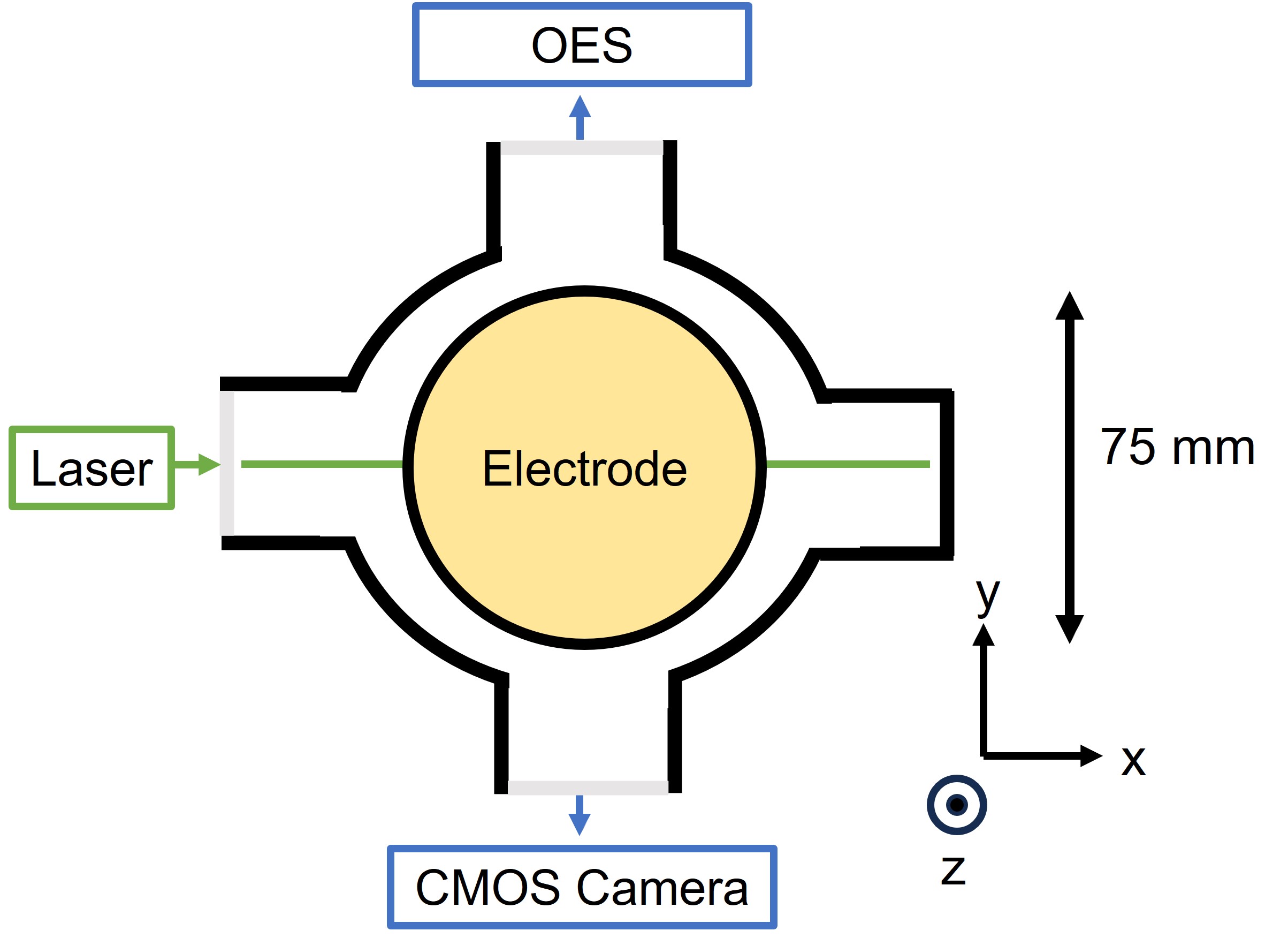}
        \caption{}
        \label{topview}
    \end{subfigure}    
    \begin{subfigure}{0.45\textwidth}
          \centering
        \includegraphics[width=\linewidth]{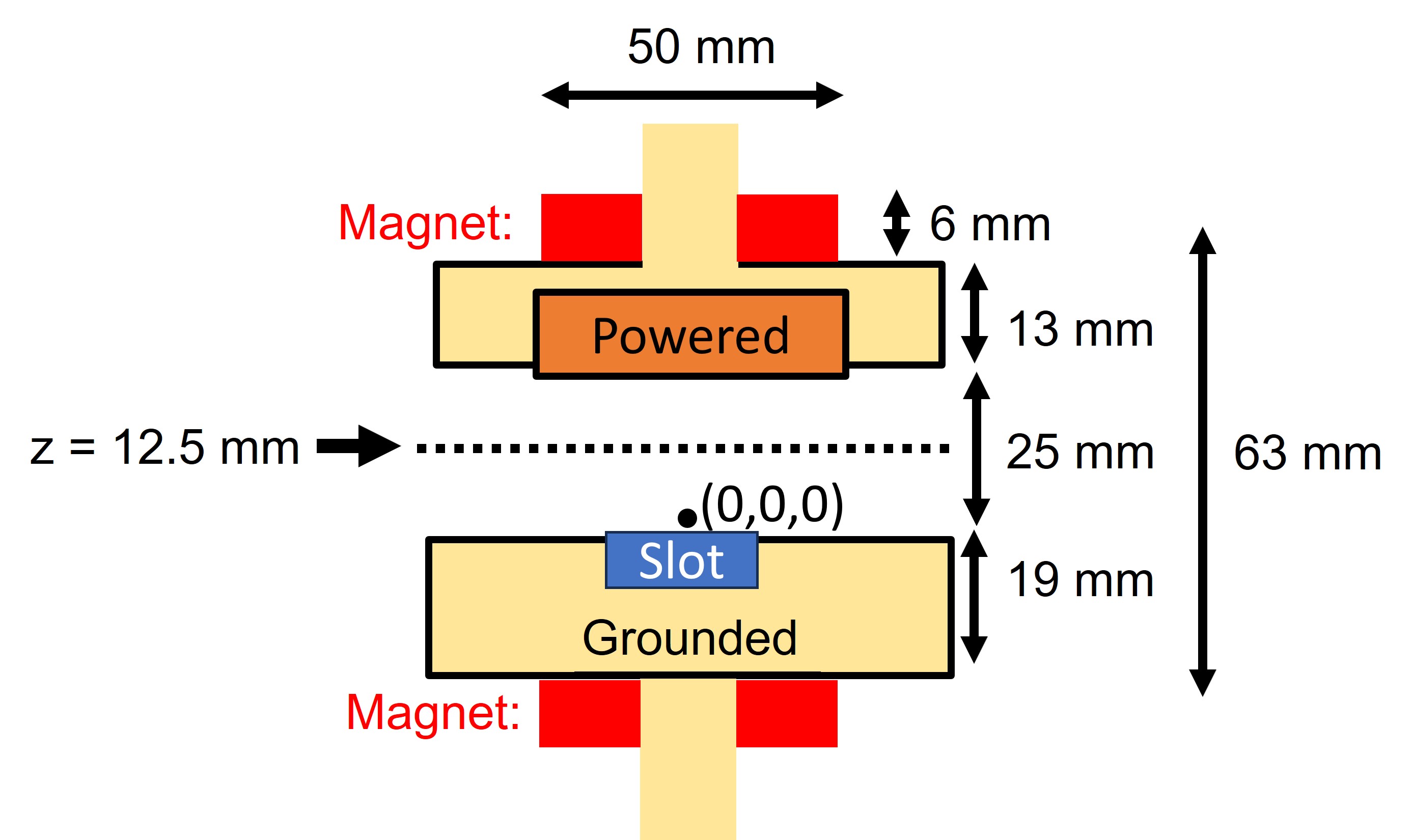}
       \caption{}
       \label{sideview_c}
    \end{subfigure}  
     %\hfill
    \begin{subfigure}{0.45\textwidth}
        \centering        \includegraphics[width=0.7\linewidth]{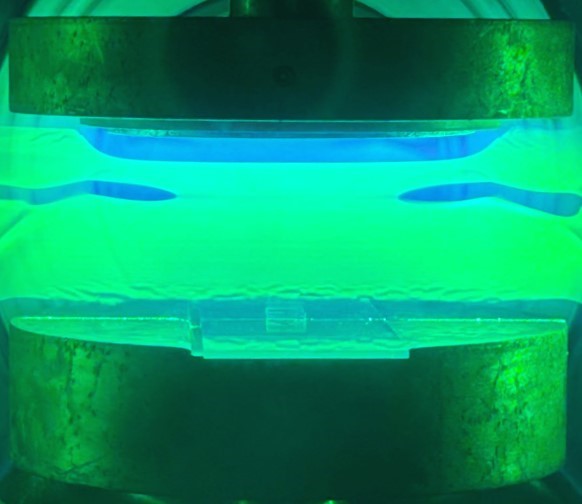}
       \caption{}
       \label{real_picture}
    \end{subfigure}    
    \caption{(Schematics not drawn to scale) (a) Side view showing chamber configurations and laser sheet. (b) Top view showing the laser, and measurement by camera and OES. (c) Side view showing the placement of the permanent magnet rings above and below the top and bottom electrodes respectively. The mid plane between the electrodes at z = 12.5 mm is shown. The coordinate origin (0,0,0) is at the centre of the surface of bottom electrode. (d) A photograph of an Ar/TTIP dusty plasma. Laser light scattering show a dust cloud between the electrode.}
    \label{experimentalsetup}
\end{figure*}

\begin{figure}[ht]
\centering
\begin{tikzpicture}[node distance=0.5cm, auto]
  % Define nodes
  \node [rectangle, draw, minimum width=6.3cm] (start) {Base pressure = 3 $\pm$ 0.3 mTorr};
  \node [rectangle, draw, minimum width=6.3cm, below=of start] (OpenTTIP) {Open TTIP valve or $C_2H_2$ MFC};
  \node [rectangle, draw, minimum width=6.3cm, below=of OpenTTIP] (GasPressure) {Chamber pressure = 35 $\pm$ 3  mtorr};  
  \node [rectangle, draw, below=of GasPressure, minimum width=6.3cm] (OpenAr) {Open argon MFC};
  \node [rectangle, draw, minimum width=6.3cm, below=of OpenAr] (GasPressure2) {Chamber pressure = 45 $\pm$ 3  mTorr};  
  \node [rectangle, draw, below=of GasPressure2, minimum width=6.3cm] (AdjustValve) {Adjust isolation valve};
  \node [diamond, aspect=3.5, draw, below=of AdjustValve] (Decision) {Pressure = 300 $\pm$ 1 mTorr?};
  \node [rectangle, draw, below=of Decision, minimum width=6.3cm] (ON) {Laser,  Camera, and OES ON};
  \node [rectangle, draw, below=of ON, minimum width=6.3cm] (P_ON) {Plasma ON};
  \node [rectangle, draw, below=of P_ON, minimum width=6.3cm] (Collect) {Collect dust when plasma OFF};
  \node [rectangle, draw, below=of Collect, minimum width=6.3cm] (Clean) {Clean chamber};
  % Connect nodes
  \draw [->] (start) -- (OpenTTIP);
  \draw [->] (OpenTTIP) -- (GasPressure);
  \draw [->] (GasPressure) -- (OpenAr);
  \draw [->] (OpenAr) -- (GasPressure2);
  \draw [->] (GasPressure2) -- (AdjustValve);
  \draw [->] (AdjustValve) -- (Decision);
  \draw [->] (Decision) -- node {Yes} (ON);
  \draw [->] (Decision) -- node[pos=0.45] {No} ++(-3.7,0) |- (AdjustValve);
  \draw [->] (ON) -- (P_ON);
  \draw [->] (P_ON) -- (Collect);
  \draw [->] (Collect) -- (Clean);
  \draw [->] (Clean) -- +(-4.2,0) |- (start);
\end{tikzpicture}
\caption{Flowchart to summarize the steps of the experiments.}
\label{flowchart}
\end{figure}
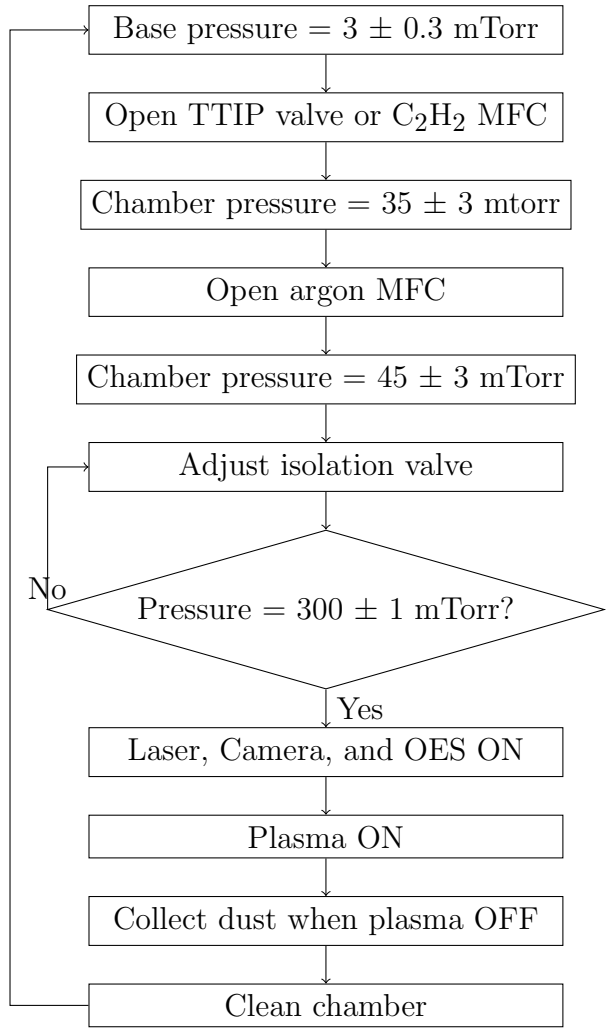

%The particle growth experiments reported in this paper were performed using the experimental setup previously reported in literature \cite{ramkorun2024introducing, Couedel19, jaiswal2020}.
The particle growth chamber is described below. A schematic drawing and photograph of the experiment are shown in Fig. \ref{experimentalsetup}. The vacuum chamber was a 6-way, stainless, ISO-100 cross. Inside was a pair of 75 mm aluminium electrodes, separated by 25 mm, in a parallel plate configuration in order to produce a capacitively coupled RF generated plasma. 50 mm of the top electrode was powered, via  a 300 W, 13.56 MHz (RF VII, Inc RF-3-XIII) fixed frequency RF generator, and a (RF VII, Inc AIM-5/10) auto matching network. This powered region was surrounded by a grounded counter ring with an inner diameter of 50 mm and an outer diameter of 75 mm as shown in Figs. \ref{sideview_a} and \ref{sideview_c}. The lower electrode, also 75 mm in diameter, was grounded. The top and bottom electrode were 13 mm and 19 mm thick respectively. The base pressure was 3.0 $\pm$  0.3  milliTorr (mTorr). The TTIP cylinder and its line connected to the chamber were heated to 75.0  $\pm$ 0.5  $\SI{}{\degreeCelsius}$. A Swagelok VCR metering valve (SS-SVR4-VH) was opened to allow the TTIP pressure to be 35 $ \pm$ 3  mTorr. When carbonaceous dust was grown, the TTIP line was replaced by a $C_2H_2$ line, which was controlled by a mass flow controller (MFC). The flow of $C_2H_2$ as set to 1 standard cubic centimetres per minute (sccm), also bringing the chamber pressure to 35 $ \pm$ 1  mTorr. After the reactive gas started flowing, a different MFC was used to flow Ar into the chamber at 7 sccm. This brought the total chamber pressure to 45 $\pm$ 3  mTorr. The total experimental pressure was then brought up to 300 $\pm$ 1 mTorr by adjusting an isolation valve (Edwards SP Speedivalve) between the chamber and vacuum pump. The RF forward (reflected) power was 60W (1W) in the first 10 seconds and 30W (1W) during the remainder of the experimental time. The higher power was necessary to initiate nucleation in the Ar/TTIP plasma, but cyclic growth was maintained at the lower power. In order to keep the experiment procedure consistent, the same power setting was used for the Ar/$C_2H_2$ plasma. Fig. \ref{flowchart} is a flowchart which summarizes the experimental procedures.

The chamber had 3 quartz viewing ports. A green laser sheet was used to illuminate the dust cloud in two dimensions (2D) from one viewing port. A complementary metal-oxide semiconductor (CMOS) camera (xiQ MQ042MG-CM) was used to record the evolution of the dust cloud during particle growth experiment up to atmost 50 frames per second (FPS) from the second viewing port. OES was used to monitor line intensities between 680 nm and 860 nm from the third viewing port. A broadband spectrometer was used (Avantes ULS4096CL). The OES resolution was 0.59 nm, the slit size was 25 $\mu$m, grating was 600 lines/mm, the integration time was 100 ms, and 5 data points were averaged. The schematics are shown in Fig. \ref{topview}.

\begin{figure}[ht]
    \centering
    
    % Subfigure (a) on the top spanning the full width
    \begin{subfigure}{\linewidth}
        \centering
        \includegraphics[width=0.5\linewidth]{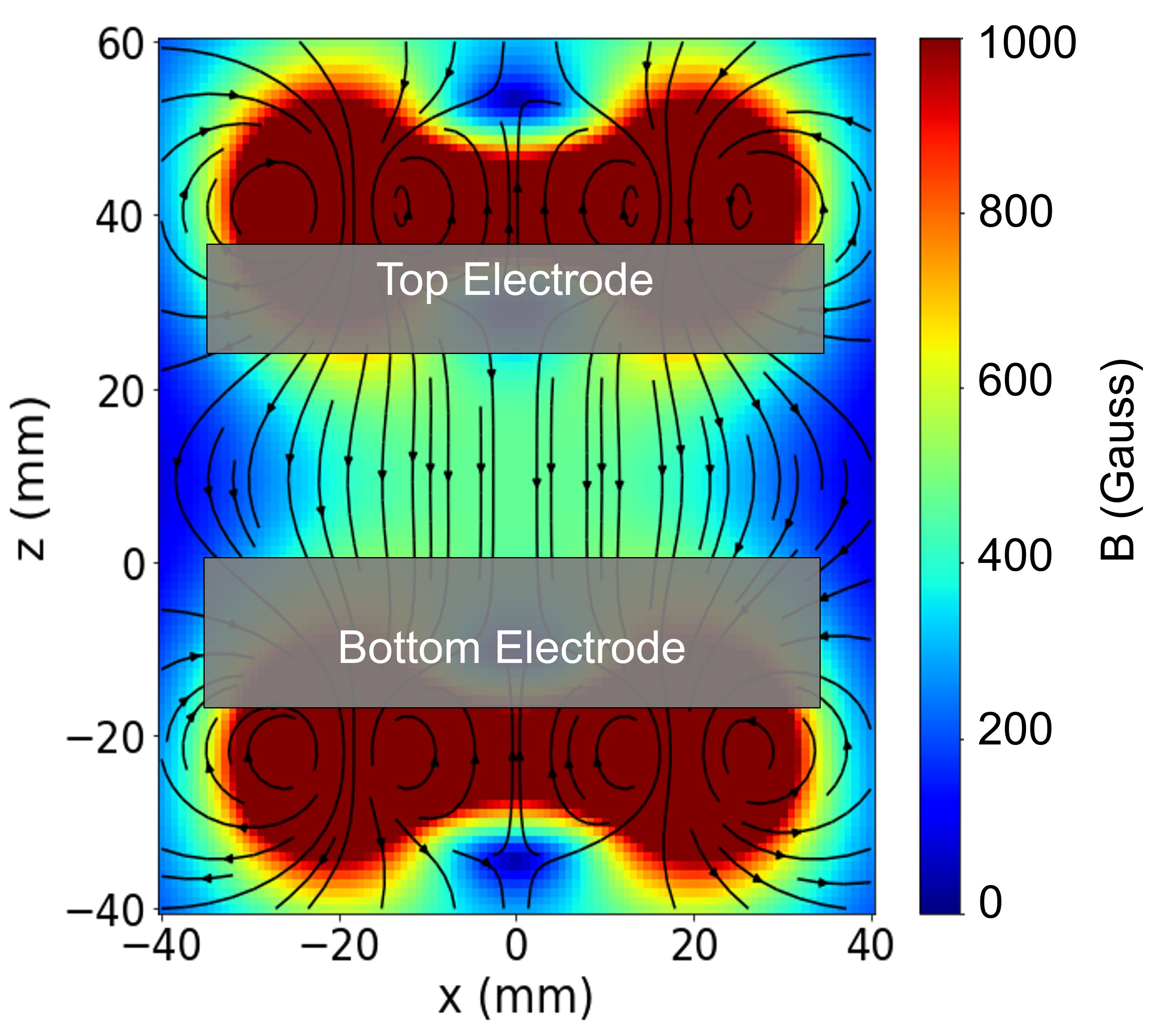}
        \caption{}
        \label{contour_field}
    \end{subfigure}
    
   \vfill % Adds vertical space between the rows
    
    % Subfigures (b) and (c) on the bottom next to each other
    \begin{subfigure}{0.5\linewidth}
        \centering
    \includegraphics[width=\linewidth]{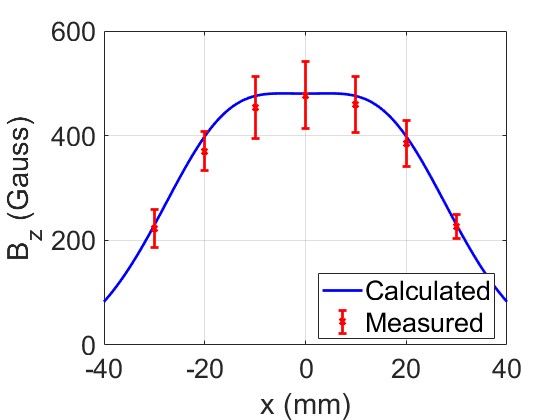}
        \caption{}
        \label{radial_B}
    \end{subfigure}%
    %\hfill
    \begin{subfigure}{0.5\linewidth}
        \centering
    \includegraphics[width=\linewidth]{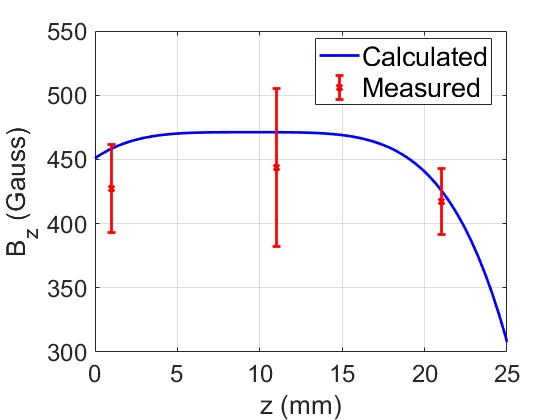}
        \caption{}
        \label{BvsZ}
    \end{subfigure}
    
    \caption{Magnetic field lines between the two permanent magnet rings in the plasma region. (a) Two-dimensional plane at the center of the electrodes. (b) Radial variation of $B_z$ vs x at z = 12.5 mm. (c) Transverse variation of $B_z$ vs z at x = 0 mm.}
    \label{Bfield}
\end{figure}

When experiments were performed using magnetic field, a pair of axially magnetized permanent magnet rings were used.  They were made from neodymium-iron-boron (NdFeB), of grade 42 with a 51 mm outer diameter, 25 mm inner diameter, and 6 mm thickness.  The magnets were placed above and below the plasma electrodes as shown in Fig. \ref{sideview_c}. This produced a nominal magnetic field at the mid-point between the electrodes of $\approx 500$ Gauss. The magnetic field lines distribution in a two-dimensional plane at the center of the electrode is shown in Fig. \ref{contour_field}. The coordinate origin (0,0,0) is at the center of the surface of bottom electrode. The field lines were calculated using magpylib library in Python \cite{ortner2020, pizzey2021}. Moreover, Fig. \ref{radial_B} shows the radial profile of $B_z$ at the middle of the two electrodes, i.e. z = 12.5 mm, and Fig. \ref{BvsZ}  shows the profile of $B_z$ as a function of z at x = 0 mm.  The red values are measured using F.W. Bell Gauss/Teslameter Model 5080 and blue lines are calculated using magpylib. In each individual plasma, the partial pressure of the precursor gas was constant, at $\sim$ 35 mTorr. We assume that within each experiment, the plasma parameters are not changing due to variation in the initial density of precursors. Any changes that will eventually be detected should be due to the magnetic field.

%%%%%%%%%%%%%%%%%%%%%%%%%%%%%%%%%%%%%%%%%%%

\section{Results} 
\label{sec:results}

Material characterisation were not performed on the samples grown. Instead, we rely on knowledge that was previously reported from the same chamber using either Ar/$C_2H_2$ or Ar/TTIP. Material grown in Ar/$C_2H_2$ plasma are purely carbonaceous with aromatic signatures \cite{Couedel19}. Material grown in Ar/TTIP plasma are initially amorphous, but they eventually crystallize into either anatase or rutile after air-annealing at high temperature \cite{ramkorun2024introducing, 10481106}. Carbon was present on the particles grown from Ar/TTIP, most likely due to the organic nature of TTIP \cite{fictorie1994kinetic}; however, there was no indication of carbon bonding with titanium. Therefore, we assume that these dusty plasma are amorphous titania with superficial carbon contamination.

The growth cycles were assessed through three distinct methods: firstly, by measuring the line intensity of Ar I at 760.5 nm via OES; secondly, by capturing images of laser light scattered from the dust cloud using a CMOS camera; and thirdly, by analyzing the dust size distribution for particles grown at different length of time. Each of these processes is described below as they are applied to this study.

\subsection{Optical Emission Spectroscopy}

\begin{figure*}[ht]
  \begin{minipage}{0.48\textwidth}
    \centering
    \begin{subfigure}{\linewidth}
      \includegraphics[width=\linewidth]{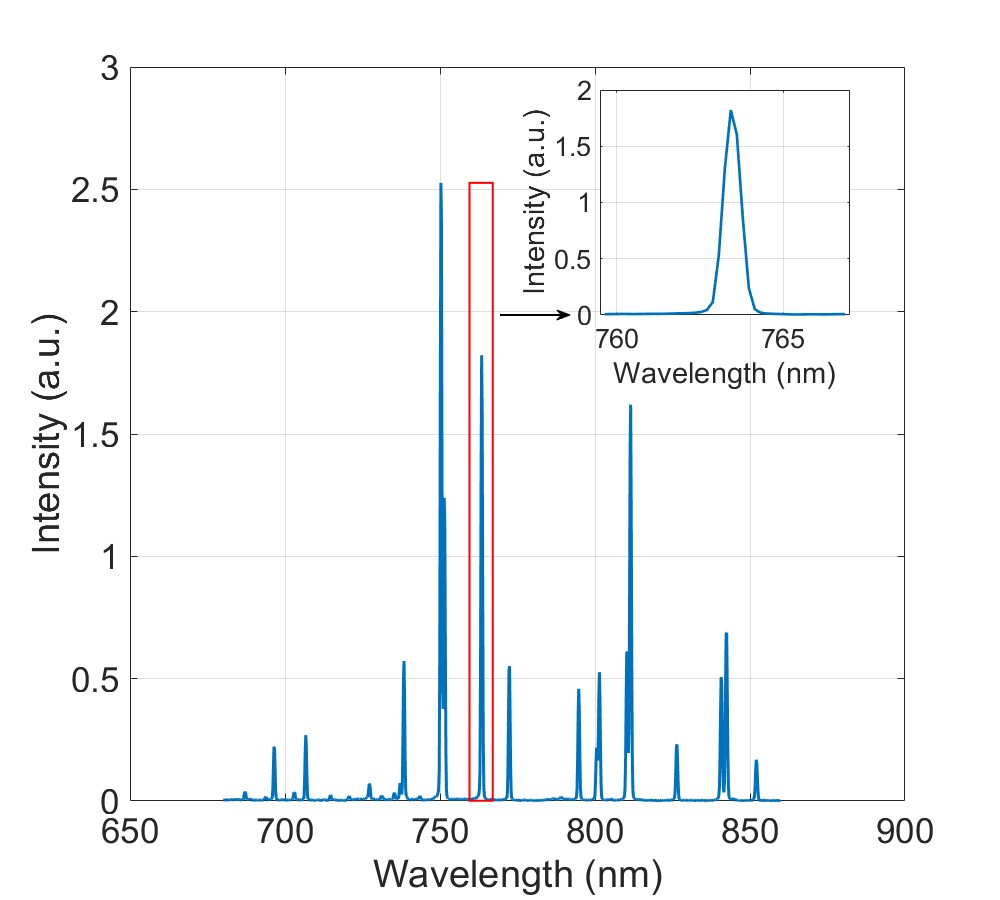}
      \caption{}
      \label{surveyOES}
    \end{subfigure}
  \end{minipage}%
  \begin{minipage}{0.51\textwidth}
    \begin{subfigure}{0.49\linewidth}
      \centering
      \includegraphics[width=\linewidth]{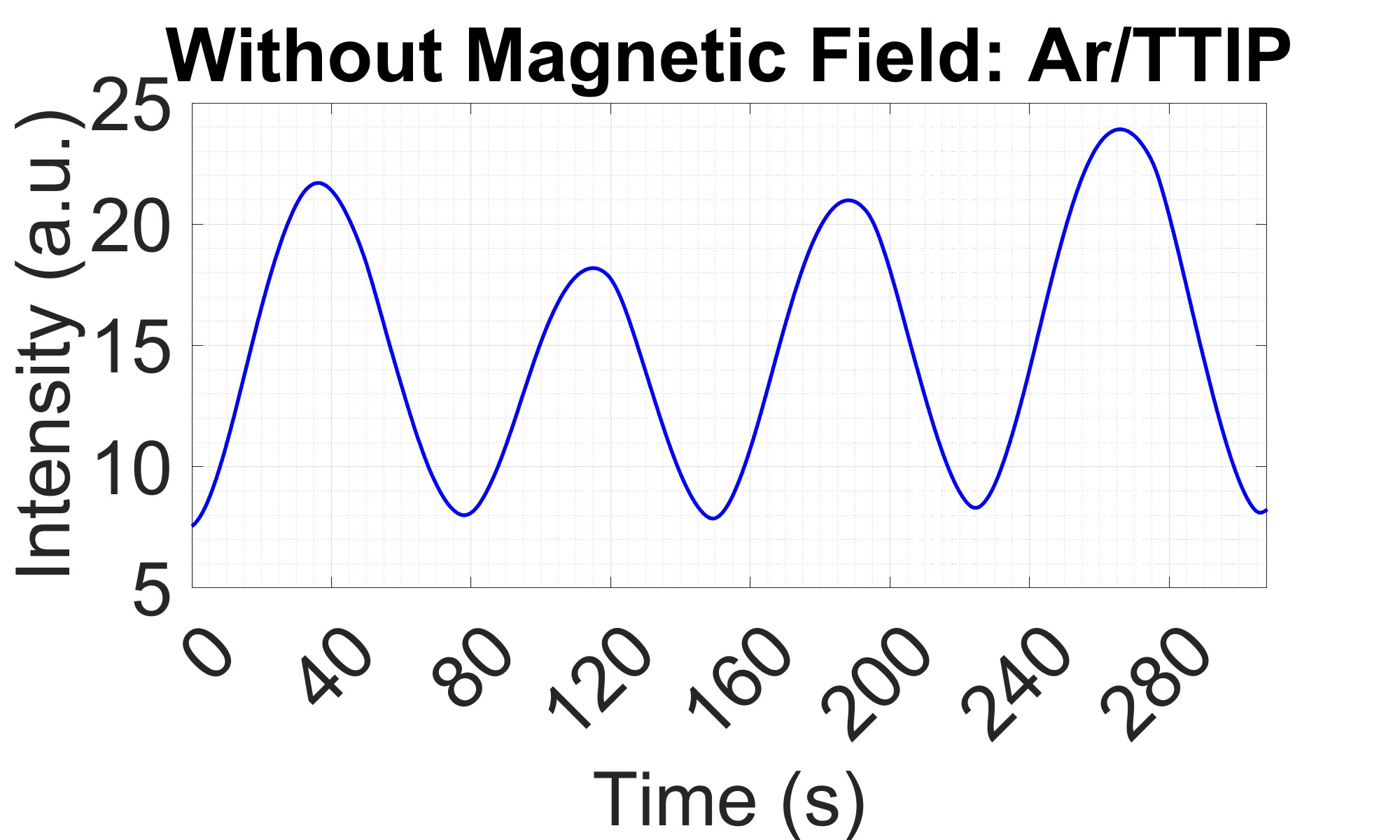}
      \caption{}
      \label{OES_TTIP_B0}
    \end{subfigure}%
    \begin{subfigure}{0.49\linewidth}
      \centering
      \includegraphics[width=\linewidth]{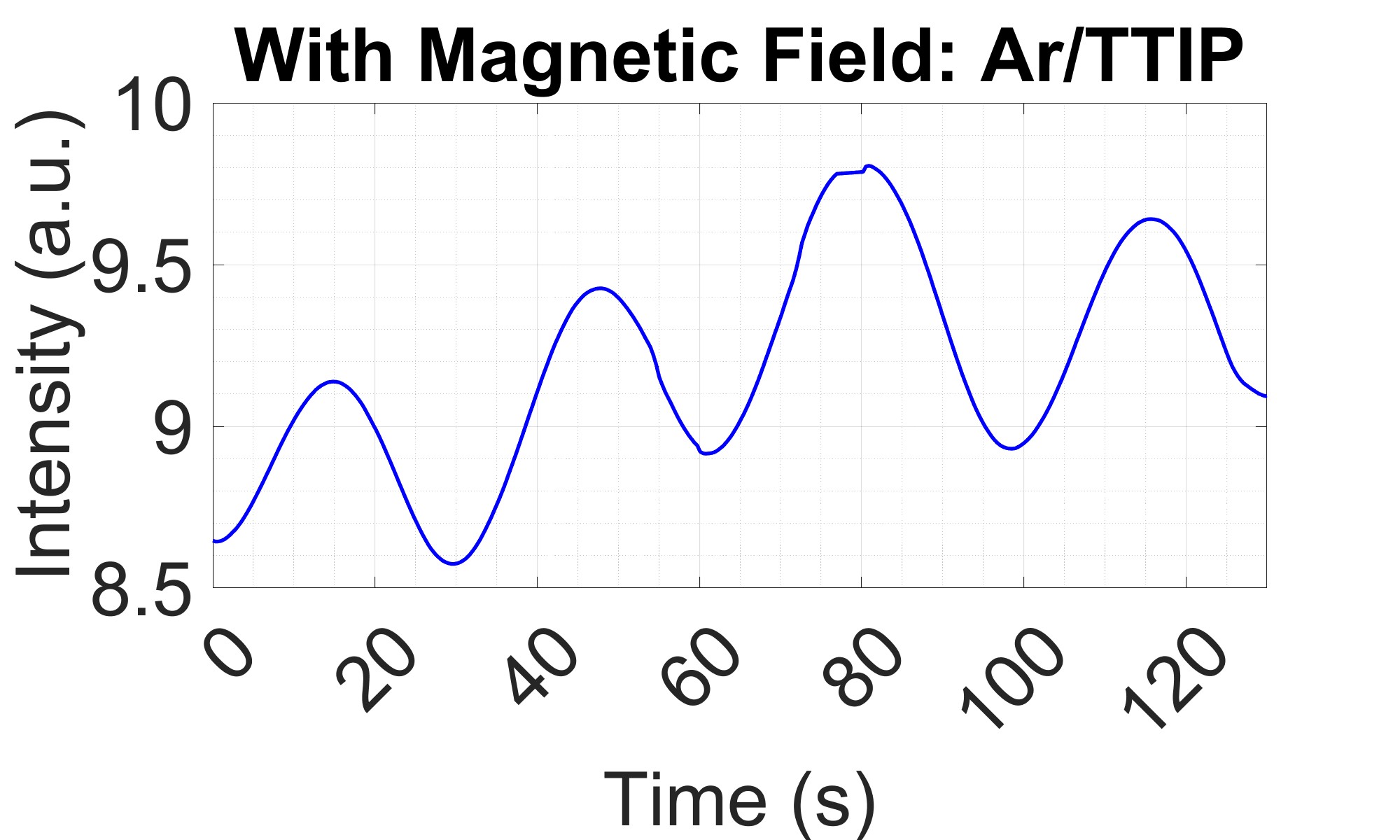}
      \caption{}
      \label{OES_TTIP_B500}
    \end{subfigure}
    
    \vspace{1em} % Adjust the vertical space between rows
    
    \begin{subfigure}{0.49\linewidth}
      \centering
      \includegraphics[width=\linewidth]{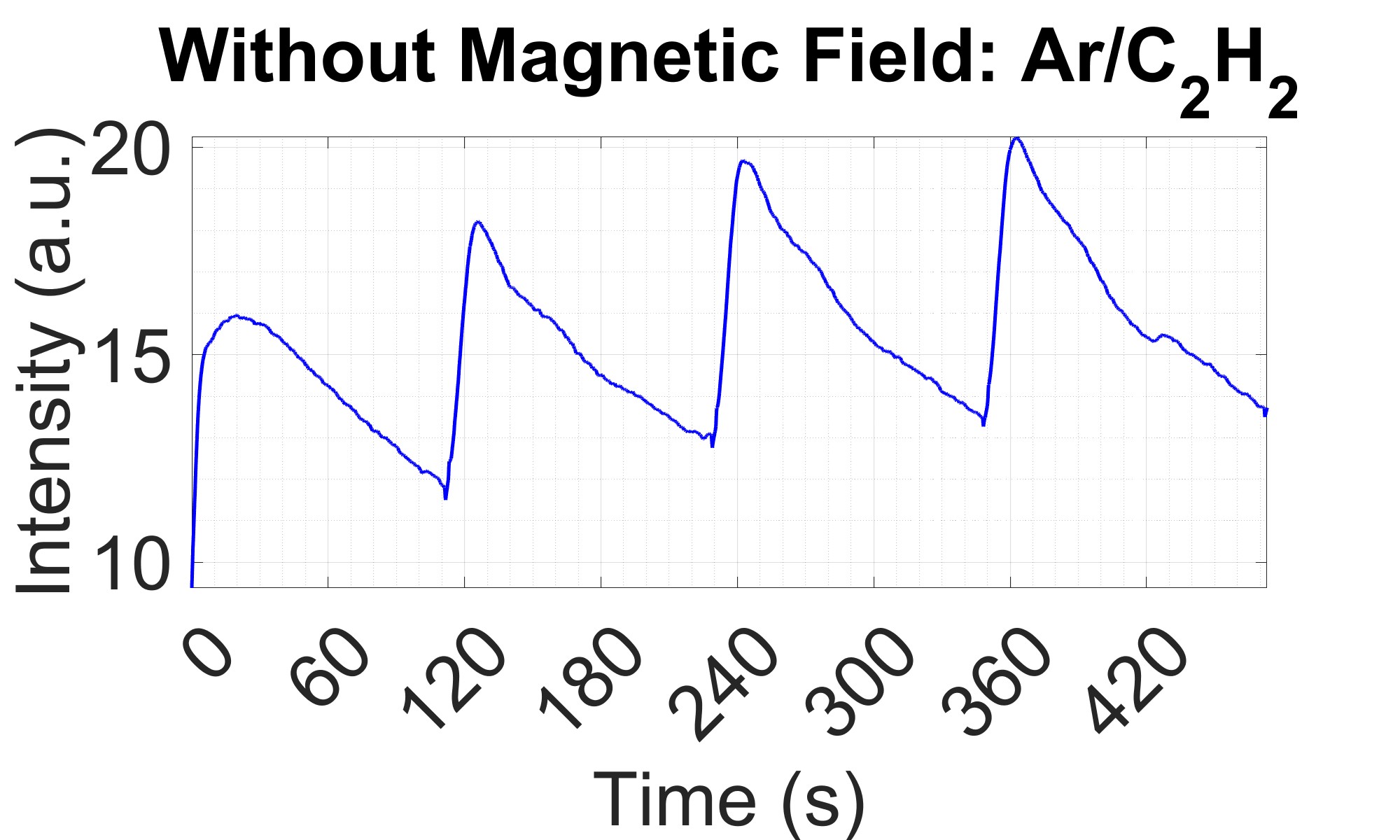}
      \caption{}
      \label{OES_C2H2_B0}
    \end{subfigure}%
    \begin{subfigure}{0.49\linewidth}
      \centering
      \includegraphics[width=\linewidth]{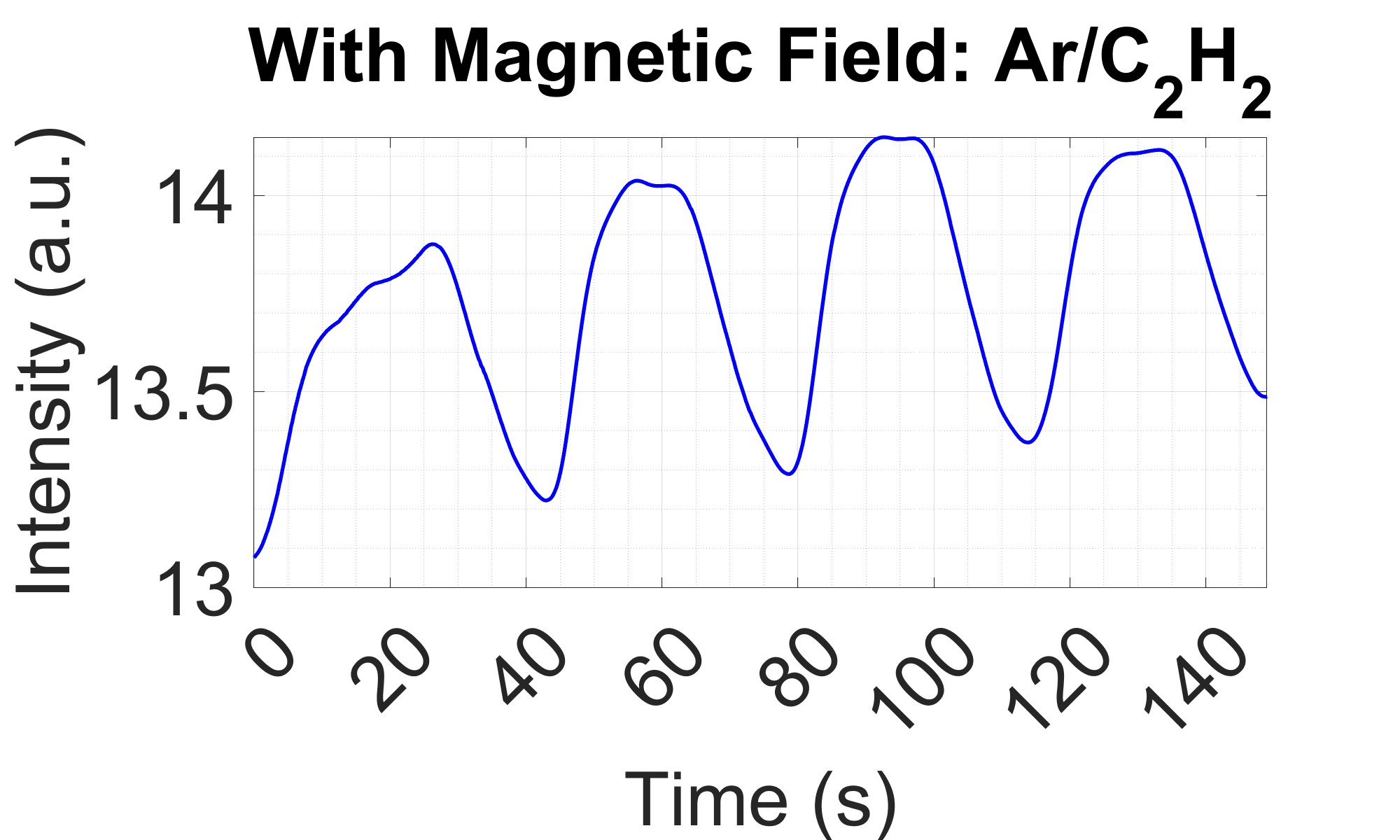}
      \caption{}
      \label{OES_C2H2_B500}
    \end{subfigure}
  \end{minipage}
  \caption{ OES of the plasma. No new lines were seen during the presence of either TTIP or $C_2H_2$ in the plasma. (a) Survey scan between 680 and 860 nm revealing several Ar I peaks. The  line at 763.5 nm is an isolated peak.  Cyclic variation of the intensity of Ar I (763.5 nm) line: Ar/TTIP dusty plasma (b) without magnetic field $(77 \pm 4 s)$ and (c) with magnetic field $(32 \pm 3 s)$, Ar/$C_2H_2$ dusty plasma (d) without magnetic field $(115 \pm  s)$  and (e) with magnetic field $(39 \pm 1 s)$.}
\end{figure*}

The OES is obtained in our experiments using a broadband spectrometer. An example of a time-resolved spectra as shown in Fig. \ref{surveyOES} whereby several Ar peaks are identified. The electron temperature could not be calculated from the line ratio of different Ar emission lines because the spectrometer has relatively low resolution (0.59 nm) causing several lines to blend as seen in Fig. 6a. Furthermore, the intensity measurements of the spectrometer have not been calibrated. Nevertheless, the cyclic variation in Ar line intensity, which mimics the particle growth cycle, is reported. In particular, while it is observed that all of the emission lines vary as a function of time during the growth process, an examination of the (NIST) database reveals that the argon neutral (Ar I) line at 763.5 nm is a promising candidate for OES analysis. This line was chosen because it was the highest intensity line according to the survey scan without any other line blending.  As seen in the inset in Fig. \ref{surveyOES} - and confirmed by NIST database - this line, which was the electric dipole transition of Ar I from $3s^23p^5\left(^2P^0_{3/2}\right)4p$ to $3s^23p^5\left(^2P^0_{3/2}\right)4s$ was relatively isolated \cite{norlen1973, wiese1989, NIST_ASD}.   When the reactive gas was present, no new lines corresponding to emission of new species were seen. For example, it was anticipated to detect lines corresponding to either carbon and/or hydrogen arising from the reactive gas(es). The spectrometer may have not detected these lines either due its relatively low resolution and/or due to the higher flow rate of Ar increasing its intensity when compared to other species in the plasma \cite{ramkorun2019effect}.

The timeframe of the OES was set from its integration time and number of data averaged. Data was collected roughly every 600 ms. The time when the plasma was turned on was set to zero. For the Ar/TTIP dusty plasma, without (with) the presence of the magnetic field, the measured cycle time was 77 $\pm$ 4 s (32 $\pm$ 3 s), as shown in Fig. \ref{OES_TTIP_B0} and \ref{OES_TTIP_B500} respectively. For the Ar/$C_2H_2$ dusty plasma, without (with) the presence of the magnetic field, the cycle time was 115 $\pm$ 5 s (39 $\pm$ 1 s), as shown in Fig. \ref{OES_C2H2_B0} and \ref{OES_C2H2_B500} respectively. Studies have shown that the growth rate of dusty nanoparticles depends on the gas temperature of the reactive gas(es) \cite{stoffels2008charge, beckers2009temperature, beckers2011surprising}. Here, the gas temperature of $C_2H_2$ and TTIP are $\sim \SI{20}{\degreeCelsius}$  (room temperature), and $\sim \SI{75}{\degreeCelsius}$ respectively. Thus, the growth rate, and consequently the cycle time, of the two dusty plasma can not be directly compared to each other. Nevertheless, the individual gas temperatures are unchanged during the experiments both with and without the presence of the magnetic field. Therefore, we believe that the change in the cycle time is predominantly due to the presence of the magnetic field.

Literature suggests that instabilities, in the form of filamentary modes and azimuthally rotating voids, may arise towards the end of the growth cycle \cite{samsonov1999instabilities}  or from dust particles laying on the surface of the electrodes \cite{mikikian2003formation}. Here, the instabilities were not present. Were instabilities to arise, they would have been detected as rapid fluctuations in OES intensity.

\subsection{Laser light scattering}

\begin{figure*}[!ht]
    \centering
    % Column headings
    \begin{minipage}[b]{0.19\textwidth}
        \centering
        \textbf{$\mathbf{T_c}$/4}
    \end{minipage}
    \begin{minipage}[b]{0.19\textwidth}
        \centering
        \textbf{$\mathbf{T_c}$/2}
    \end{minipage}
    \begin{minipage}[b]{0.19\textwidth}
        \centering
        \textbf{3$\mathbf{T_c}$/4}
    \end{minipage}
    \begin{minipage}[b]{0.19\textwidth}
        \centering
        \textbf{$\mathbf{T_c}$}
    \end{minipage}

    \vspace{0.5cm} % Space between headings and first row of figures

    % First row with label
    \raisebox{1.7cm}{%
        \makebox[0pt][r]{%
            \parbox{80pt}{%
                \raggedleft
                \textbf{Without Magnetic} \\
                \textbf{Field:} \\
                \textbf{Ar/TTIP}
            }%
            \hspace{0.1cm}%
        }
    }
    \begin{subfigure}[b]{0.19\textwidth}
        \centering
        \includegraphics[width=\textwidth]{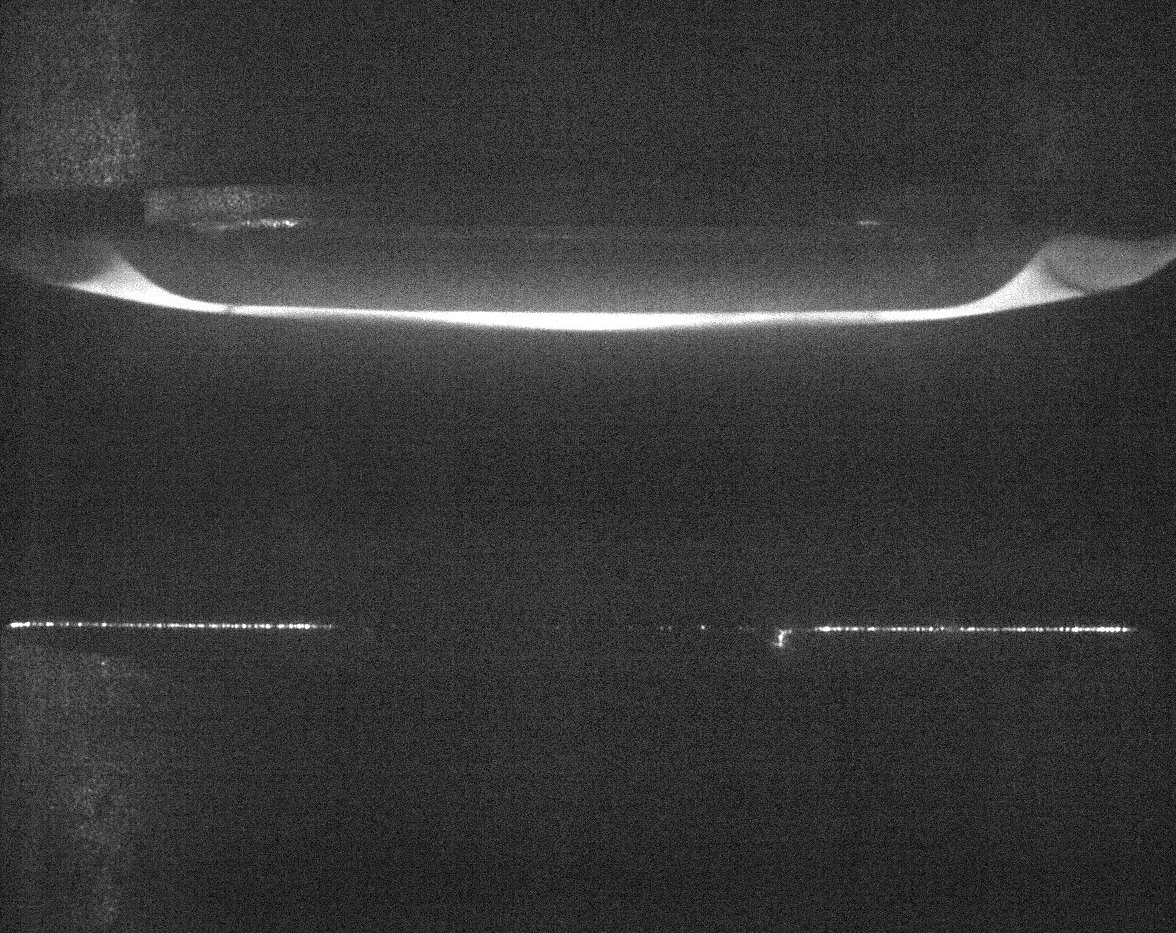}
        \caption{}
        \label{fig:t14}
    \end{subfigure}
    \begin{subfigure}[b]{0.19\textwidth}
        \centering
        \includegraphics[width=\textwidth]{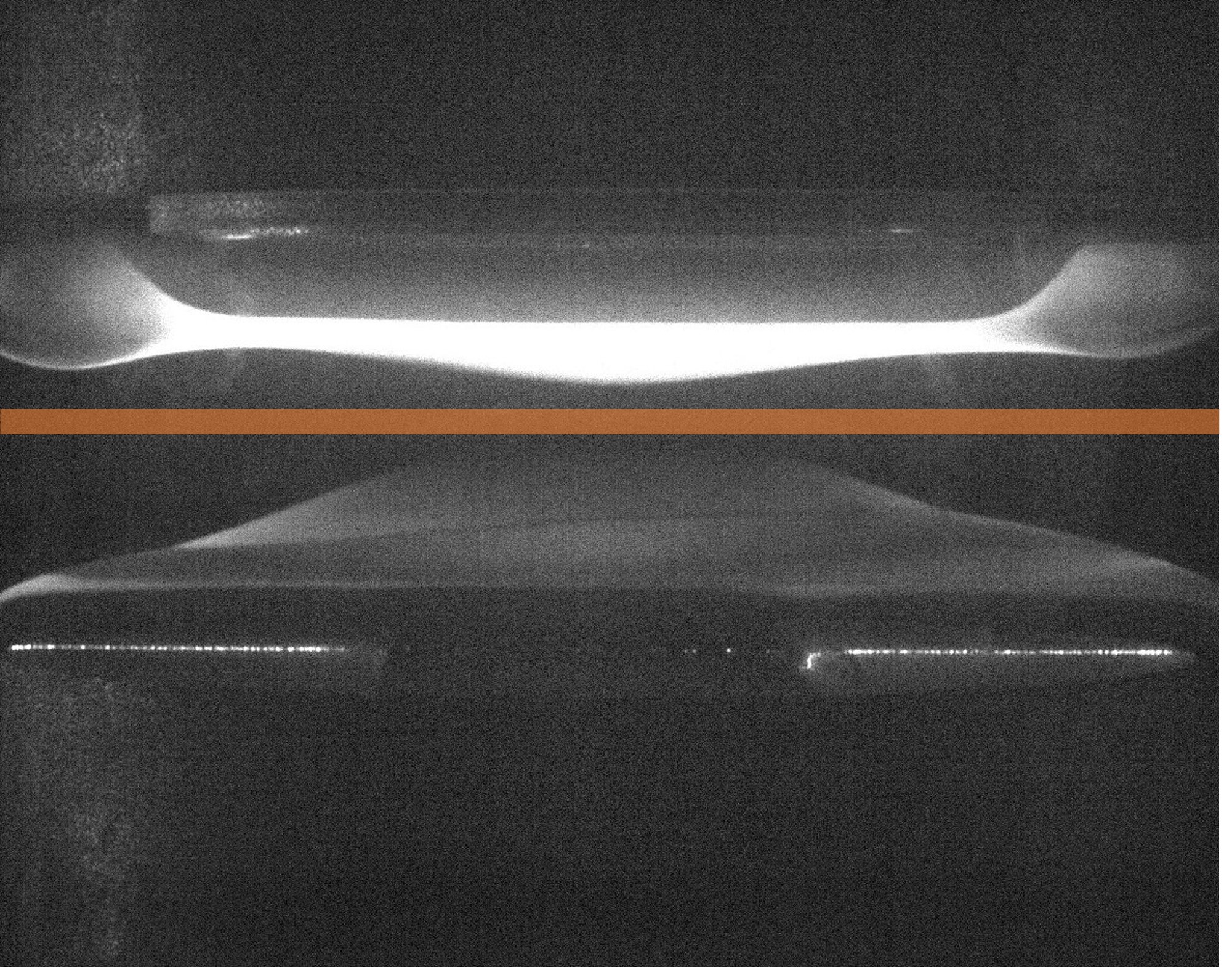}
        \caption{}
        \label{fig:t12}
    \end{subfigure}
    \begin{subfigure}[b]{0.19\textwidth}
        \centering
        \includegraphics[width=\textwidth]{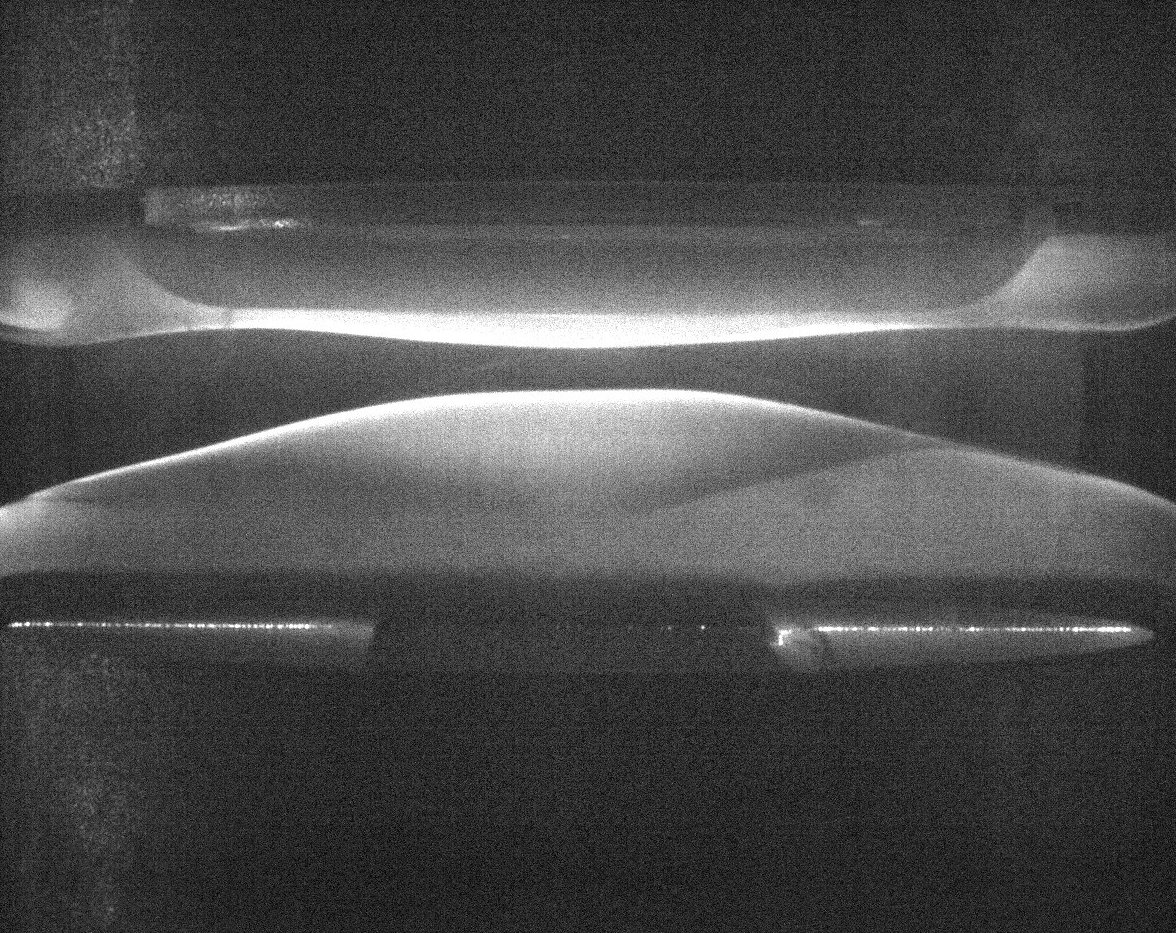}
        \caption{}
        \label{fig:t34}
    \end{subfigure}
    \begin{subfigure}[b]{0.19\textwidth}
        \centering
        \includegraphics[width=\textwidth]{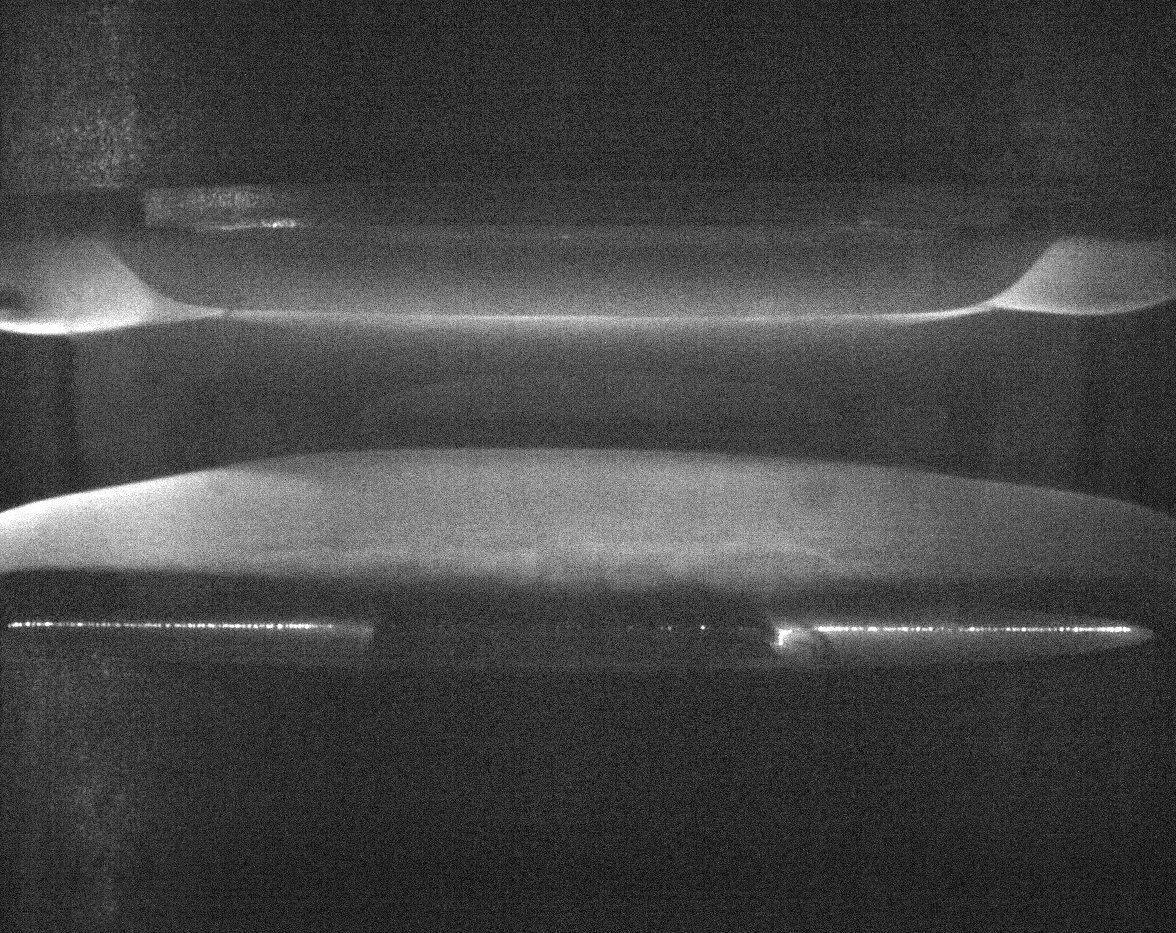}
        \caption{}
        \label{fig:t1}
    \end{subfigure}

    \vspace{0.5cm} % Space between rows

    % Second row with label
    \raisebox{1.6cm}{%
        \makebox[0pt][r]{%
            \parbox{80pt}{%
                \raggedleft
                \textbf{Without} \\
                \textbf{Magnetic} \\
                \textbf{Field:} \\
                \textbf{Ar/$C_2H_2$}
            }%
            \hspace{0.1cm}%
        }
    }
    \begin{subfigure}[b]{0.19\textwidth}
        \centering
        \includegraphics[width=\textwidth]{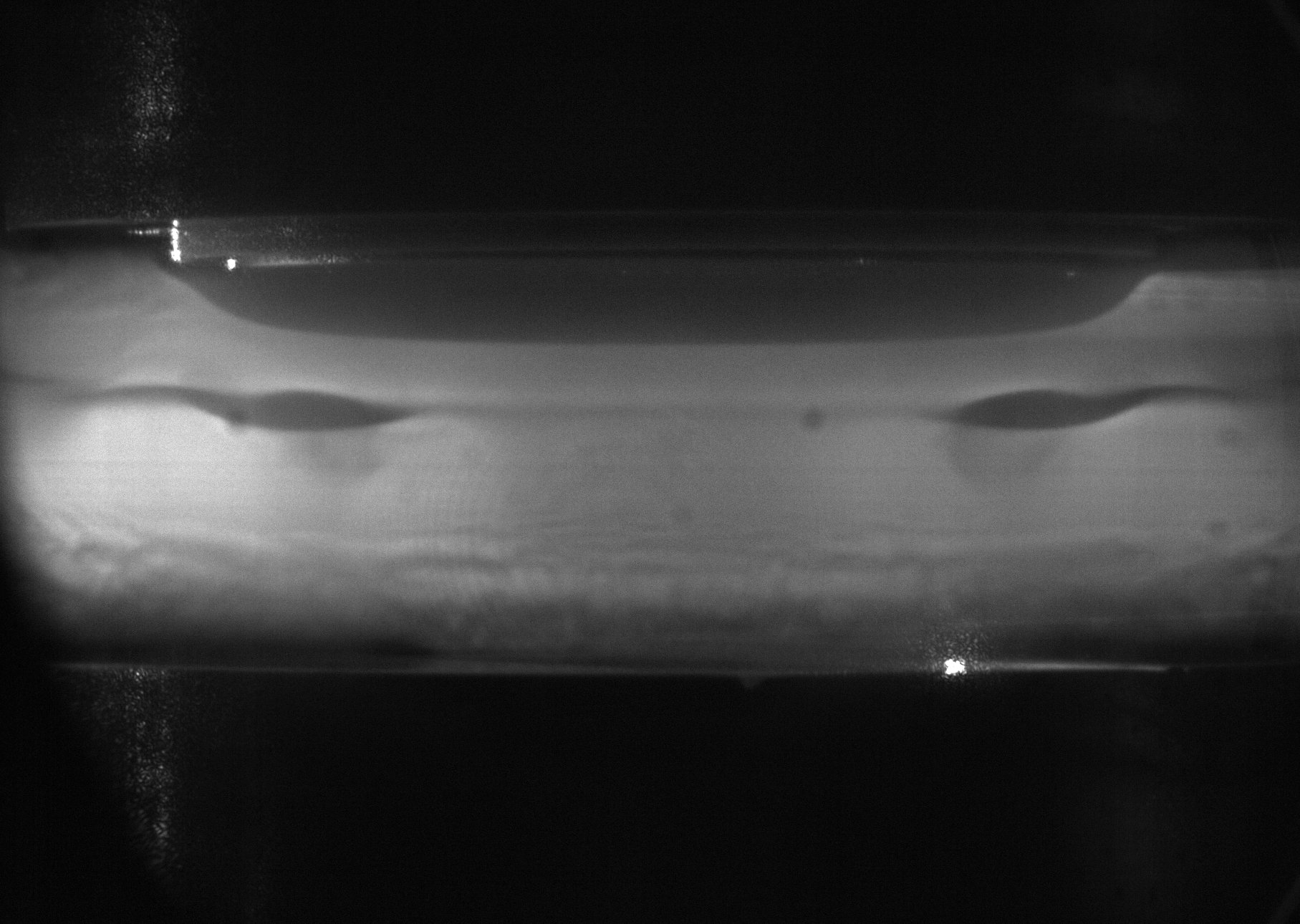}
        \caption{}
        \label{fig:c14}
    \end{subfigure}
    \begin{subfigure}[b]{0.19\textwidth}
        \centering
        \includegraphics[width=\textwidth]{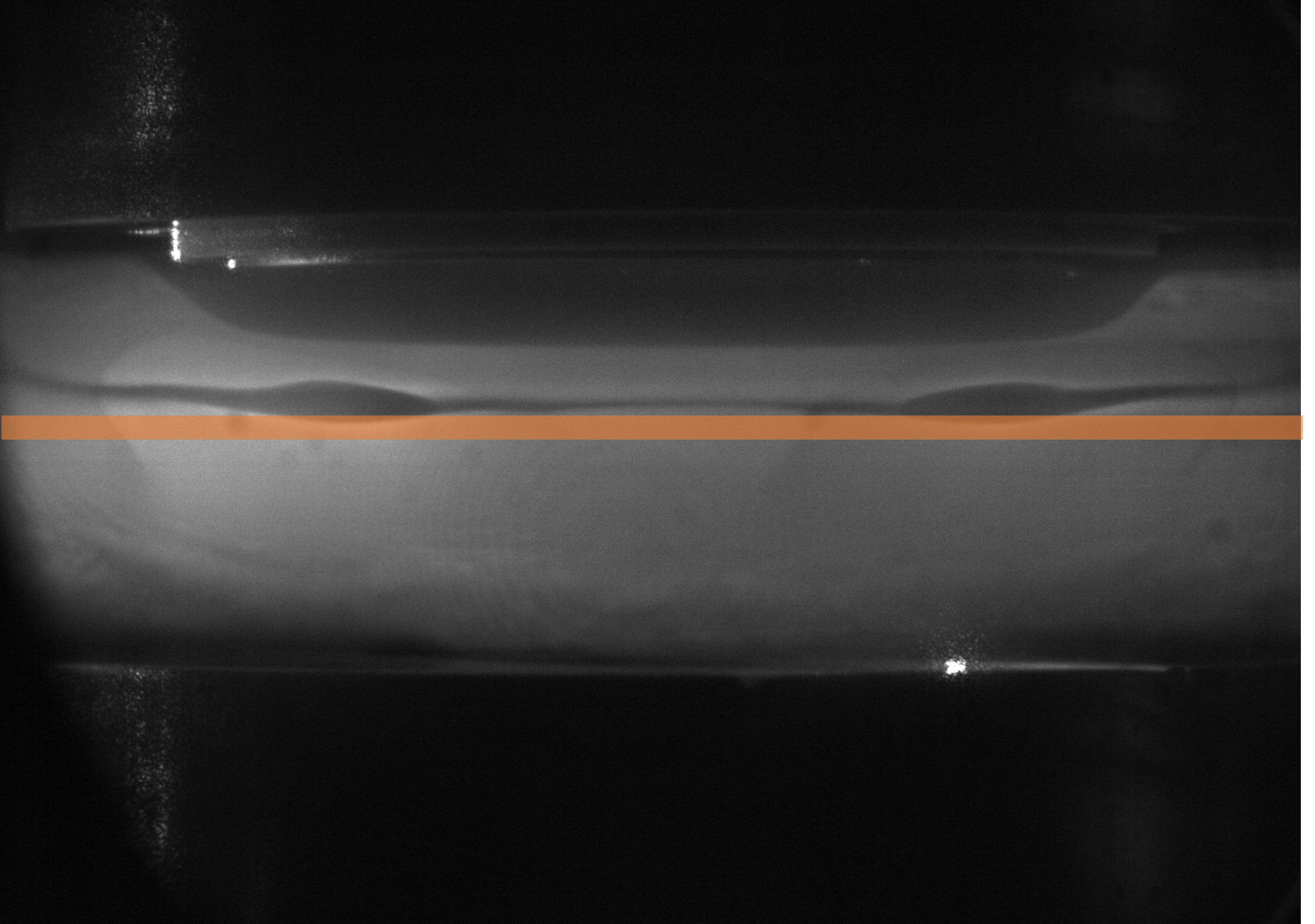}
        \caption{}
        \label{fig:c12}
    \end{subfigure}
    \begin{subfigure}[b]{0.19\textwidth}
        \centering
        \includegraphics[width=\textwidth]{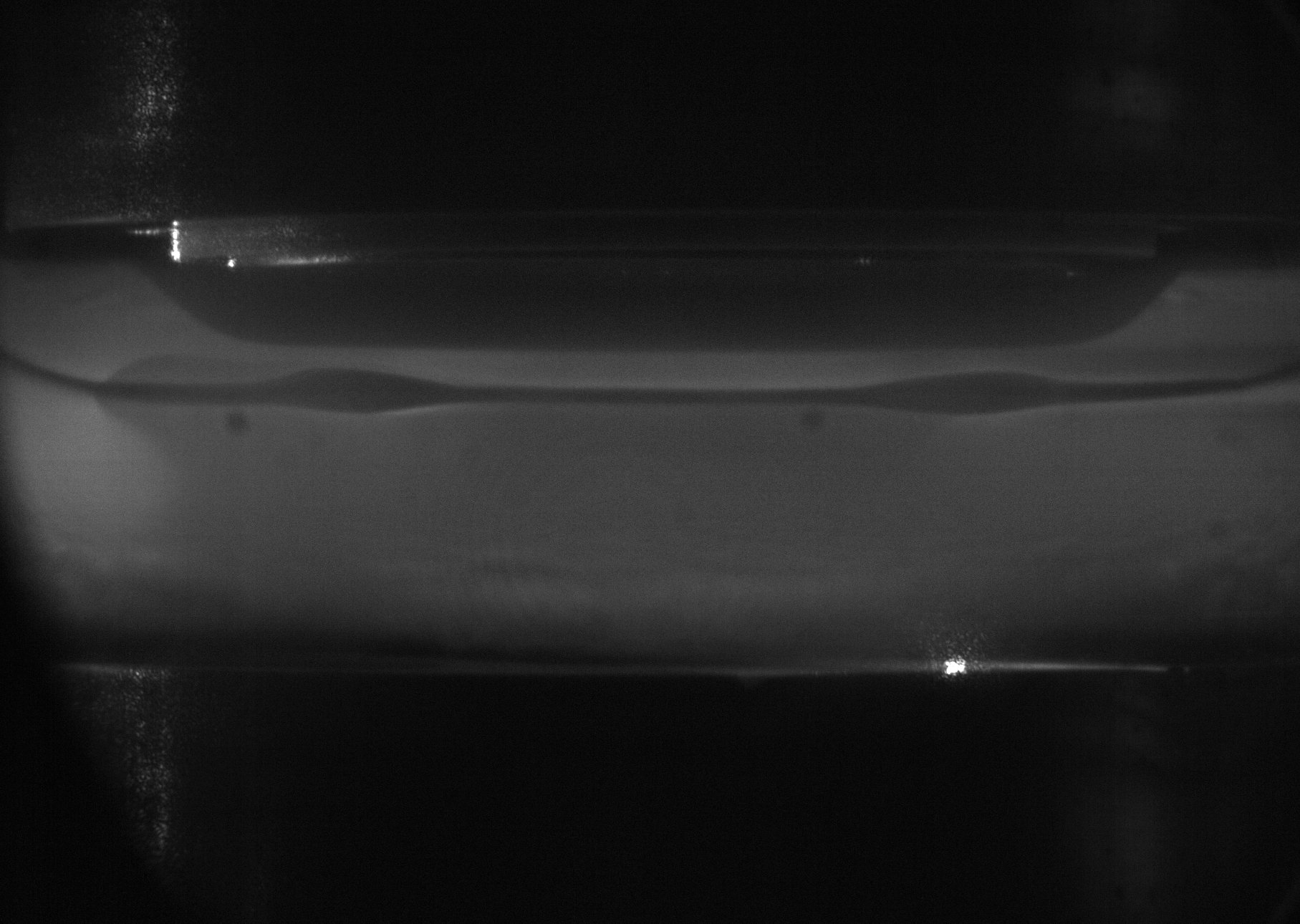}
        \caption{}
        \label{fig:c34}
    \end{subfigure}
    \begin{subfigure}[b]{0.19\textwidth}
        \centering
        \includegraphics[width=\textwidth]{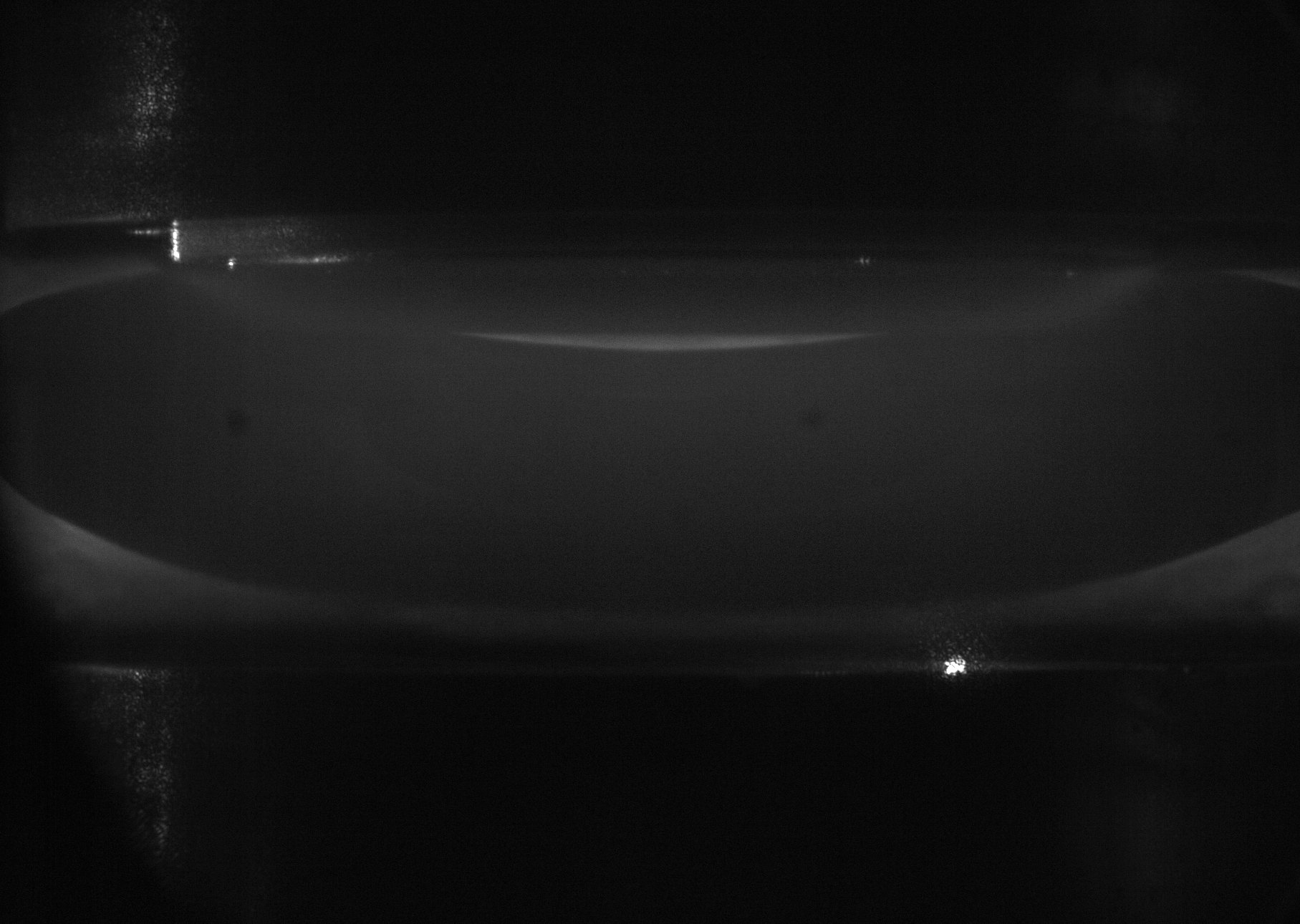}
        \caption{}
        \label{fig:c1}
    \end{subfigure}

    \vspace{0.5cm} % Space between rows

    % Third row with label
    \raisebox{1.8cm}{%
        \makebox[0pt][r]{%
            \parbox{80pt}{%
                \raggedleft
                \textbf{With} \\
                \textbf{Magnetic} \\
                \textbf{Field:} \\
                \textbf{Ar/TTIP}
            }%
            \hspace{0.1cm}%
        }
    }
    \begin{subfigure}[b]{0.19\textwidth}
        \centering
        \includegraphics[width=\textwidth]{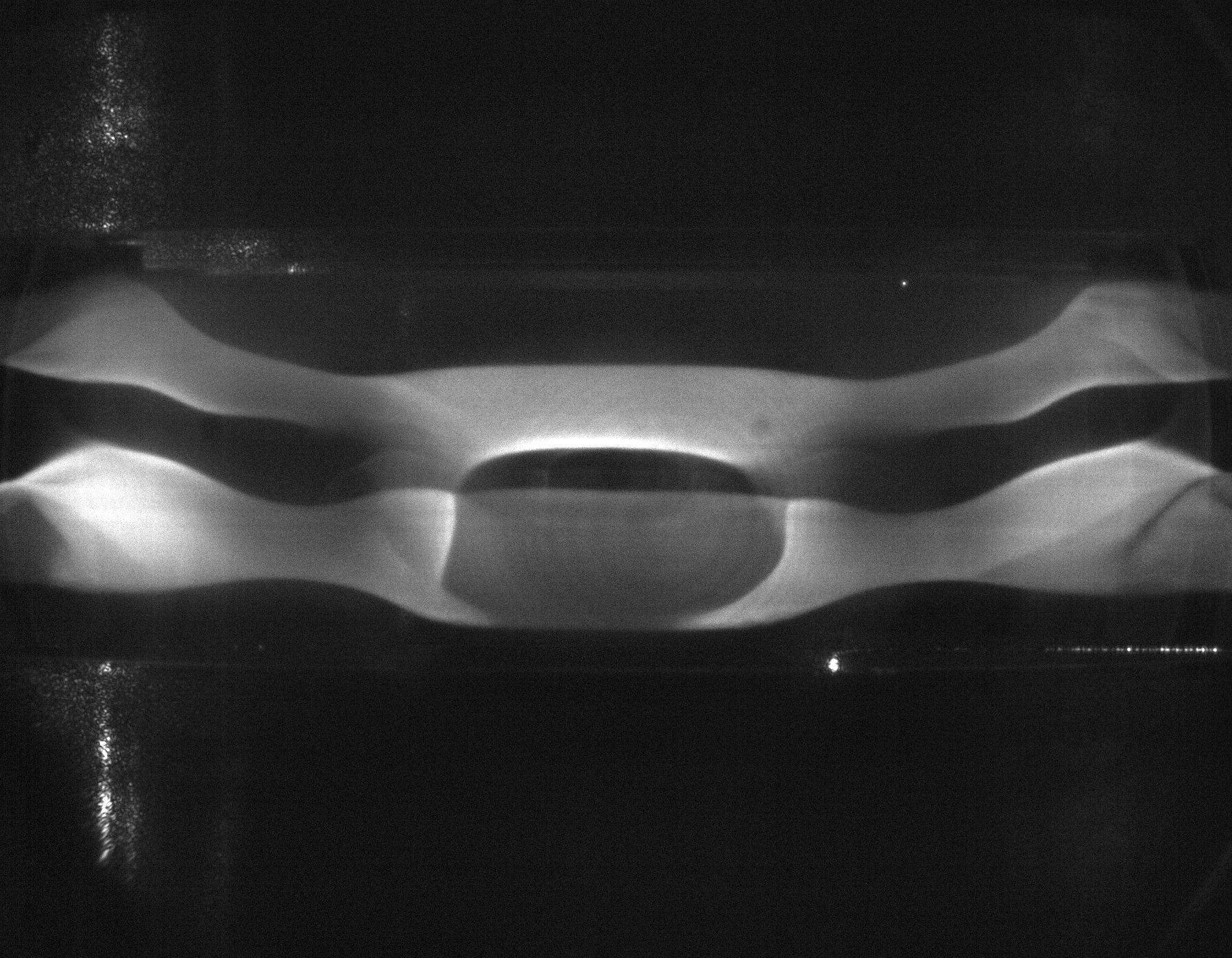}
        \caption{}
        \label{fig:bt14}
    \end{subfigure}
    \begin{subfigure}[b]{0.19\textwidth}
        \centering
        \includegraphics[width=\textwidth]{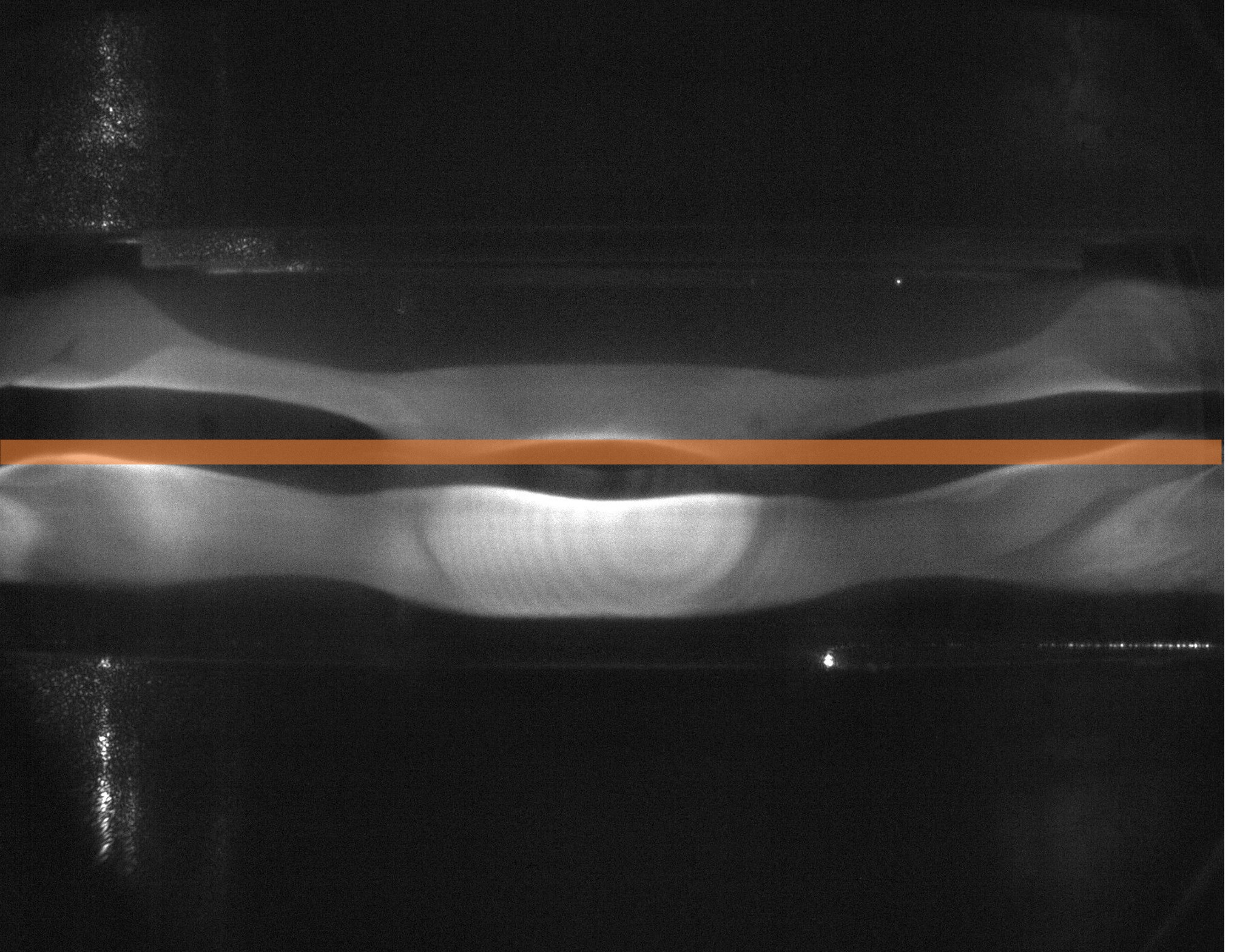}
        \caption{}
        \label{fig:bt12}
    \end{subfigure}
    \begin{subfigure}[b]{0.19\textwidth}
        \centering
        \includegraphics[width=\textwidth]{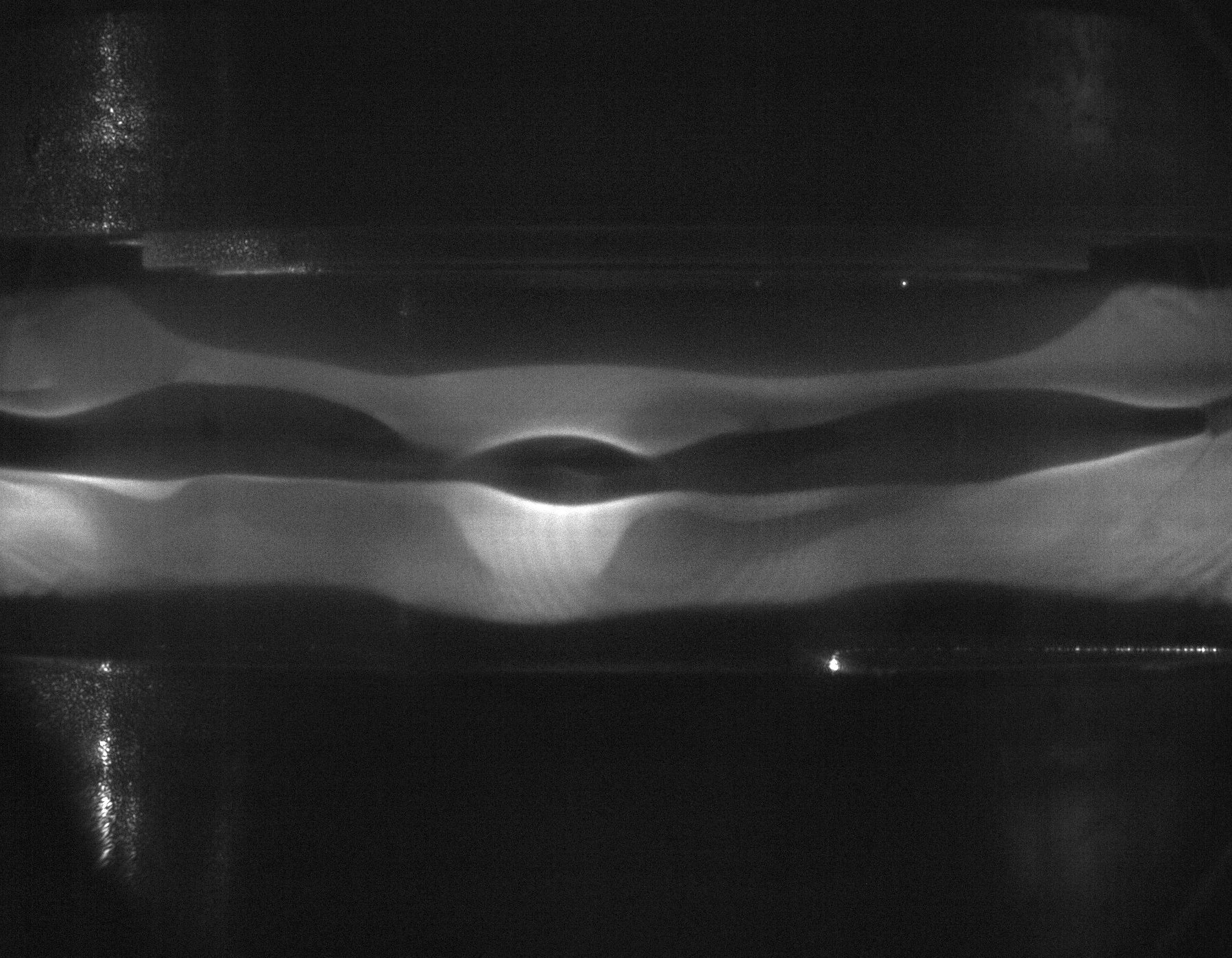}
        \caption{}
        \label{fig:bt34}
    \end{subfigure}
    \begin{subfigure}[b]{0.19\textwidth}
        \centering
        \includegraphics[width=\textwidth]{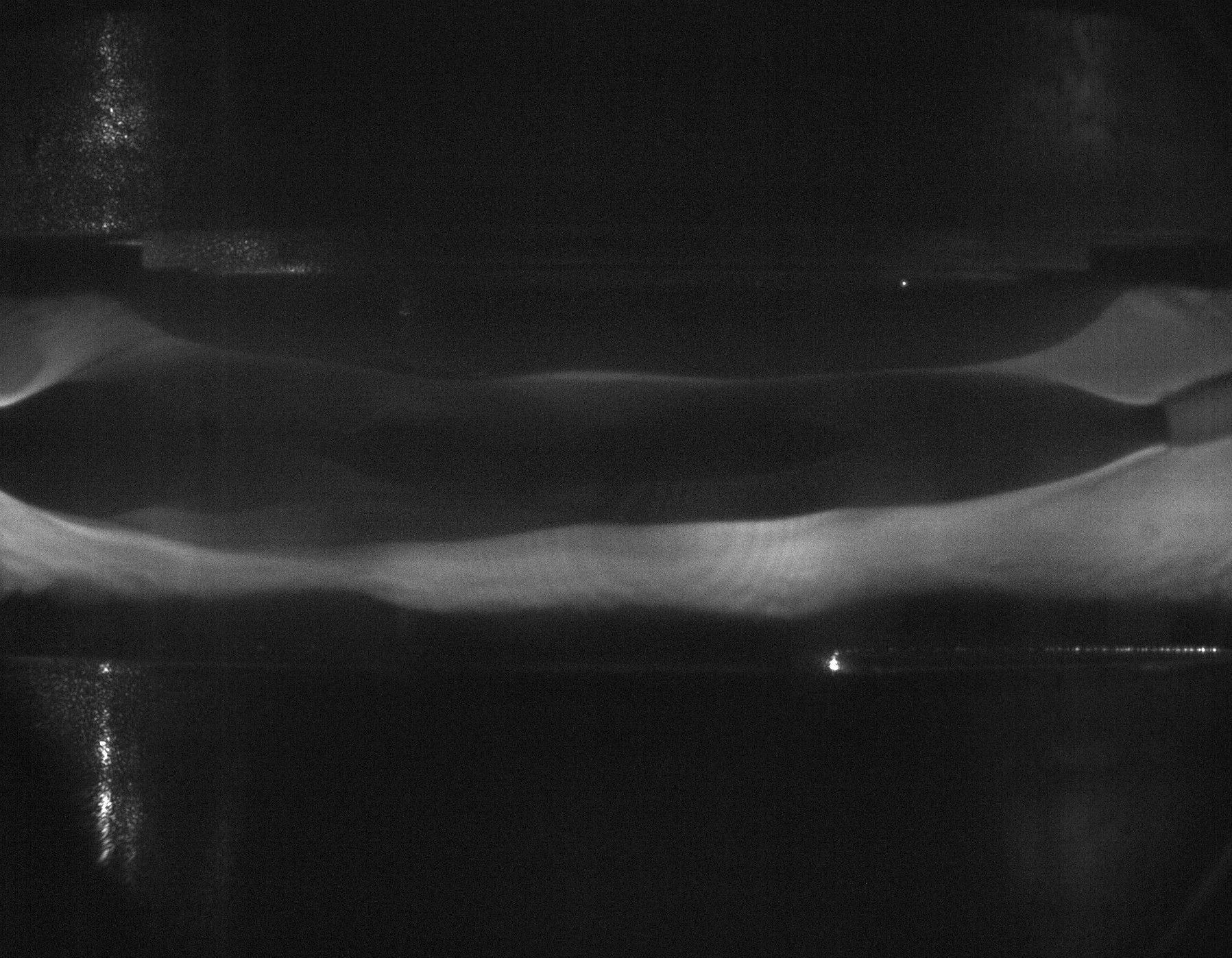}
        \caption{}
        \label{fig:bt1}
    \end{subfigure}

    \vspace{0.5cm} % Space between rows

    % Fourth row with label
    \raisebox{1.6cm}{%
        \makebox[0pt][r]{%
            \parbox{80pt}{%
                \raggedleft
                \textbf{With } \\
                \textbf{Magnetic}\\
                \textbf{Field:} \\
                \textbf{Ar/$C_2H_2$}
            }%
            \hspace{0.1cm}%
        }
    }
    \begin{subfigure}[b]{0.19\textwidth}
        \centering
        \includegraphics[width=\textwidth]{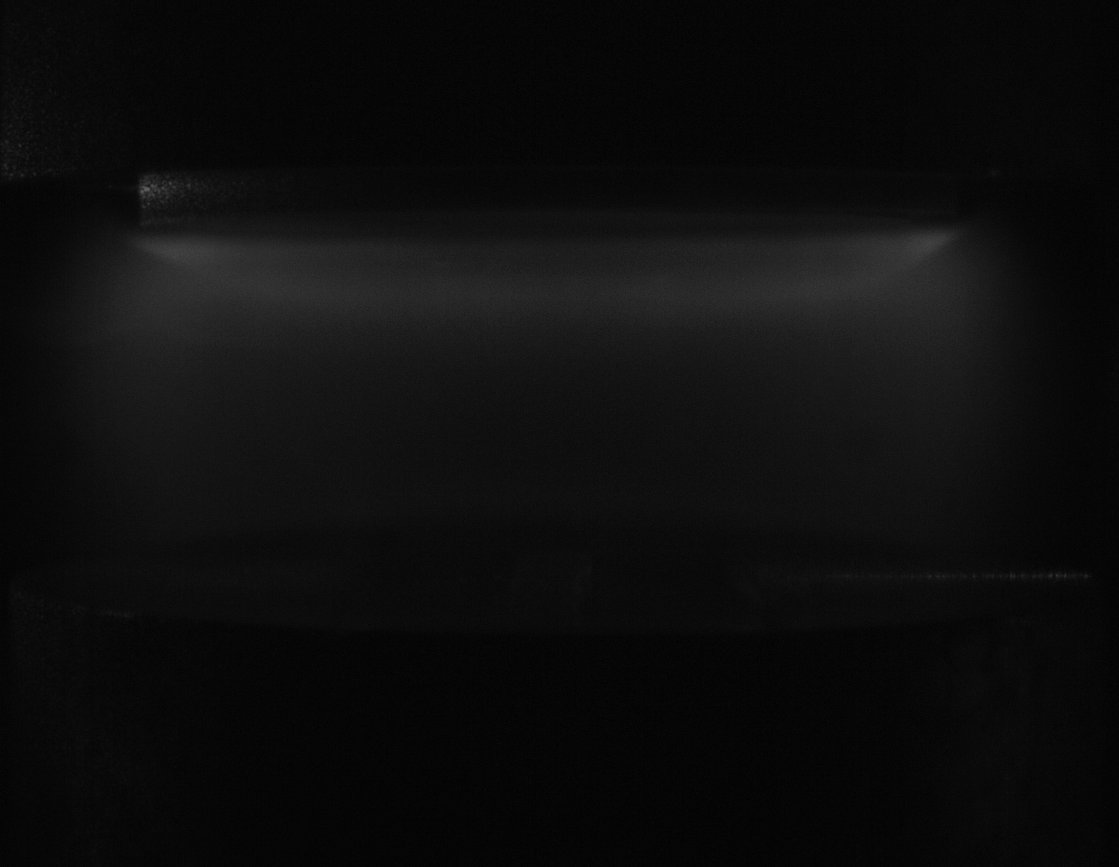}
        \caption{}
        \label{fig:bc14}
    \end{subfigure}
    \begin{subfigure}[b]{0.19\textwidth}
        \centering
        \includegraphics[width=\textwidth]{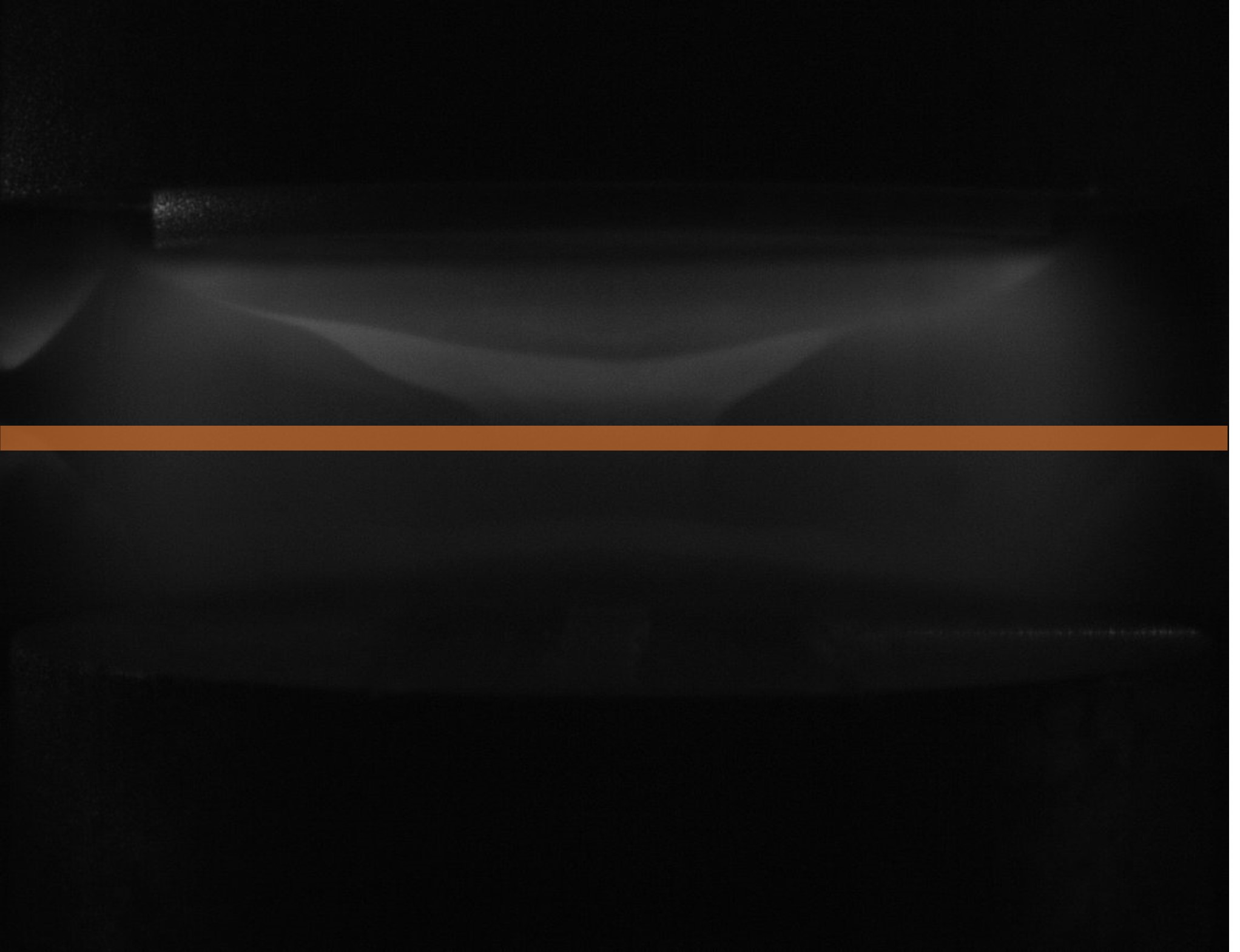}
        \caption{}
        \label{fig:bc12}
    \end{subfigure}
    \begin{subfigure}[b]{0.19\textwidth}
        \centering
        \includegraphics[width=\textwidth]{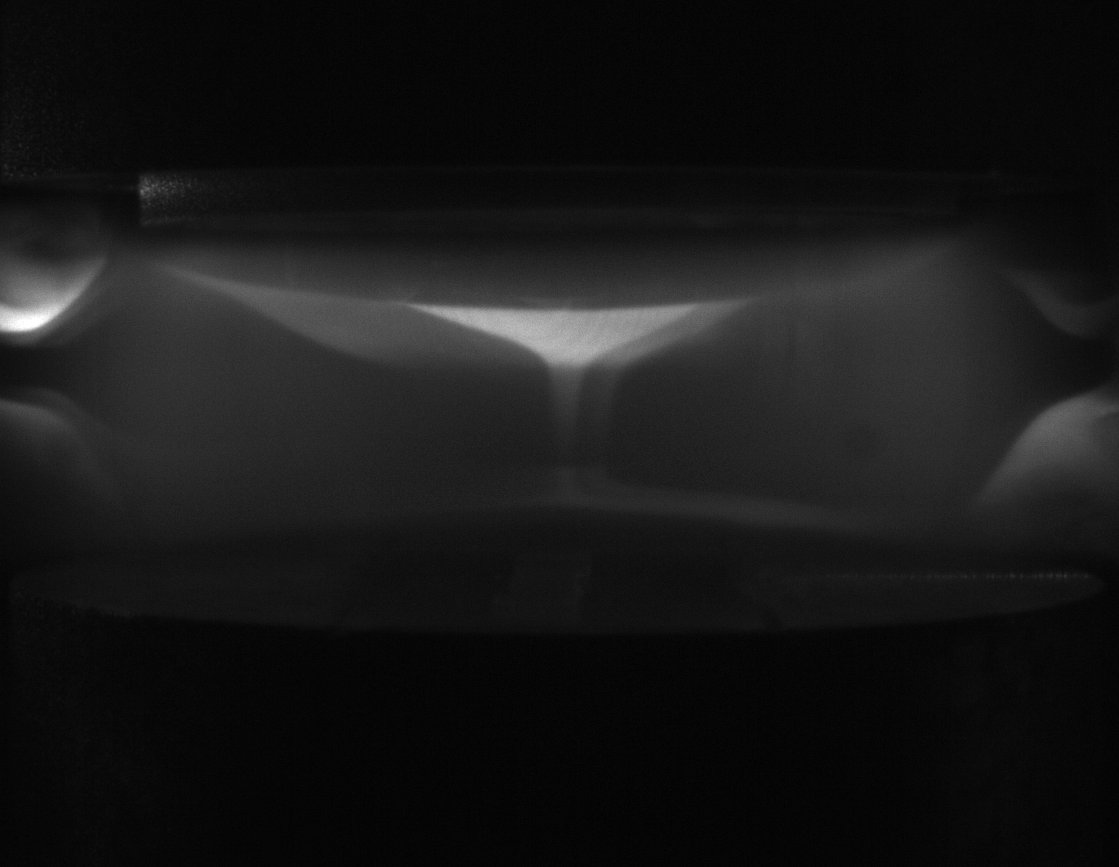}
        \caption{}
        \label{fig:bc34}
    \end{subfigure}
    \begin{subfigure}[b]{0.19\textwidth}
        \centering
        \includegraphics[width=\textwidth]{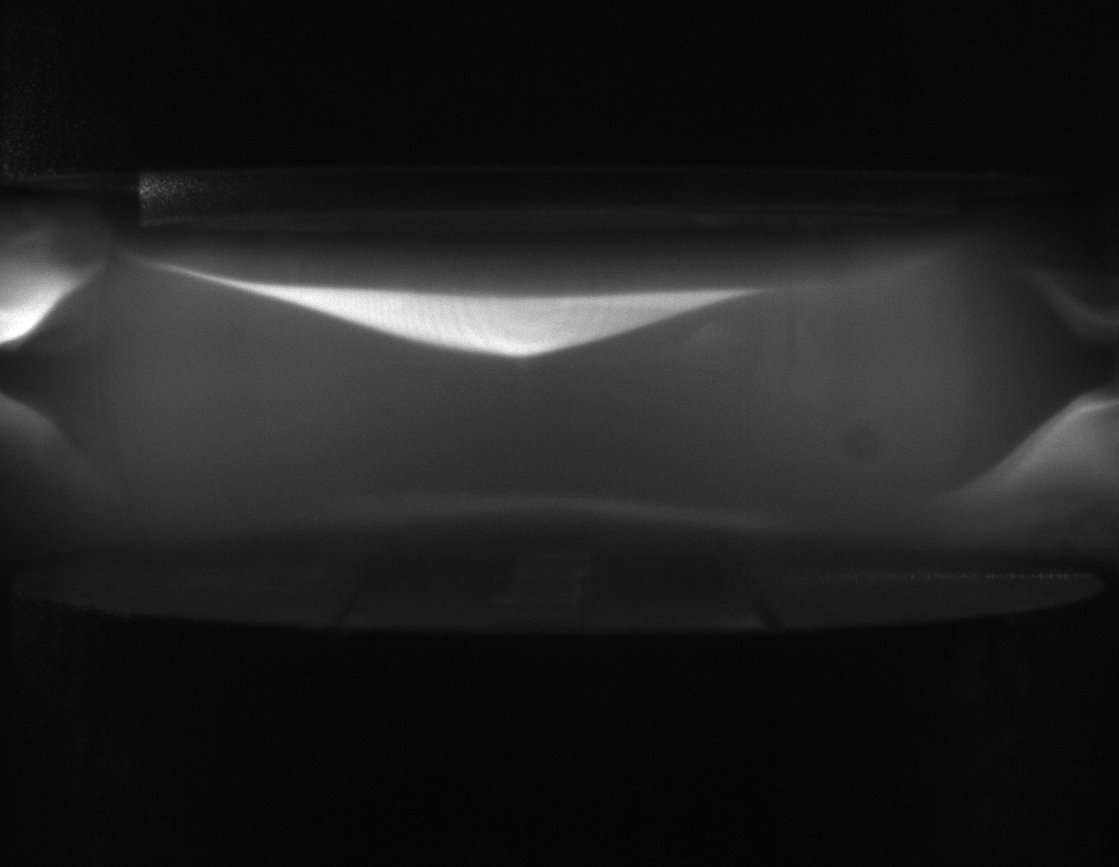}
        \caption{}
        \label{fig:bc1}
    \end{subfigure}

    \caption{Camera images of the four different dust cloud (row) at the same time (column) within each experiment, $T_c/4$, $T_c/2$, $3T_c/4$ and $T_c$, as labelled. The orange line is drawn at z = 12.5 mm of $T_c/2$, in order to compare the laser light scattering intensity in the mid-plane over 2 cycles in Fig. \ref{cloud_int}. }
    \label{fig:all_16_clouds}
\end{figure*}

The laser light scattering from the dust cloud was recorded using a CMOS camera. Dust density waves, which can be used to calculate dust density and temperature, were detected at the edge of the dusty plasma, however, they were not further studied herein \cite{tadsen2015self}. The time frame was set from the frame rate, whereby a frame was recorded every 20 ms, i.e 50 FPS.  The first frame with plasma on was set to t = 0. In order to compare the different dust cloud to each other at the same time during their cycles, 4 images are chosen and shown. The region of the dust cloud between the electrodes are shown in Fig. \ref{fig:all_16_clouds}, with each column being at $T_c/4$, $T_c/2$, $3T_c/4$, and $T_c$, where $T_c$ was the cycle time according to the OES.

\begin{figure*}[!ht]
    \centering
    % Column headings
    \begin{minipage}[b]{0.19\textwidth}
       \centering
        \textbf{30 squares}
    \end{minipage}
    \begin{minipage}[b]{0.19\textwidth}
       \centering
      \textbf{Cycles from the blue square}
    \end{minipage}
    \begin{minipage}[b]{0.19\textwidth}
      \centering
       \textbf{Mode frequency}
    \end{minipage}
    \begin{minipage}[b]{0.19\textwidth}
      \centering
      \textbf{Magnitude of mode   frequency}
    \end{minipage}

    \vspace{0.5cm} % Space between headings and first row of figures

    % First row with label
    \raisebox{1.7cm}{%
        \makebox[0pt][r]{%
            \parbox{80pt}{%
                \raggedleft
                \textbf{Without Magnetic} \\
                \textbf{Field:} \\
                \textbf{Ar/TTIP}
            }%
            \hspace{0.1cm}%
        }
    }
    \begin{subfigure}[b]{0.19\textwidth}
        \centering
        \includegraphics[width=\textwidth]{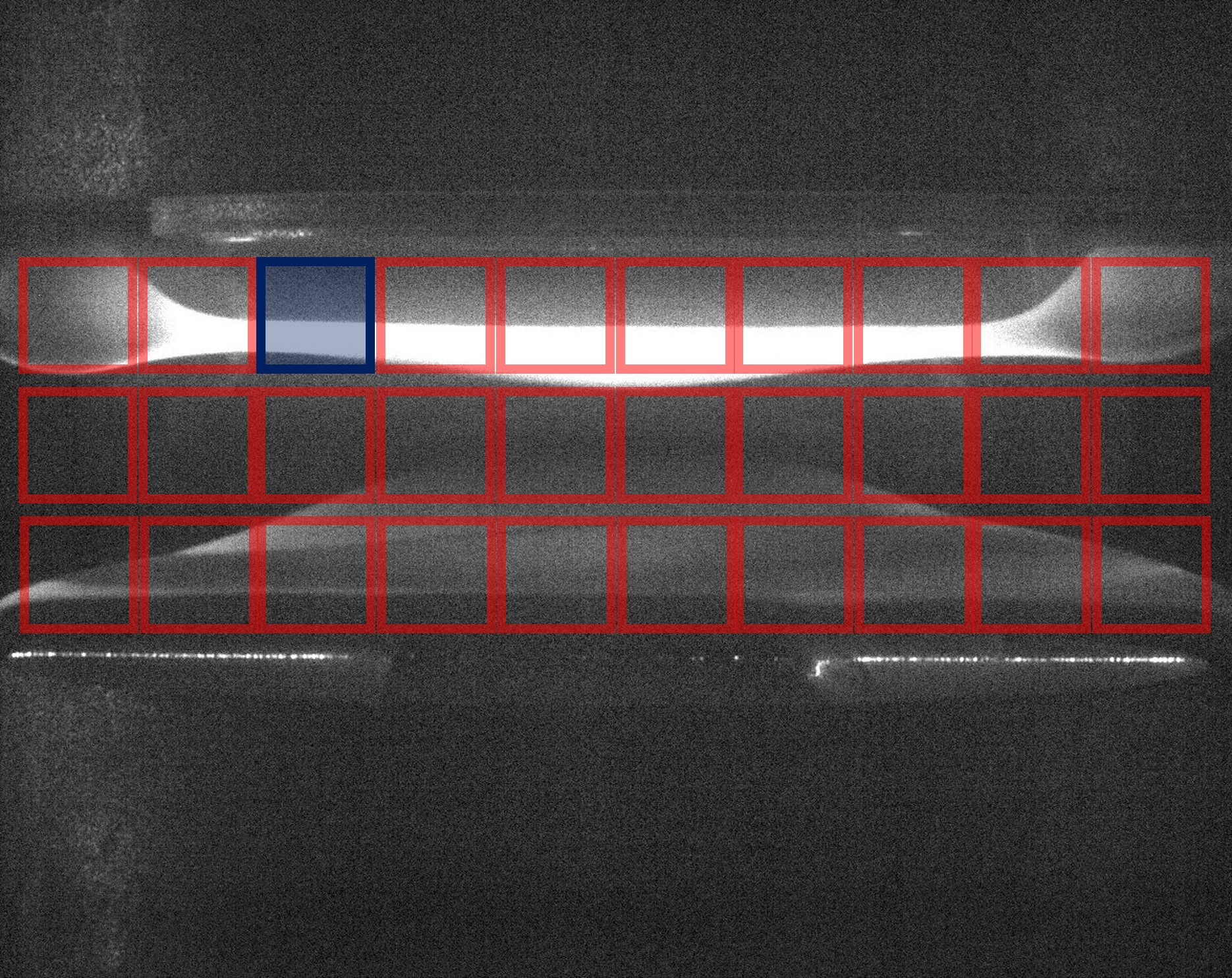}
        \caption{}
        \label{fig:t0_30_boxes}
    \end{subfigure}
    \begin{subfigure}[b]{0.19\textwidth}
        \centering
        \includegraphics[width=\textwidth]{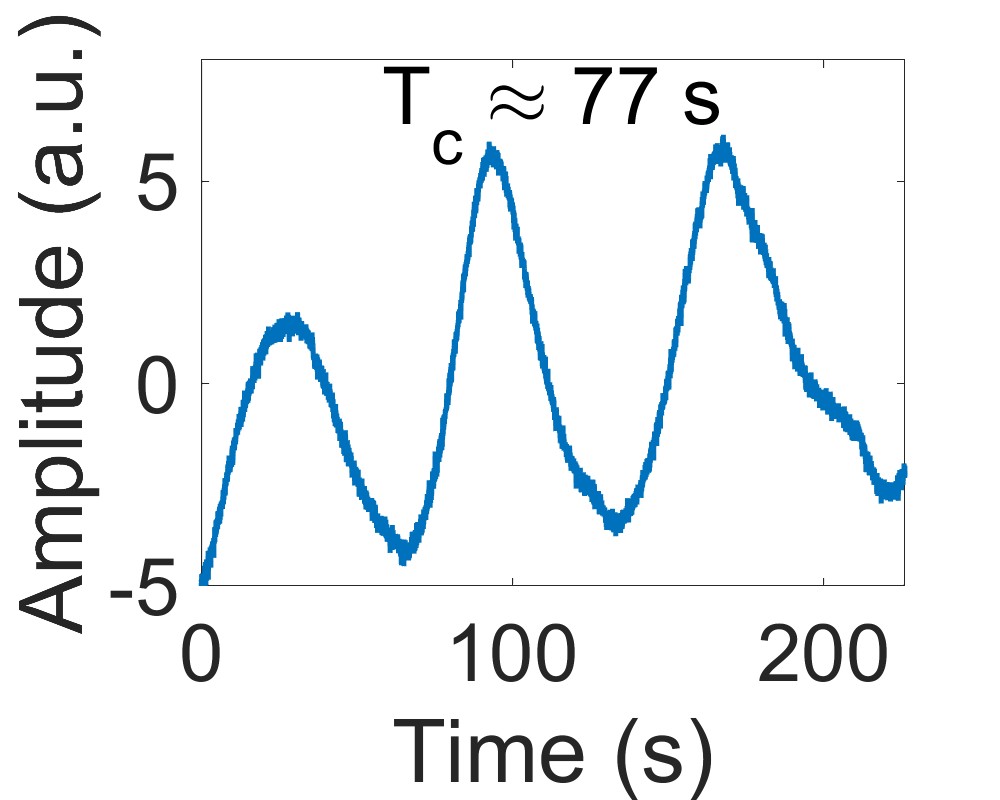}
        \caption{}
        \label{fig:NFT_cycle_t0}
    \end{subfigure}
    \begin{subfigure}[b]{0.19\textwidth}
        \centering
        \includegraphics[width=\textwidth]{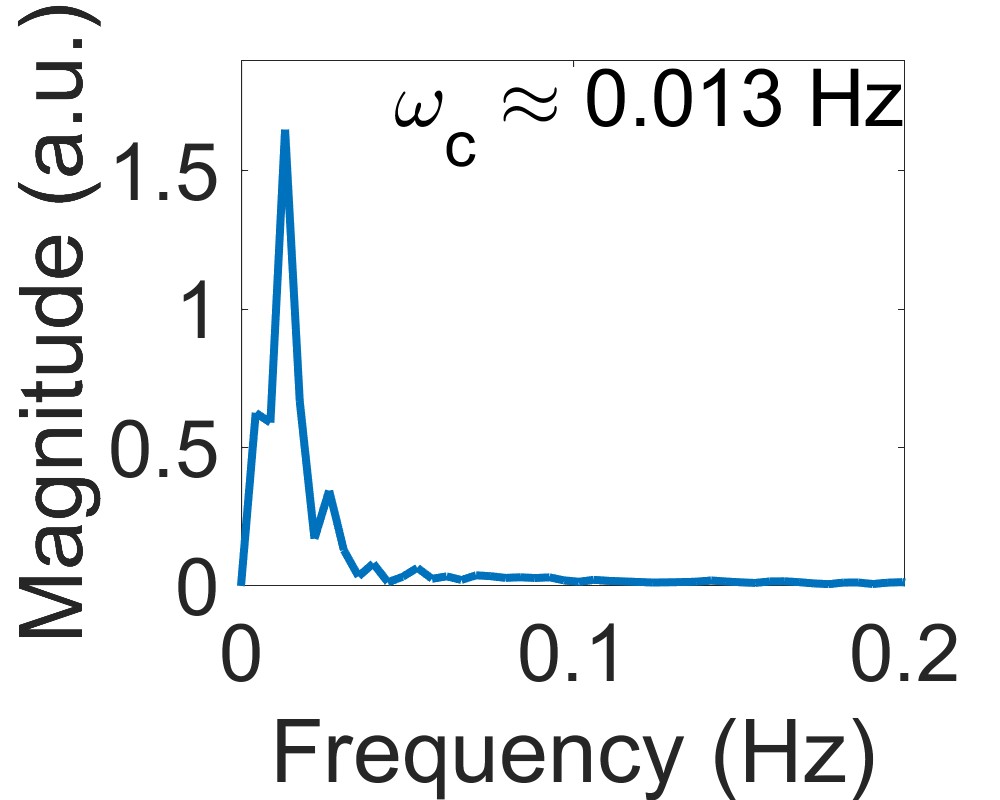}
        \caption{}
        \label{fig:NFT_mag_t0}
    \end{subfigure}
    \begin{subfigure}[b]{0.23\textwidth}
        \centering
        \includegraphics[width=\textwidth]{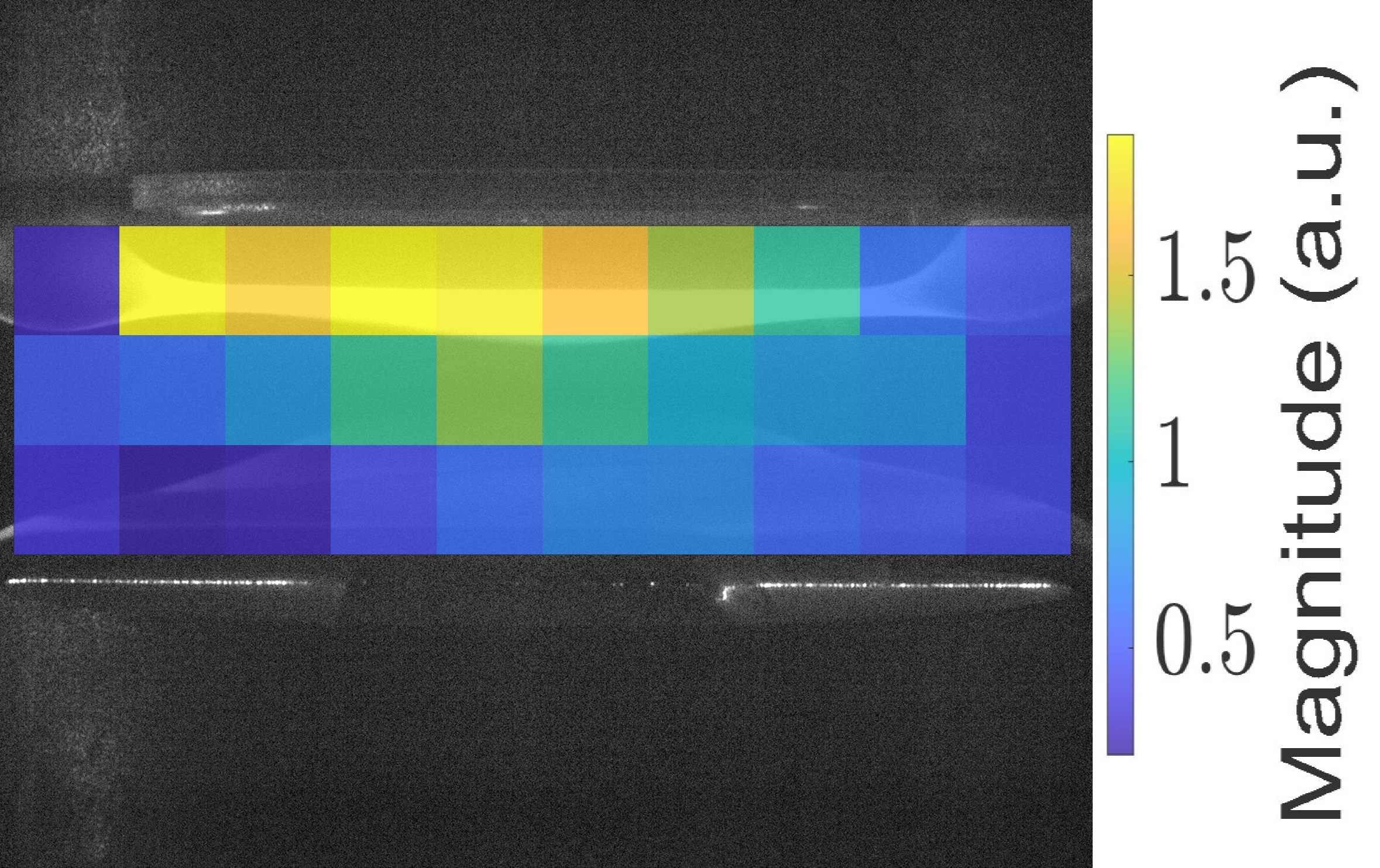}
        \caption{}
        \label{fig:t0_30_boxes_colored}
    \end{subfigure}

    \vspace{0.5cm} % Space between rows

    % Second row with label
    \raisebox{1.6cm}{%
        \makebox[0pt][r]{%
            \parbox{80pt}{%
                \raggedleft
                \textbf{Without} \\
                \textbf{Magnetic} \\
                \textbf{Field:} \\
                \textbf{Ar/$C_2H_2$}
            }%
            \hspace{0.1cm}%
        }
    }
    \begin{subfigure}[b]{0.19\textwidth}
        \centering
        \includegraphics[width=\textwidth]{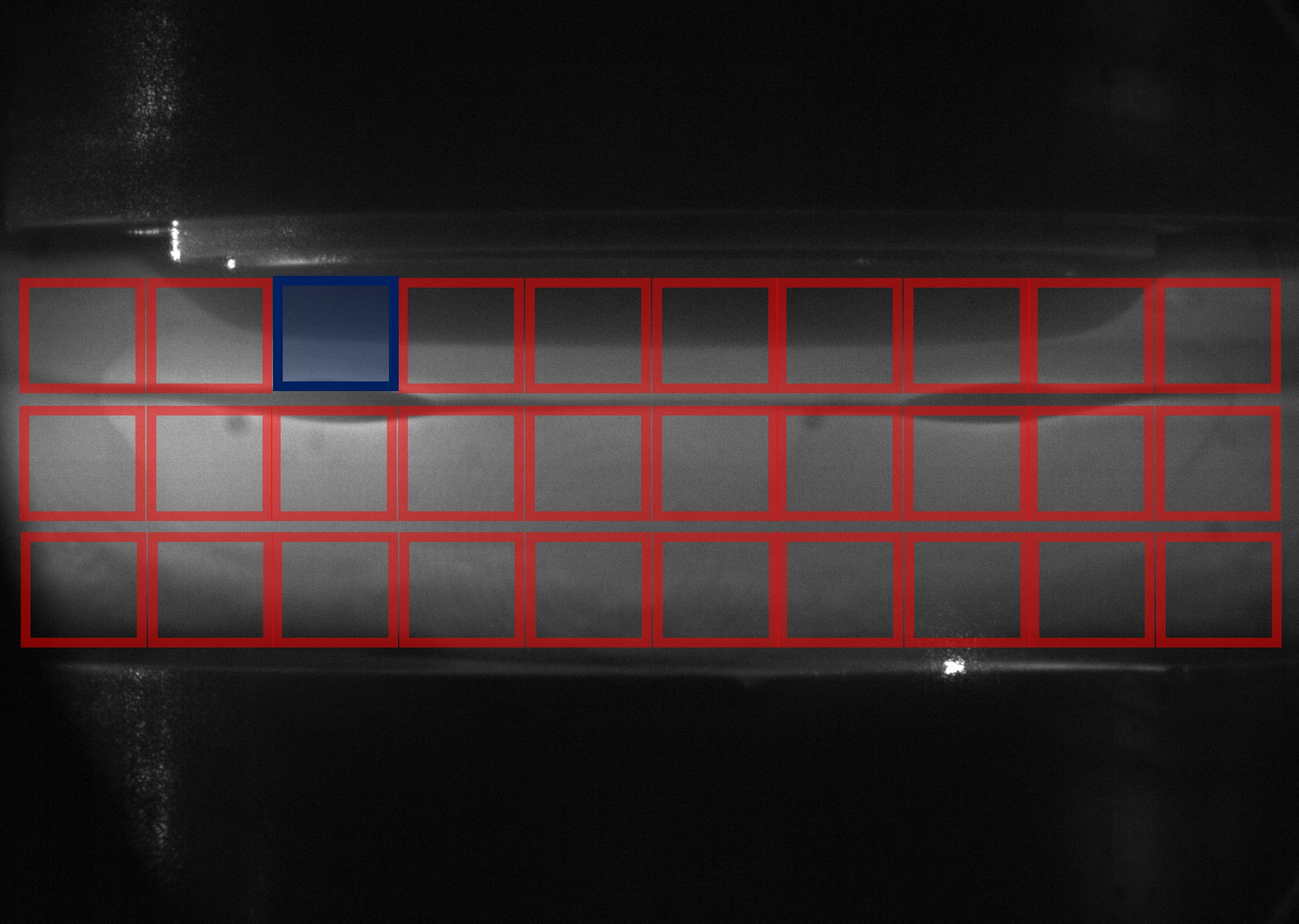}
        \caption{}
        \label{fig:c0_30_boxes}
    \end{subfigure}
    \begin{subfigure}[b]{0.19\textwidth}
        \centering
        \includegraphics[width=\textwidth]{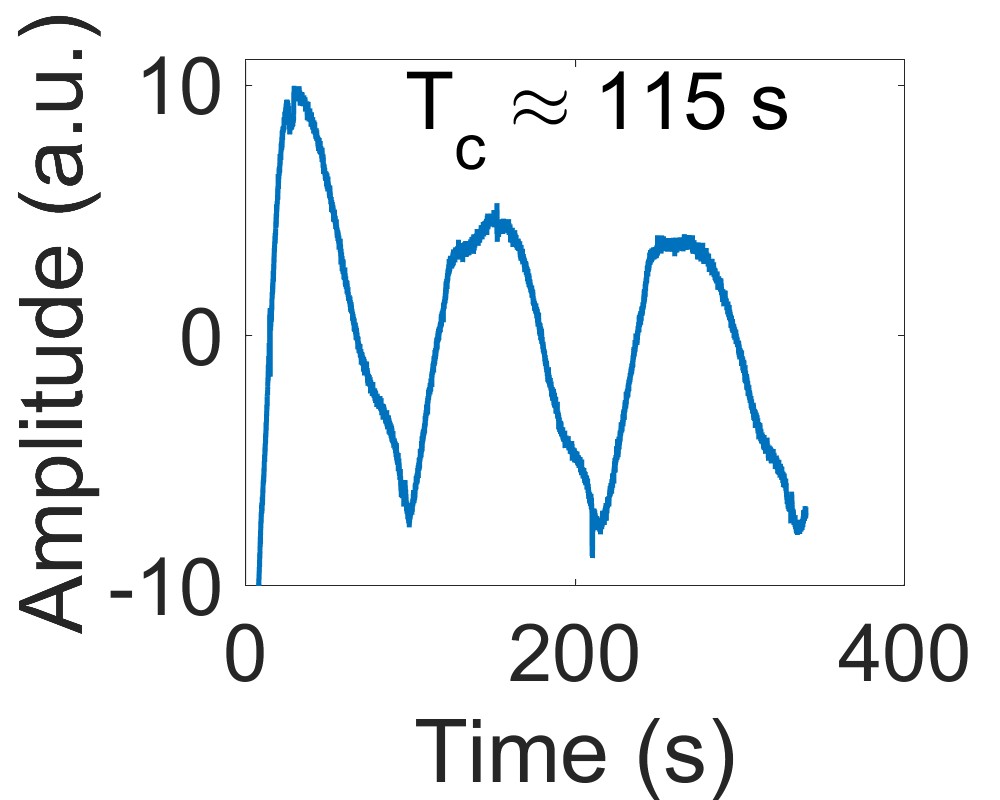}
        \caption{}
        \label{fig:NFT_cycle_c0}
    \end{subfigure}
    \begin{subfigure}[b]{0.19\textwidth}
        \centering
        \includegraphics[width=\textwidth]{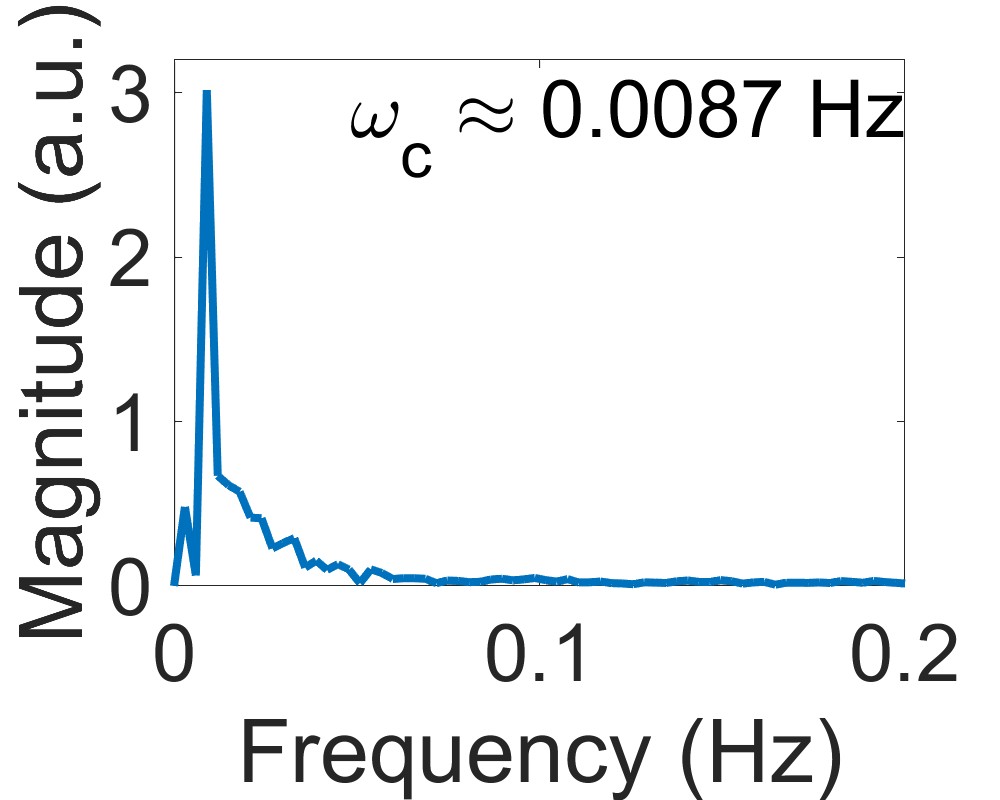}
        \caption{}
        \label{fig:NFT_mag_c0}
    \end{subfigure}
    \begin{subfigure}[b]{0.23\textwidth}
        \centering
        \includegraphics[width=\textwidth]{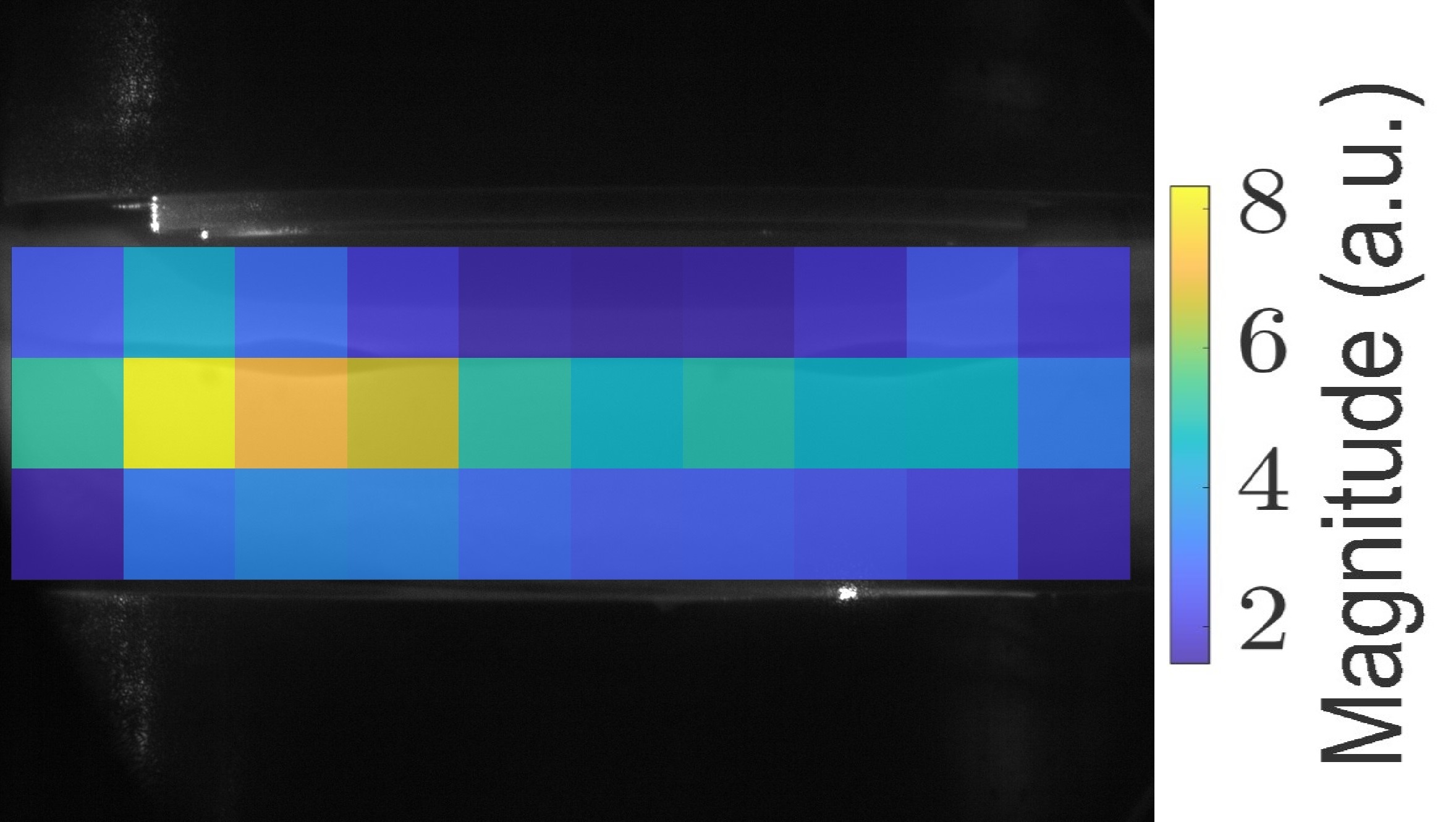}
        \caption{}
        \label{fig:c0_30_boxes_colored}
    \end{subfigure}

    \vspace{0.5cm} % Space between rows

    % Third row with label
    \raisebox{1.8cm}{%
        \makebox[0pt][r]{%
            \parbox{80pt}{%
                \raggedleft
                \textbf{With } \\
                \textbf{Magnetic}\\
                \textbf{Field:} \\
                \textbf{Ar/TTIP}
            }%
            \hspace{0.1cm}%
        }
    }
    \begin{subfigure}[b]{0.19\textwidth}
        \centering
        \includegraphics[width=\textwidth]{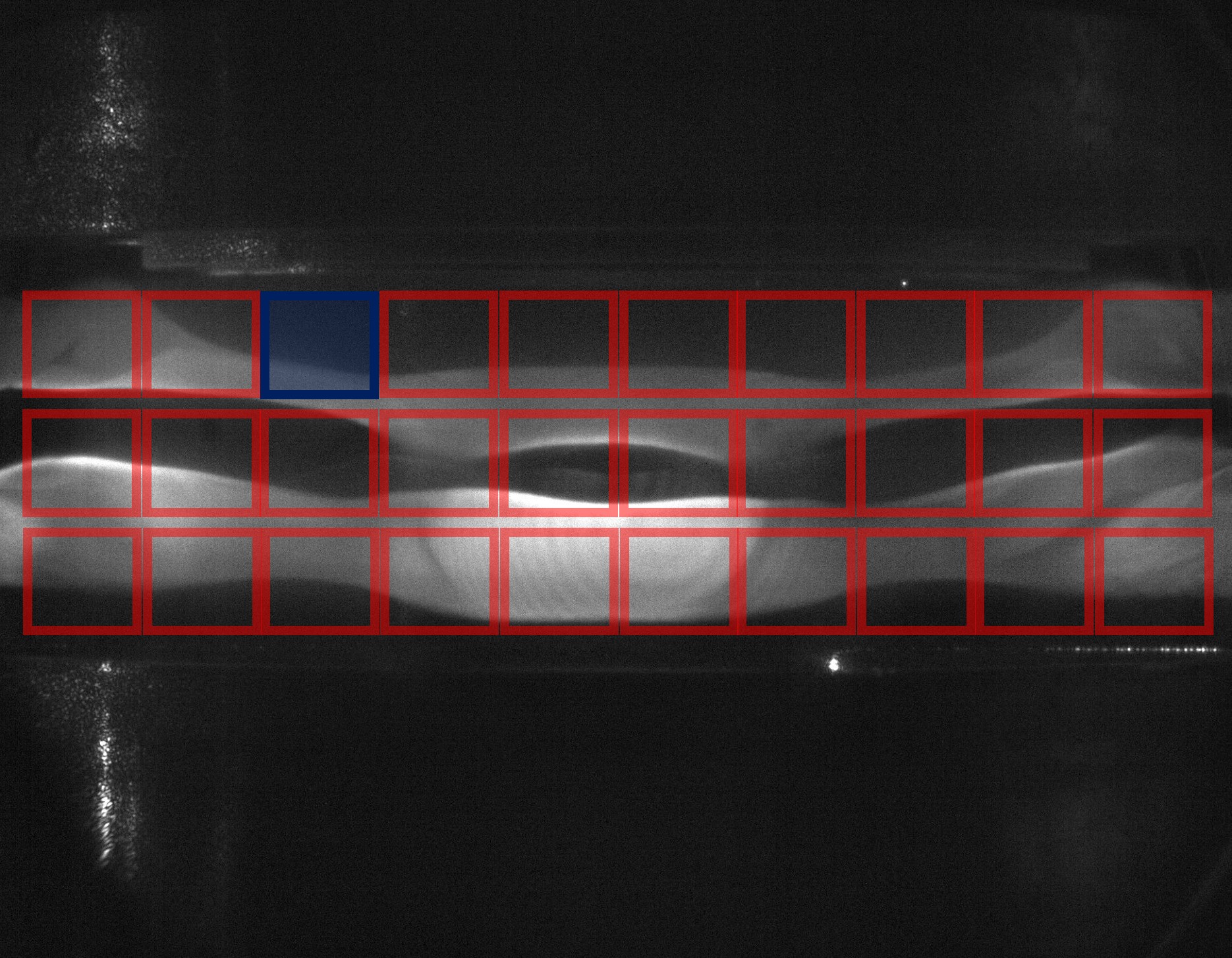}
        \caption{}
        \label{fig:t500_30_boxes}
    \end{subfigure}
    \begin{subfigure}[b]{0.19\textwidth}
        \centering
        \includegraphics[width=\textwidth]{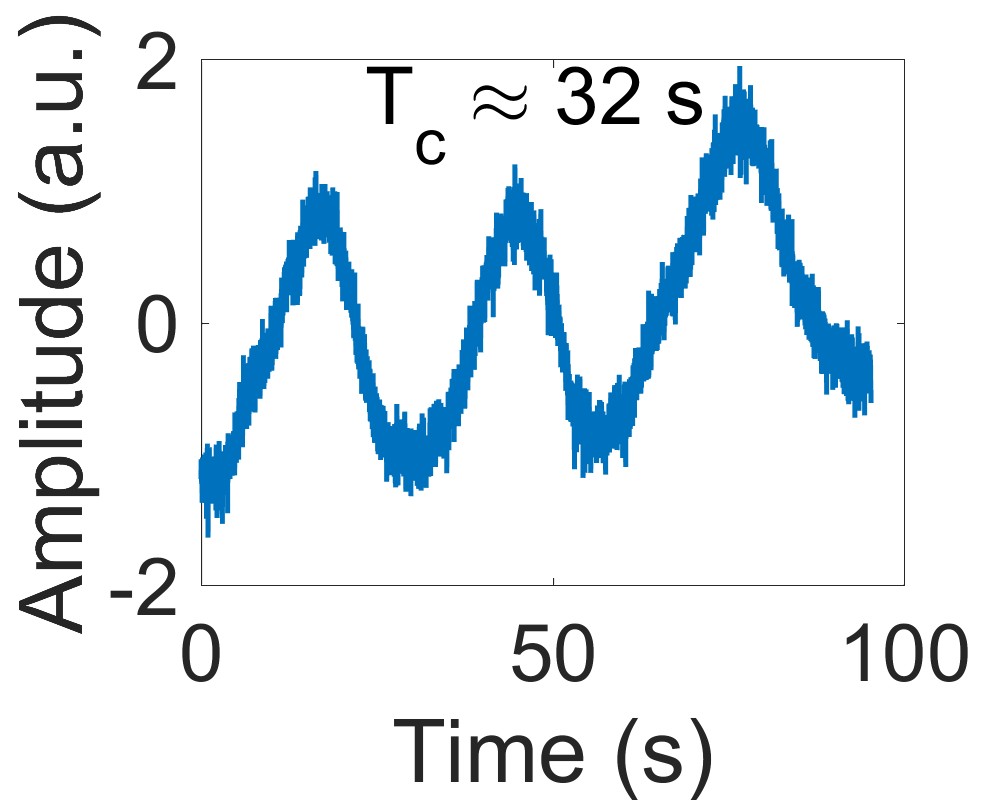}
        \caption{}
        \label{fig:NFT_cycle_t500}
    \end{subfigure}
    \begin{subfigure}[b]{0.19\textwidth}
        \centering
        \includegraphics[width=\textwidth]{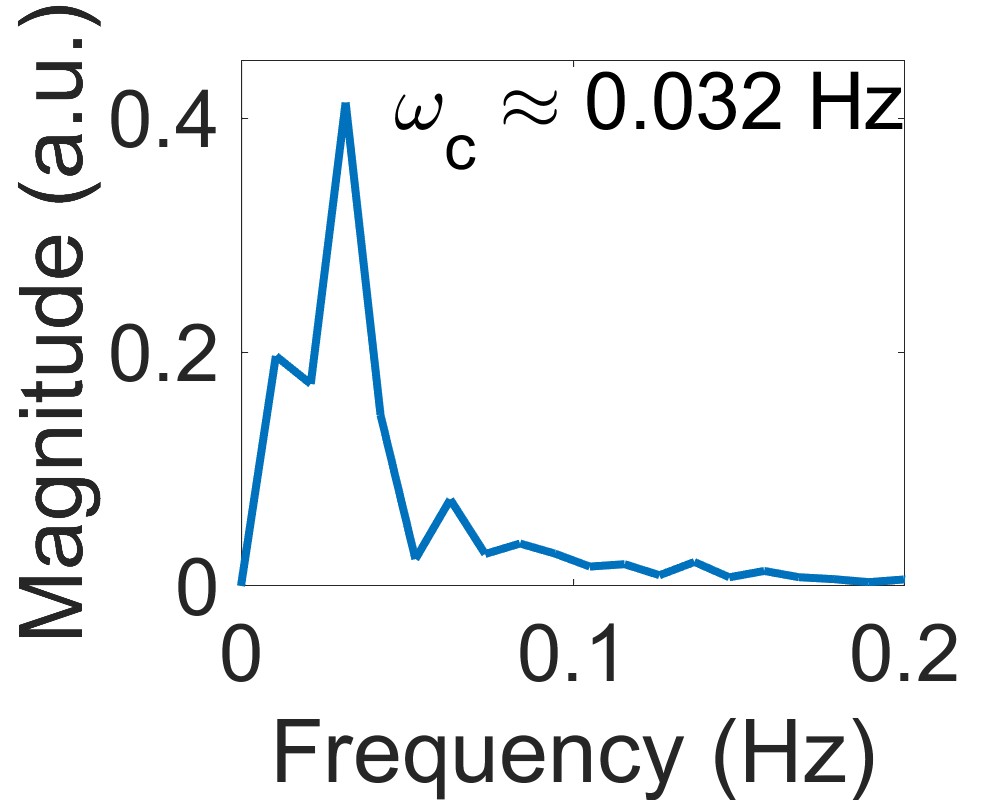}
        \caption{}
        \label{fig:NFT_mag_t500}
    \end{subfigure}
    \begin{subfigure}[b]{0.235\textwidth}
        \centering
        \includegraphics[width=\textwidth]{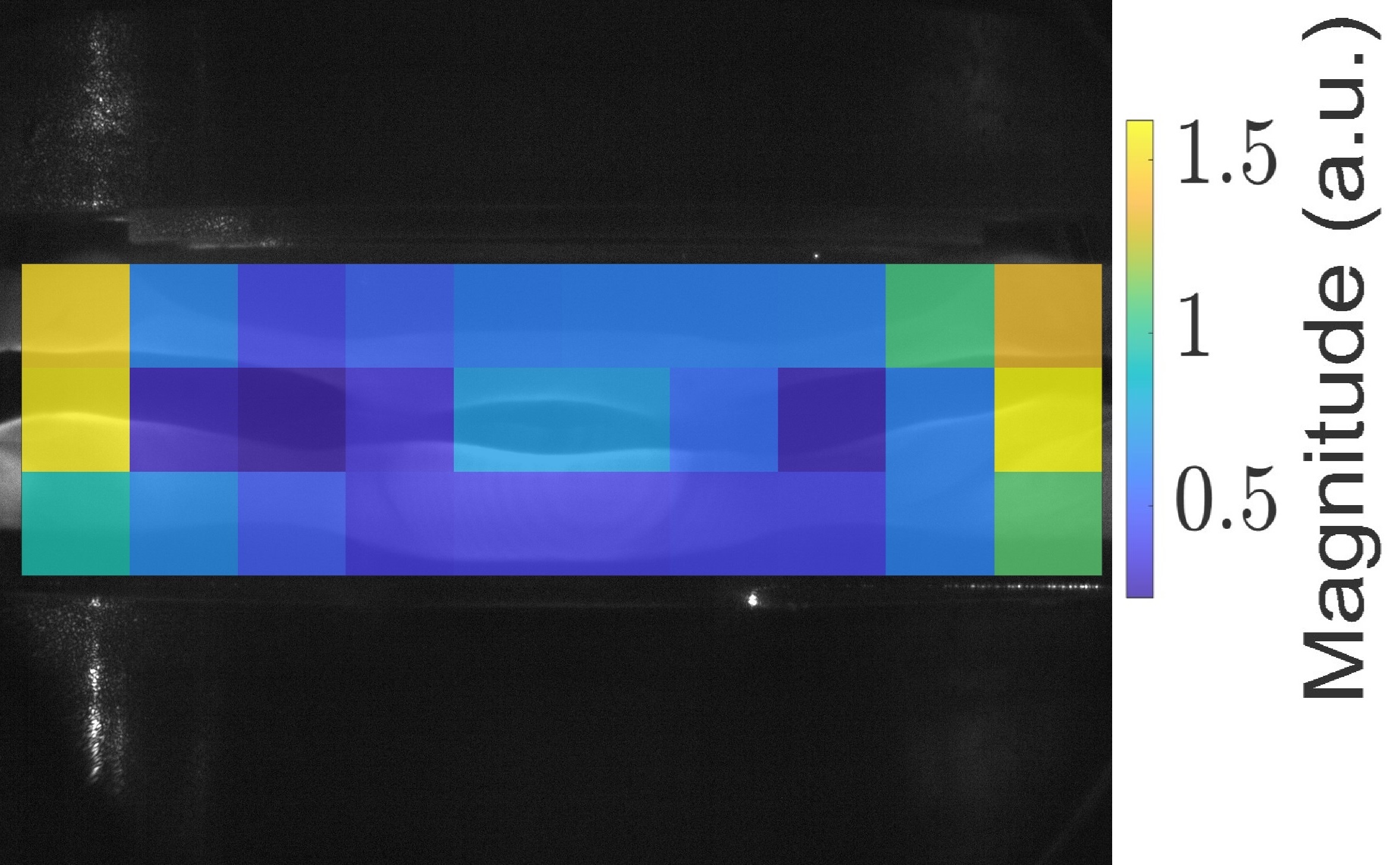}
        \caption{}
        \label{fig:t500_30_boxes_colored}
    \end{subfigure}

    \vspace{0.5cm} % Space between rows

    % Fourth row with label
    \raisebox{1.6cm}{%
        \makebox[0pt][r]{%
            \parbox{80pt}{%
                \raggedleft
                \textbf{With} \\
                \textbf{Magnetic}\\
                \textbf{Field:} \\
                \textbf{Ar/$C_2H_2$}
            }%
            \hspace{0.1cm}%
        }
    }
    \begin{subfigure}[b]{0.19\textwidth}
        \centering
        \includegraphics[width=\textwidth]{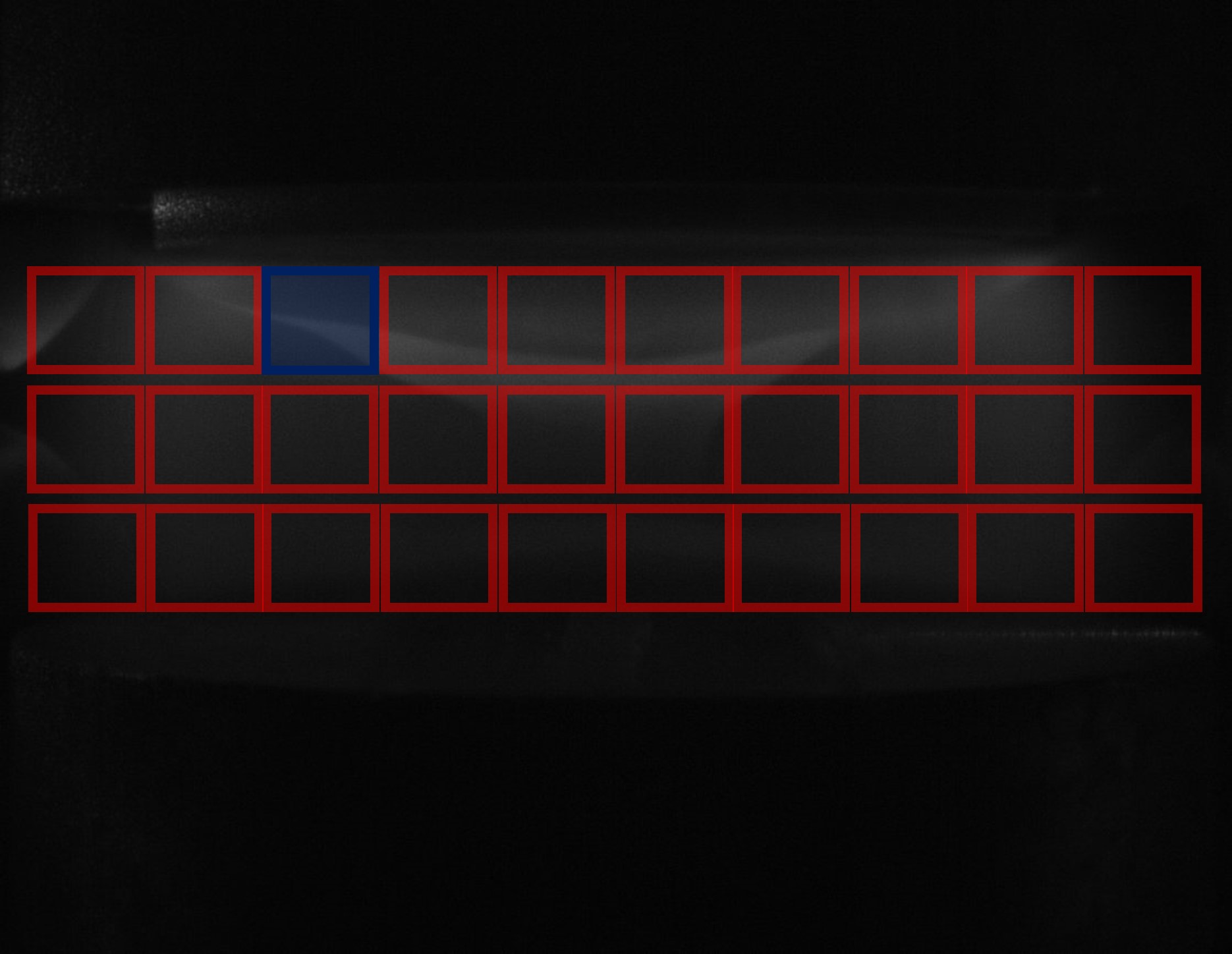}
        \caption{}
        \label{fig:c500_30_boxes}
    \end{subfigure}
    \begin{subfigure}[b]{0.19\textwidth}
        \centering
        \includegraphics[width=\textwidth]{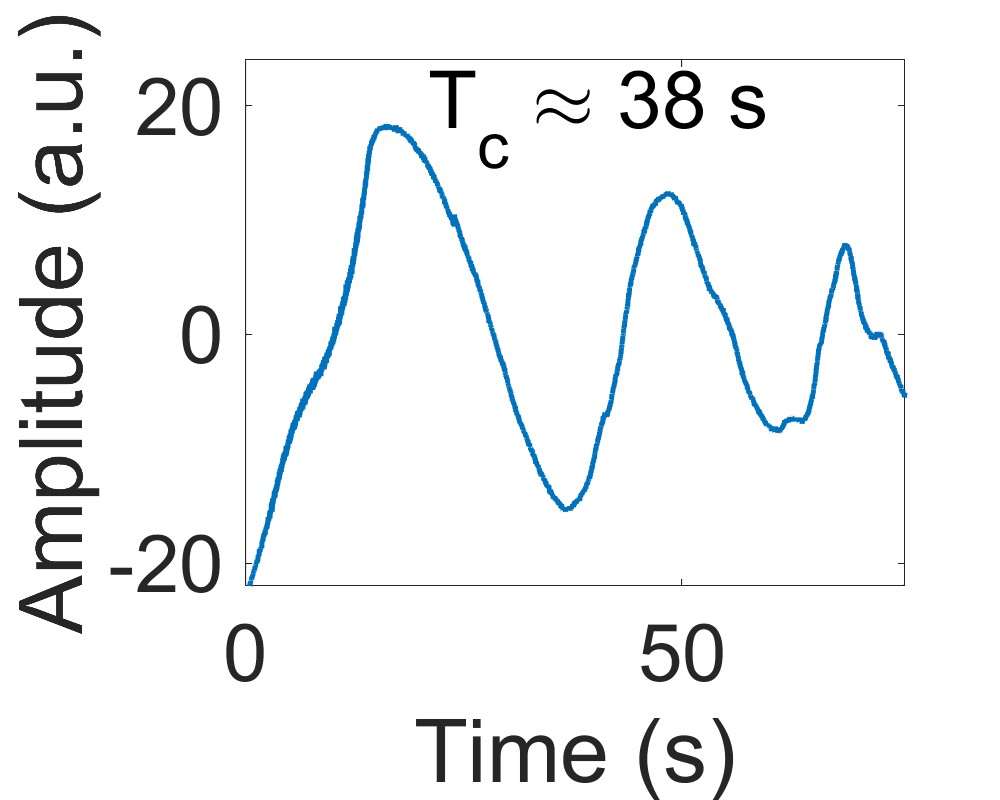}
        \caption{}
        \label{fig:NFT_cycle_c500}
    \end{subfigure}
    \begin{subfigure}[b]{0.19\textwidth}
        \centering
        \includegraphics[width=\textwidth]{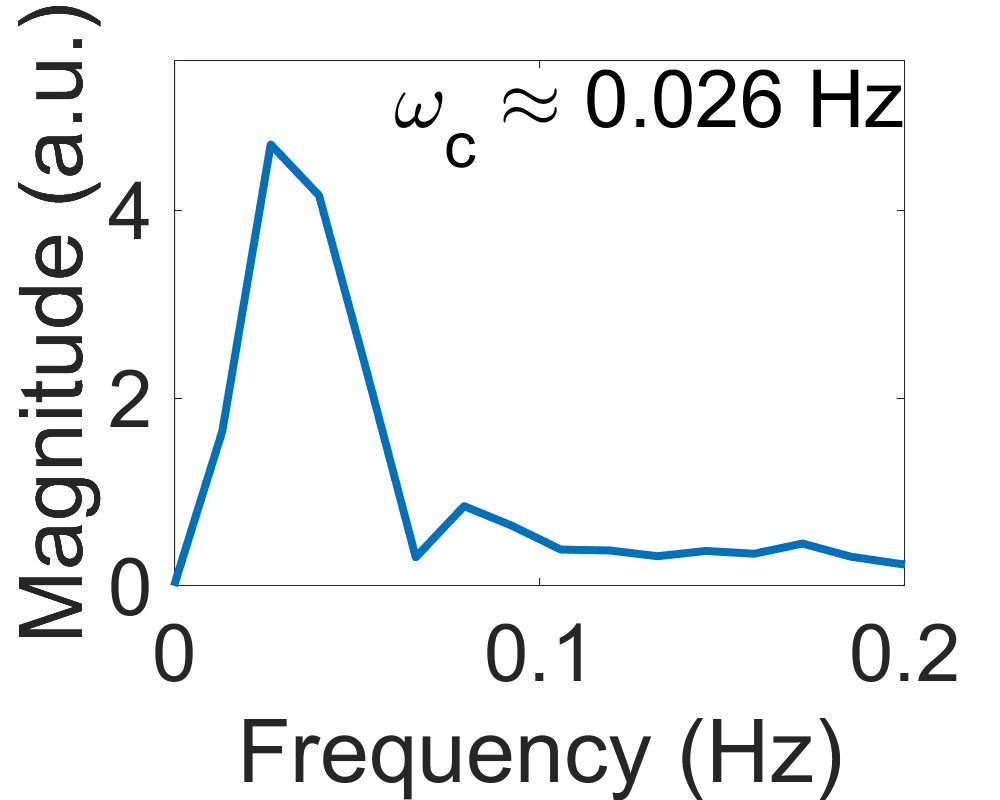}
        \caption{}
        \label{fig:NFT_mag_c500}
    \end{subfigure}
    \begin{subfigure}[b]{0.23\textwidth}
        \centering
        \includegraphics[width=\textwidth]{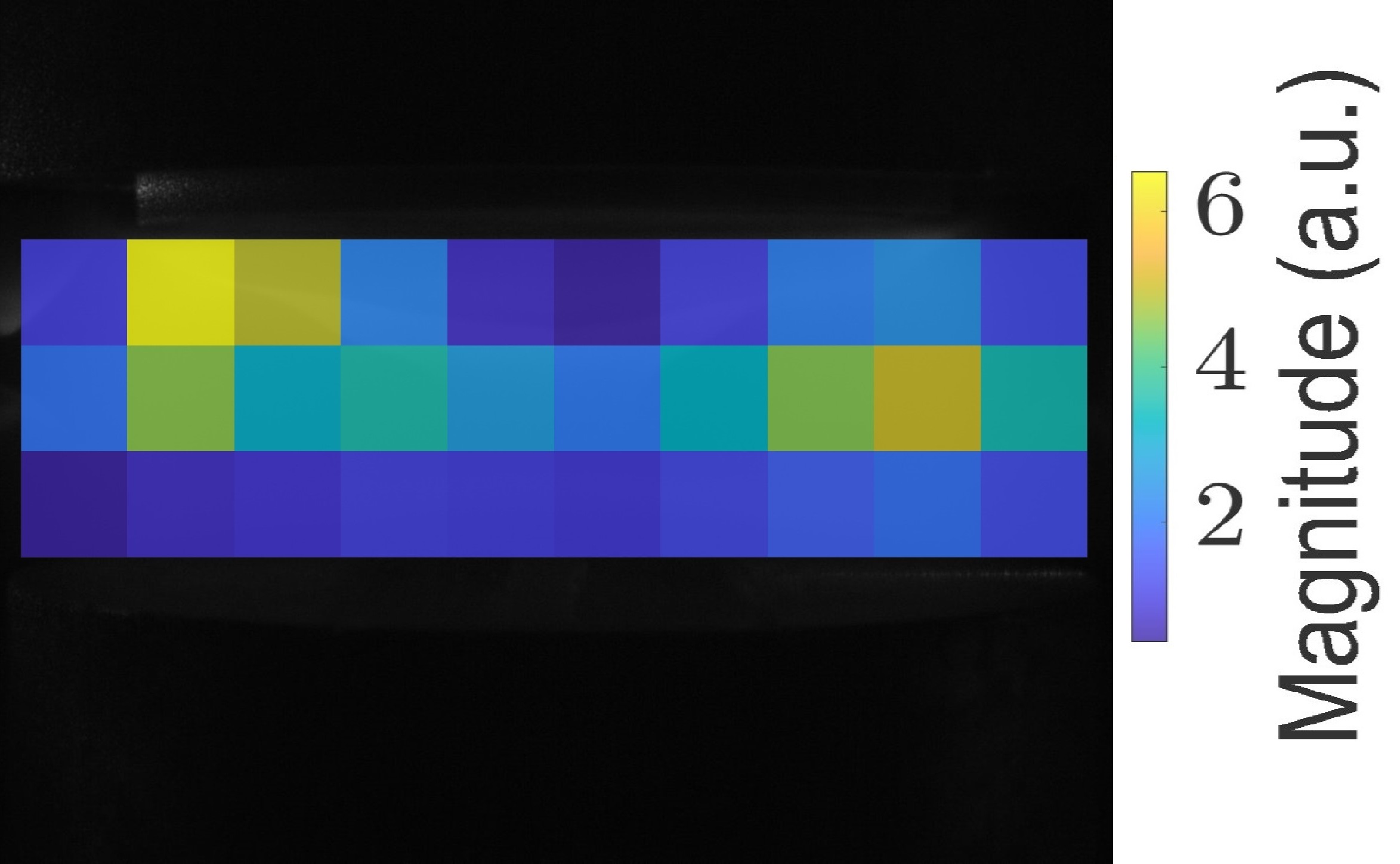}
        \caption{}
        \label{fig:c500_30_boxes_colored}
    \end{subfigure}

    \caption{Each row represent an individual dusty plasma as labelled. The first column shows 30 squares of $\sim$ 100 $\times$ 100 pixels, in 3 rows and 10 columns, over the dust cloud at $T_C/2$. The second column shows the cyclic variation in light intensity at row 1 and column 3 (blue square). The third column shows the mode frequency obtained from all 30 square. The fourth column shows the magnitude of the mode frequency in all 30 squares overlayed over the dust cloud at $T_C/2$. }
    \label{fig:all_figures_NFT}
\end{figure*}

Unlike OES which scans light emission across a volume, the laser light scattering occurs in a 2D plane. In order to consistently compare the cycle time, 30 squares measuring $\sim$ 100 $\times$ 100 pixels each were drawn on each dust cloud, 10 in 3 different rows as shown in the first column of Fig. \ref{fig:all_figures_NFT}. They were drawn to cover the entire region of the cloud between the electrodes where light scattering from dust particles is observed.  Within each analysis box, the average light intensity was calculated from all of the pixels.  A time series, spanning over three growth cycles, was then created  by recording the light intensity as a function of time for each box. A numerical Fourier transform (NFT) of the time series was performed for each box to identify the dominant frequencies. The most occurring dominant frequency, i.e. mode frequency, measured from the 30 NFT was chosen to represent and calculate the growth cycles. Examples of the cyclic variation of the light emission intensity from the first row and third column (blue box) is shown in the second column of Fig. \ref{fig:all_figures_NFT}. The average intensity over 3 cycles is subtracted from the data before calculating NFT, hence the negative magnitudes. The X-axis of indicates the time calculated from the camera frames. The mode frequency from all 30 boxes is shown in the third column of Fig. \ref{fig:all_figures_NFT}.  Moreover, the magnitude of the mode frequency in the 30 boxes superimposed on the dust cloud at $T_c/2$ is shown in the fourth column of Fig. \ref{fig:all_figures_NFT}. For the Ar/TTIP dusty plasma, without (with) the presence of the magnetic field, the mode frequency and cycle time were 0.0130 Hz (0.0315 Hz) and 77 s (32 s) respectively. For the Ar/$C_2H_2$ dusty plasma, without (with) the presence of the magnetic field, the mode frequency and cycle time were 0.0087 Hz (0.0264 Hz) and 115 s (38 s) respectively. The mode frequency and cycle time were rounded to two significant figures and they are in agreement with the values obtained from OES.

\subsection{Size Distribution}

\begin{figure*}[ht]
  \centering
  \begin{subfigure}{0.4\textwidth}
    \begin{minipage}{\textwidth}
      \includegraphics[width=\linewidth]{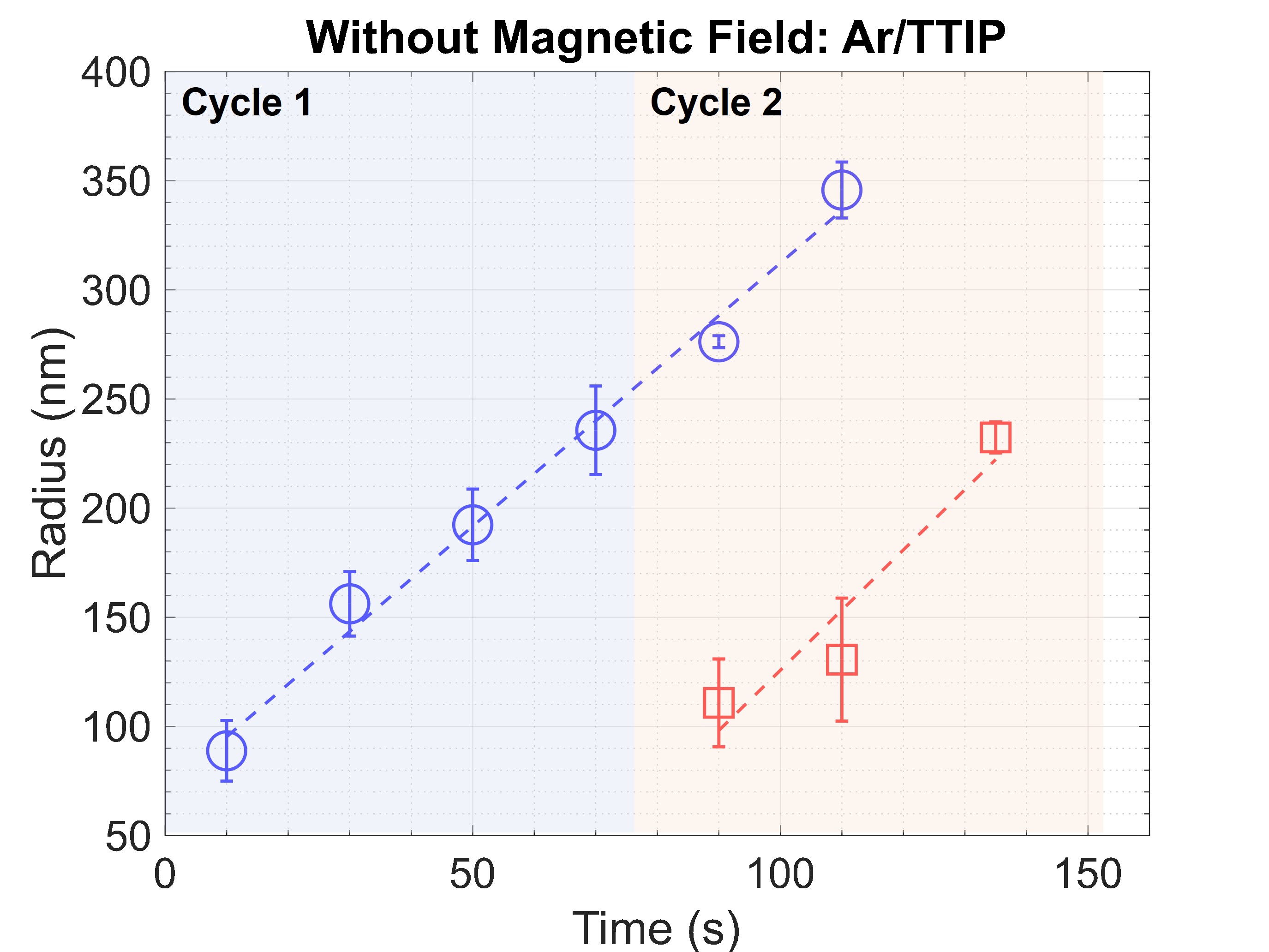}
      \subcaption{}
      \label{size_t_0}
    \end{minipage}
  \end{subfigure}
  \begin{subfigure}{0.4\textwidth}
    \begin{minipage}{\textwidth}
      \includegraphics[width=\linewidth]{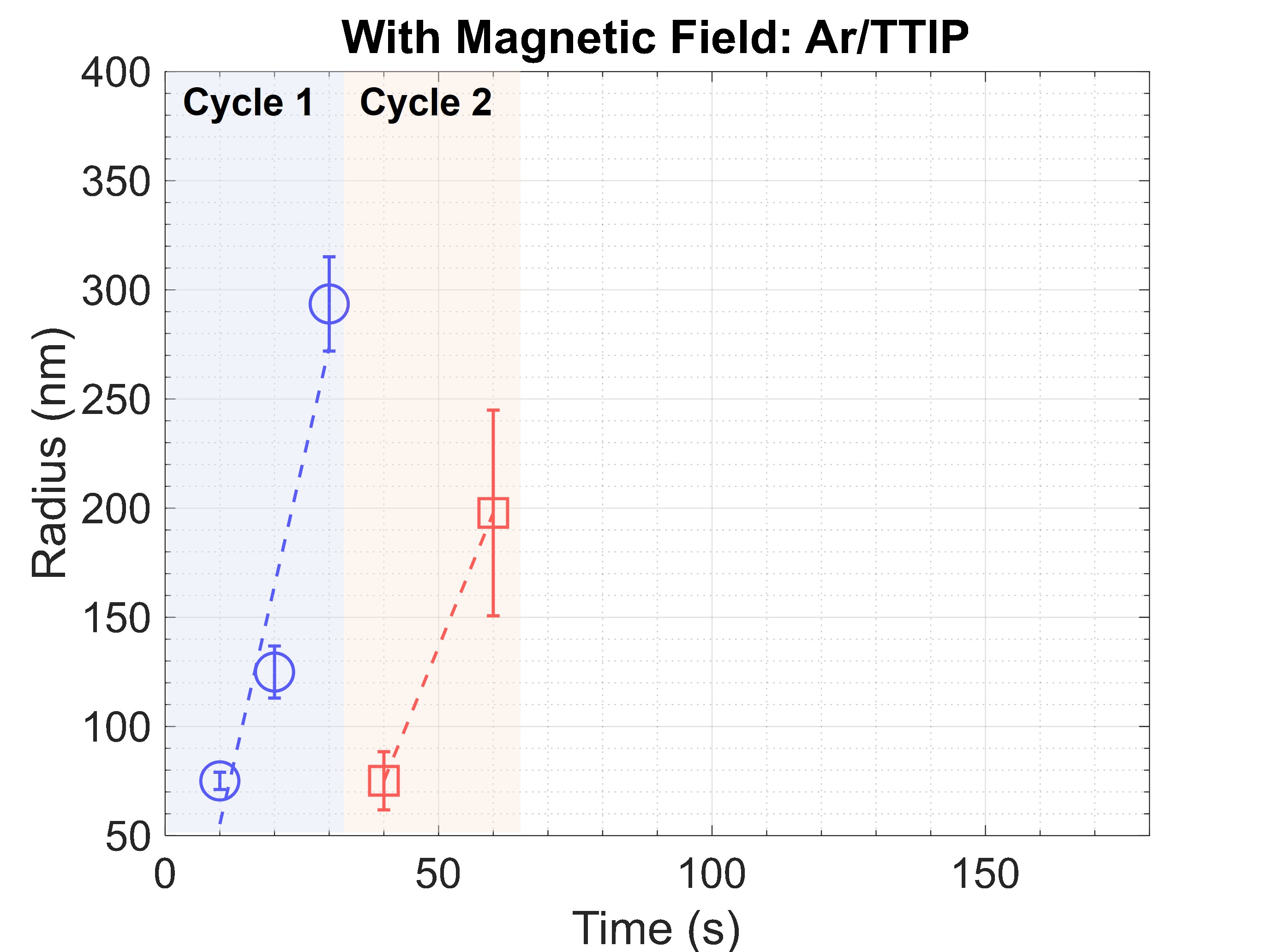}
      \subcaption{}
      \label{size_t_500}
    \end{minipage}
  \end{subfigure}
  \begin{subfigure}{0.4\textwidth}
    \begin{minipage}{\textwidth}
      \includegraphics[width=\linewidth]{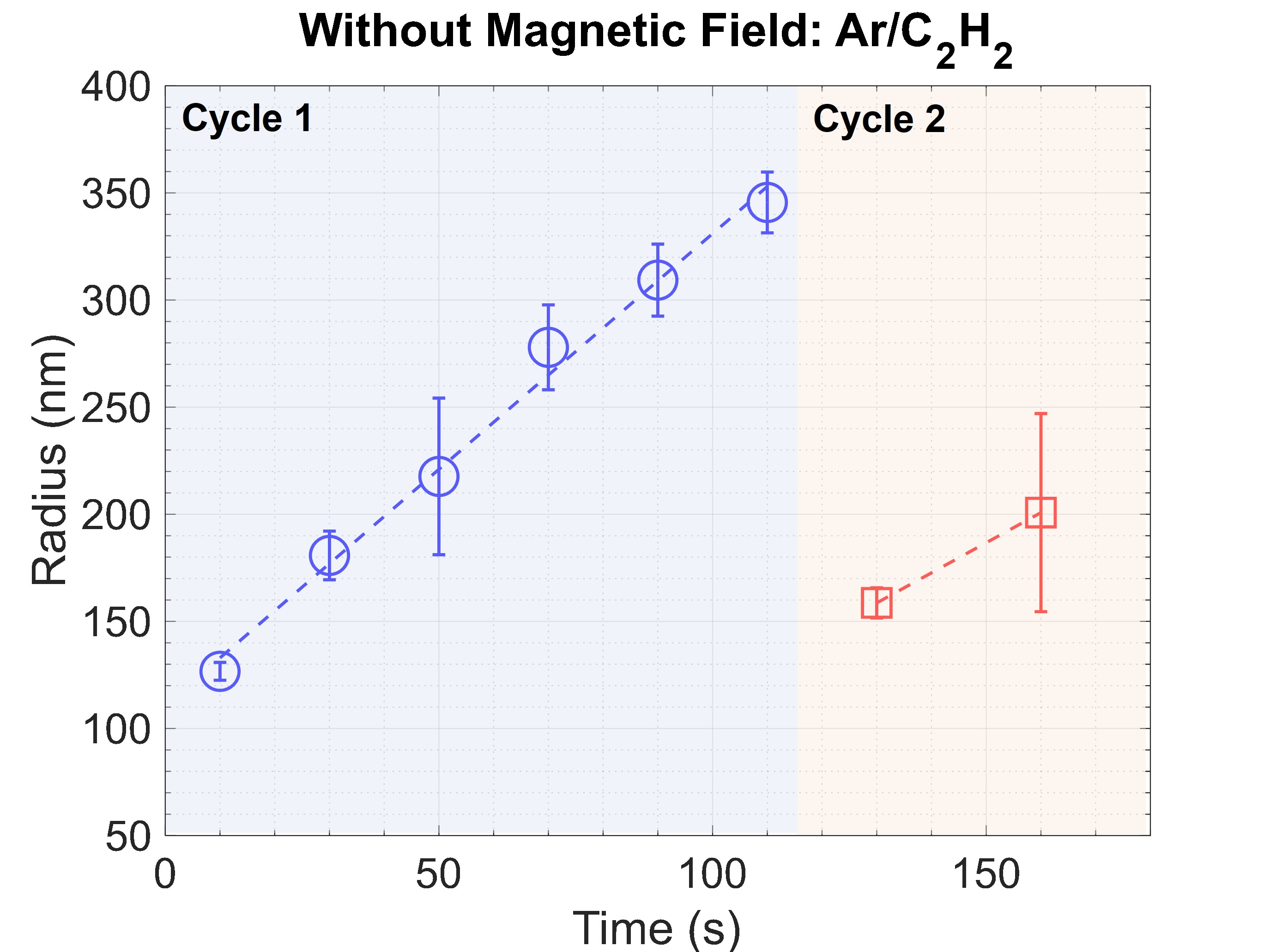}
      \subcaption{} 
      \label{size_c_0}
    \end{minipage}
  \end{subfigure}
  \begin{subfigure}{0.4\textwidth}
    \begin{minipage}{\textwidth}
      \includegraphics[width=\linewidth]{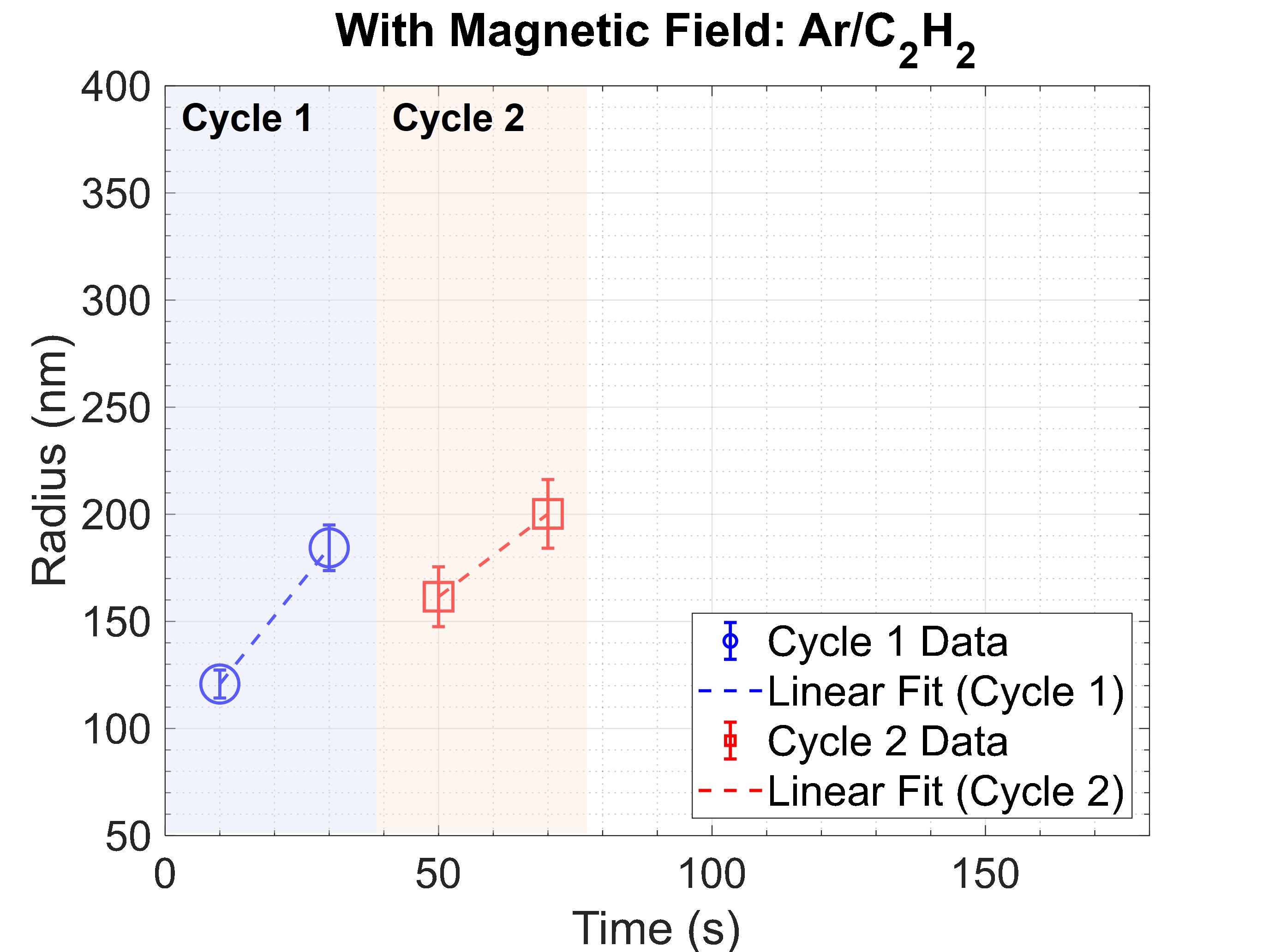}
      \subcaption{} 
      \label{size_c_500}
    \end{minipage}
  \end{subfigure}
    \caption{Linear growth of nanoparticles' radii within two growth cycles. Legend of (d) applies to all the subfigures. (a)  Ar/TTIP without magnetic field, with a bimodal size distribution at 90s and 110s (also reported in \cite{ramkorun2024introducing}). (b) Ar/TTIP with magnetic field. (c) Ar/$C_2H_2$ without magnetic field (d) Ar/$C_2H_2$ with magnetic field. }
  \label{Size}
\end{figure*}

In order to obtain the size of the nanoparticles, the individual experiments were ran for a certain instant of time and the nanoparticles subsequently were collected on a glass slide at the bottom electrode. The substrates were sputter coated with 15 nm of gold before being analyzed by Scanning Electron Microscope (SEM) (JEOL 7200F Field Emission SEM). The particles reached their maximum size by the end of each cycle. When a new cycle started, new particles initiated their growth. This growth exhibited a consistent linear progression in size throughout two cycle, as depicted in Figure \ref{Size}. The time frame for the size distribution analysis was determined by referring to the frames from CMOS camera images. Each particle growth experiment was meticulously recorded and timed according to the number of frames, spanning from the activation of plasma to its deactivation. 

At the end of the first cycle, the void expands, followed by a newer cycle which eventually replaces the precious one. Therefore, the size distribution of nanoparticles is reset in the second cycle. The growth of titania without magnetic field is the only exception, whereby, the dust from the previous cycle moves relatively slower out of the plasma and hence a bimodal size distribution is collected at 90 and 110 seconds. The maximum radius of the particles after 77 s were 235 $\pm$ 20 nm, however, the particles' radii up grew to 355 $\pm$ 12 nm at 110 s, as shown in Fig. \ref{size_t_0}. The carbonaceous particles reached their maximum radius of  346 $\pm$ 14 after 110 s, shortly after which the new cycle started, as shown in Fig. \ref{size_c_0}. The maximum radii of titania and carbonaceous dust particles grown during the presence of magnetic field were 294 $\pm$ 22 nm and 200 $\pm$ 16 nm, respectively, as shown in Figs. \ref{size_t_500} and \ref{size_c_500}.

%%%%%%%%%%%%%%%%%%%%%%%%%%%%%%%%%%%%%%%%%%%

\section{Discussion}
\label{sec:Discussion}

In prior investigations of dusty plasma particle growth, it was shown that a weak magnetic field of 320 Gauss reduced the  growth cycle time \cite{Couedel19}. Our findings are consistent with these observations for both types of dust particles, as summarized in Table \ref{ResultsTable}. Literature has shown that a weak magnetic field as low as 50 Gauss can affect the plasma potential, and hence the electric field in capacitively coupled Ar plasmas \cite{kushner2003modeling}. The smaller radii of nanoparticles grown during the presence of magnetic field suggest that there might be a reduction in the electric field. This is consistent with preliminary measurement of our electric field, however a full description the effect of magnetic field on the electric field is requires a separate study. Towards the end of the growth cycle, the dominant forces acting on the nanoparticles are gravitational, electric and ion-drag. Therefore, a possible decrease in the electric force during the presence of the magnetic field is probably reducing the force balance, thus causing the faster growth cycle and reduced nanoparticles’ size.

\begin{table*}[h]
  \caption{Summary of the results from the two dusty plasma nanoparticle growth. With the magnetic field, the maximum radii are smaller and the cycle ends faster as measured by OES and NFT of the dust cloud.}
  \label{ResultsTable}
  \footnotesize\rm
  \begin{tabular*}{\textwidth}{@{}l*{2}{@{\extracolsep{0pt plus12pt}}ll}*{2}{@{\extracolsep{0pt plus12pt}}ll}@{}}
    \br
    \textbf{Dust} & \multicolumn{3}{c}{\textbf{Without magnetic field }} & \multicolumn{3}{c}{\textbf{With magnetic field }} \\
    
    & Max Radius (nm) & \multicolumn{2}{c} {Cycle Time (s)} &  Max Radius (nm) & \multicolumn{2}{c}{Cycle Time (s)} \\
    \cmidrule{3-4} \cmidrule{6-7}
    & & OES & NFT && OES & NFT \\
    \mr
    
    %titania
    Titania & 355 $\pm$ 12 & 77 $\pm$ 4 & 77 & 294 $\pm$ 22 & 32 $\pm$ 3 & 32 \\
    
    %carbonaceous
    Carbonaceous & 346 $\pm$ 14 & 115 $\pm$ 5 & 115 & 200 $\pm$ 16 & 39 $\pm$ 1  & 38 \\
    \br
  \end{tabular*}
\end{table*}

In addition to the faster growth cycles when the magnetic field is present, there is also a noticeable impact within the spatial distribution of the dust cloud in dusty plasma during the presence of the magnetic field. Specifically, the presence of the magnetic field resulted in a larger void, for the growth of carbonaceous particles as compared to the growth of the titania particles. Variations in chamber geometry are expected to influence the dust cloud. Here, there is a groove in the lower electrode to fit in the substrate for particle collection and a protrusion of the upper rf-electrode. However, this geometry is kept constant throughout all the experiments. Therefore, any changes in the dust cloud is not attributed to changes in geometry, but instead, to the presence of the magnetic field. Moreover, when the magnets are present, they are put either above the top electrode or below the bottom electrode, so that they do not change the geometry of the surfaces which are directly in contact with  the plasma and do not interfere with the dust cloud.

\begin{figure*}[ht]
  \centering
  \begin{subfigure}{0.3\textwidth}
    \includegraphics[width=\linewidth]{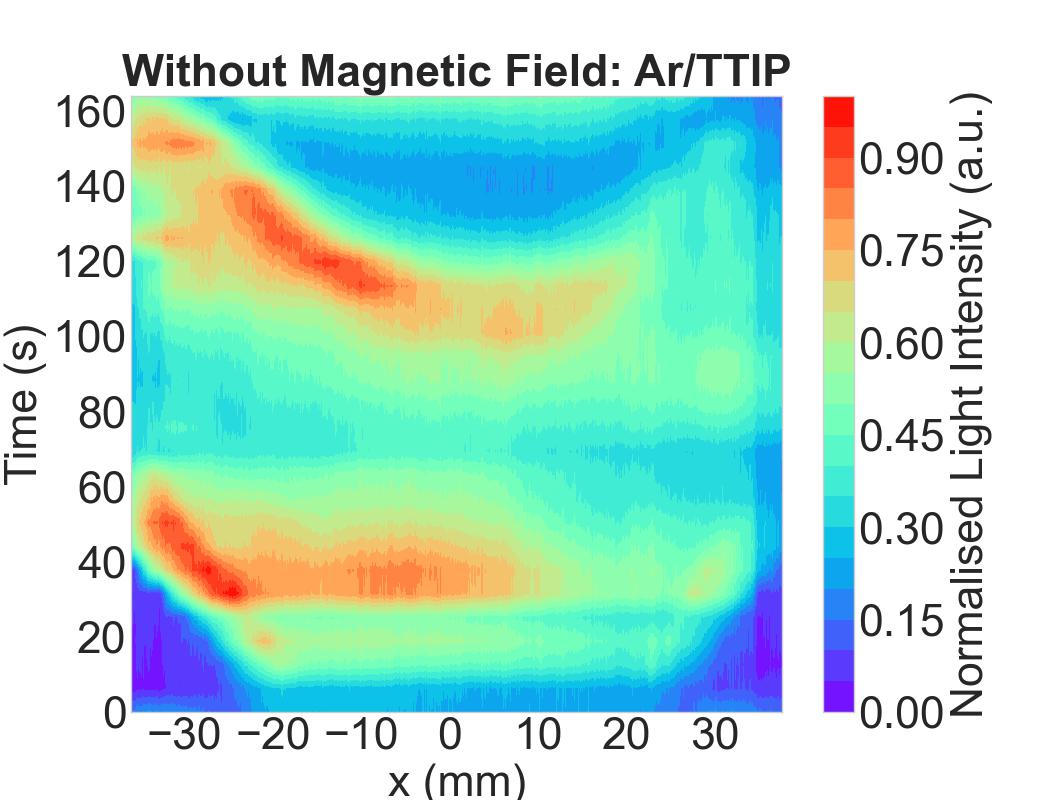}
    \caption{}
    \label{line_TTIP_B0}
  \end{subfigure}
  \begin{subfigure}{0.3\textwidth}
    \includegraphics[width=\linewidth]{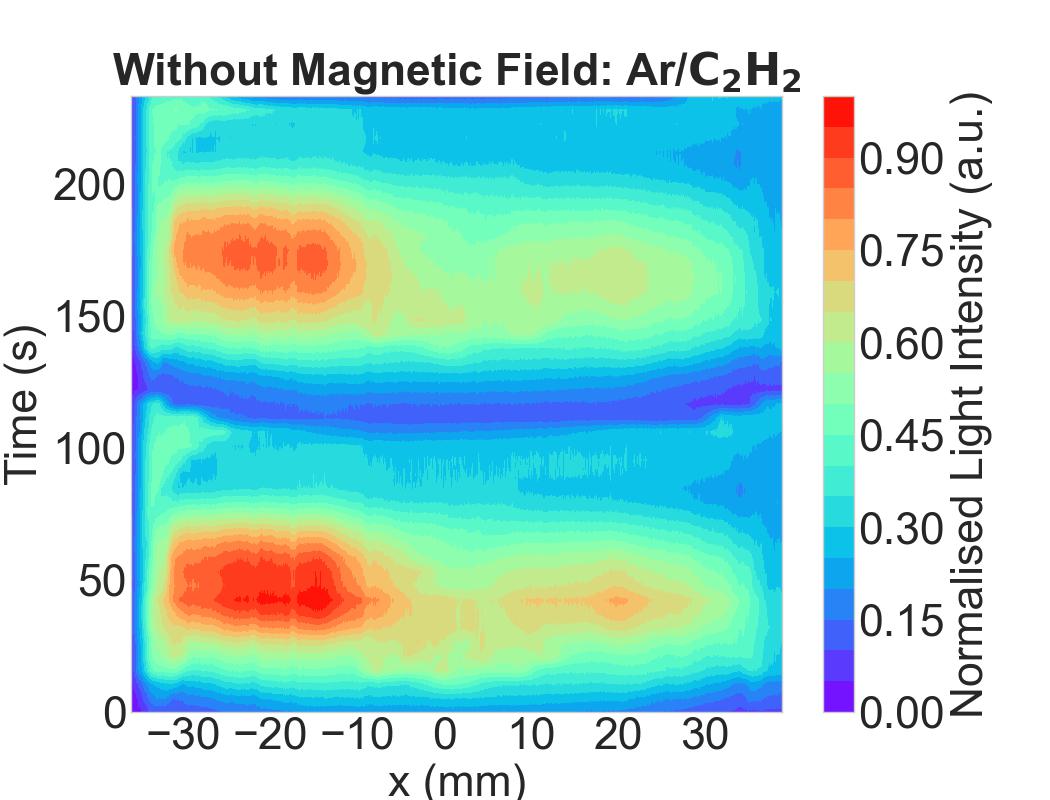}
    \caption{}
    \label{line_C2H2_B0}
  \end{subfigure}
  \begin{subfigure}{0.3\textwidth}
    \includegraphics[width=\linewidth]{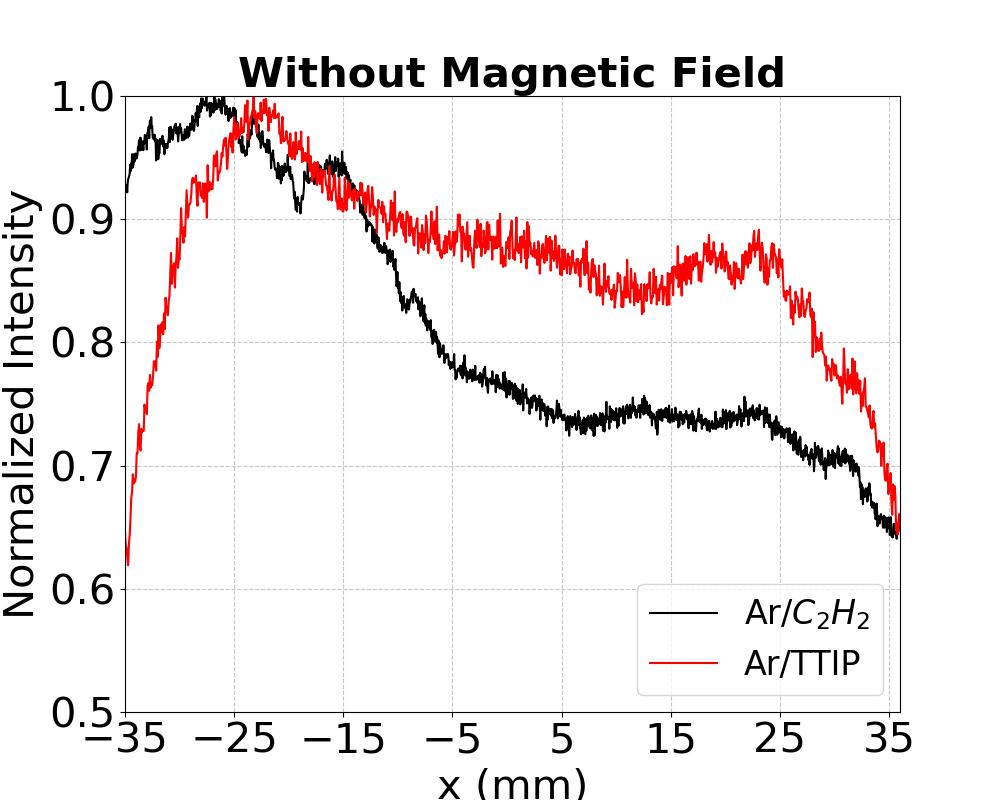} 
    \caption{}
    \label{int_line_b0}
  \end{subfigure}

  \begin{subfigure}{0.3\textwidth}
    \includegraphics[width=\linewidth]{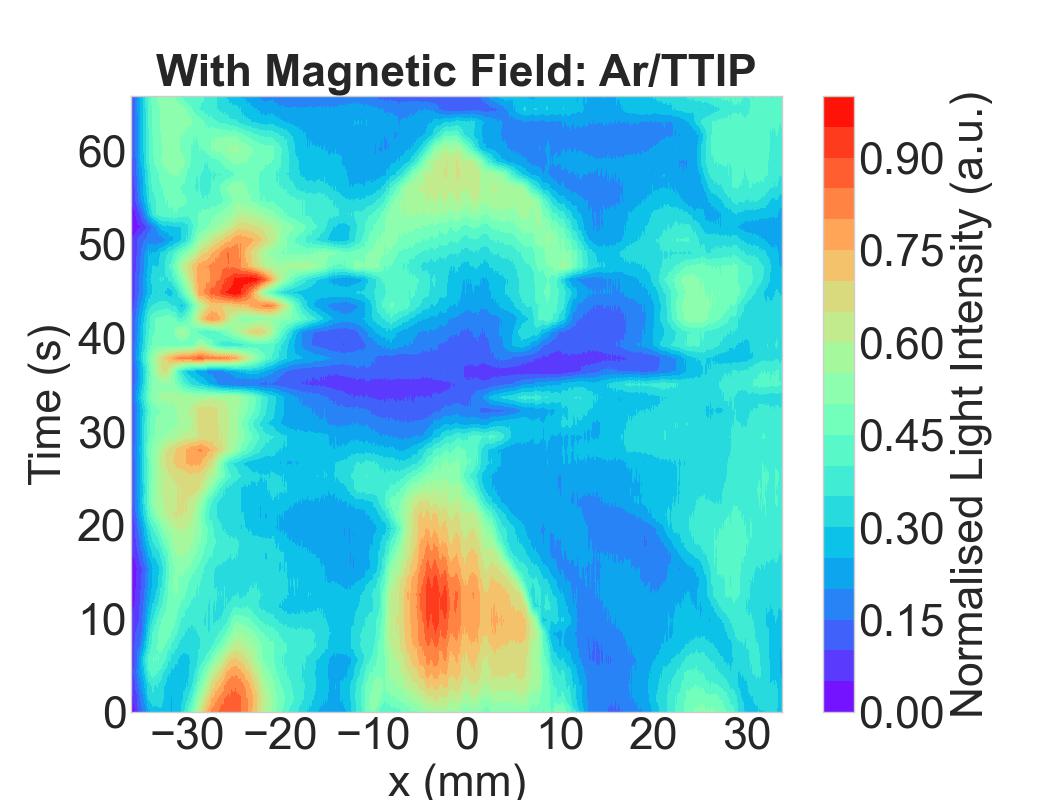} 
    \caption{}
    \label{line_TTIP_B500}
  \end{subfigure}
  \begin{subfigure}{0.3\textwidth}
    \includegraphics[width=\linewidth]{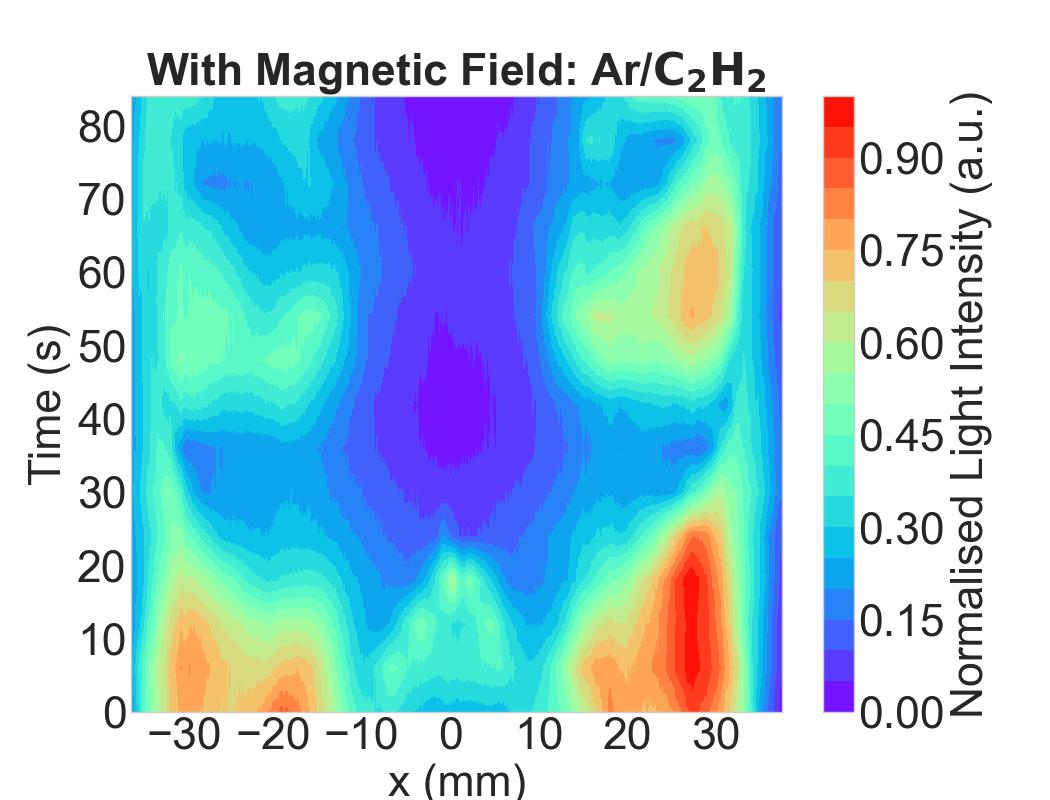}
    \caption{}
    \label{line_C2H2_B500}
  \end{subfigure}
  \begin{subfigure}{0.3\textwidth}
    \includegraphics[width=\linewidth]{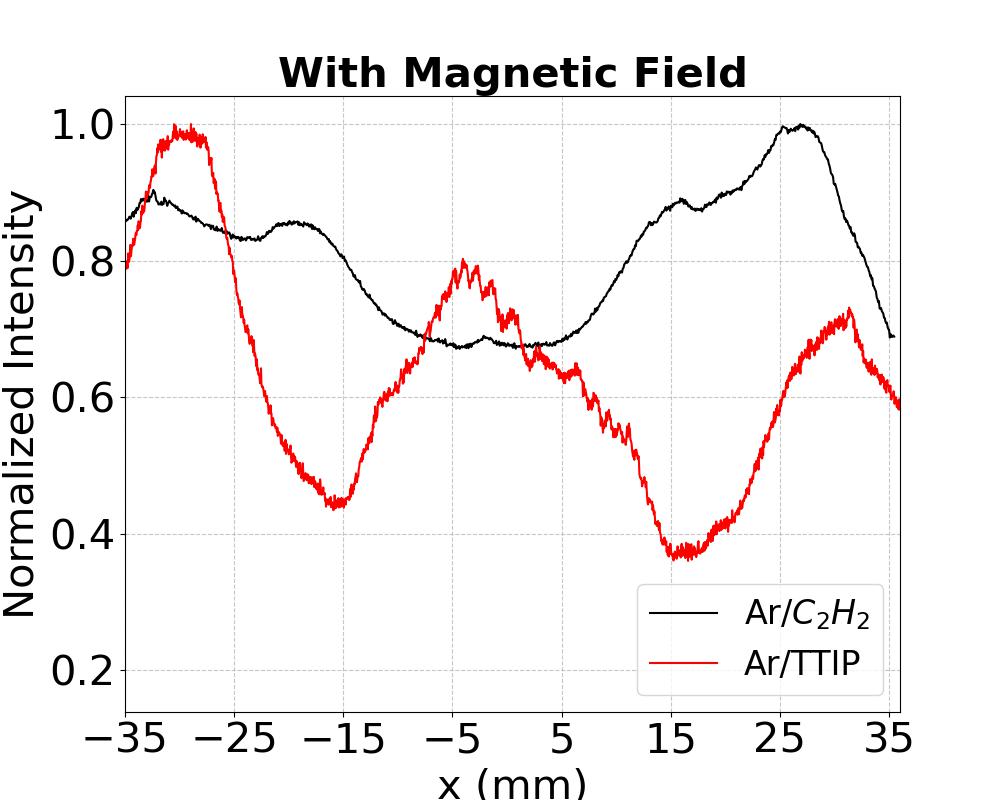}
    \caption{}
    \label{int_line_b500}
  \end{subfigure}
  \caption{ Temporal evolution of the light emission intensity from a line at z = 12.5 mm across the diameter of the dust cloud, as drawn in Fig. \ref{fig:all_16_clouds}. Without magnetic field: (a) Ar/TTIP and (b) Ar/$C_2H_2$. With magnetic field: (d) Ar/TTIP and (e) Ar/$C_2H_2$.
  Sum of intensity of both dusty plasma from light emitted intensity due to laser scattering across two growth cycles: (c) without magnetic field, (f) with magnetic field.}
  \label{cloud_int}
\end{figure*}

To visualize the overall radial distribution of the dust particles, the light intensity along the orange line of $T_C/2$ of Fig. \ref{fig:all_16_clouds} was summed over all frames for  horizontal positions, in the range of $-37.5 \leq$ x (mm) $\leq 37.5$, positioned at the midplane between the two electrodes at z = 12.5 mm, as shown in Fig. \ref{sideview_c}. The variation in intensity along the line, 50 pixels thick, represents the variation in spatial dust cloud distribution. The light intensity obtained was stacked vertically to show the temporal evolution of the dust cloud. Fig. \ref{line_TTIP_B0} and \ref{line_C2H2_B0} illustrate this for Ar/TTIP and Ar/$C_2H_2$ dusty plasmas without a magnetic field, while Fig. \ref{line_TTIP_B500} and \ref{line_TTIP_B500} show the same for Ar/TTIP and Ar/$C_2H_2$ dusty plasmas with a magnetic field, respectively. The intensity for each dusty was normalized to its highest number to enable a more direct comparison to be made between the particle growth of the two dust particle types. The temporal intensity profile appears to be similar for both kinds of dusty plasma without magnetic field and dissimilar during the presence of the magnetic field.  Line intensity plots are shown in Figs. \ref{int_line_b0} and \ref{int_line_b500} which shows the overall spatial distribution of the grown dust particles over two growth cycles without and with the magnetic field. These plots serve as a quantitative comparison of the two dust clouds to reveal any similarities and differences. In the absence of a magnetic field, the radial intensity variation of the two dusty plasmas exhibits similarities. However, with a magnetic field present, the Ar/TTIP dusty plasma shows peaks at the centre and edges of the electrode, whereas the Ar/$C_2H_2$ dusty plasma exhibits a peak only at the edges and a reduced intensity, due to a void, in the middle of the plasma. This suggests that there was more titiania dust in the middle of the plasma than there was carbonaceous dust; thus, there was a difference between the spatial distribution of the two dust clouds at 500 Gauss.

\begin{figure}[ht]
  \centering
   \begin{subfigure}{0.45\textwidth}
    \includegraphics[width=\textwidth]{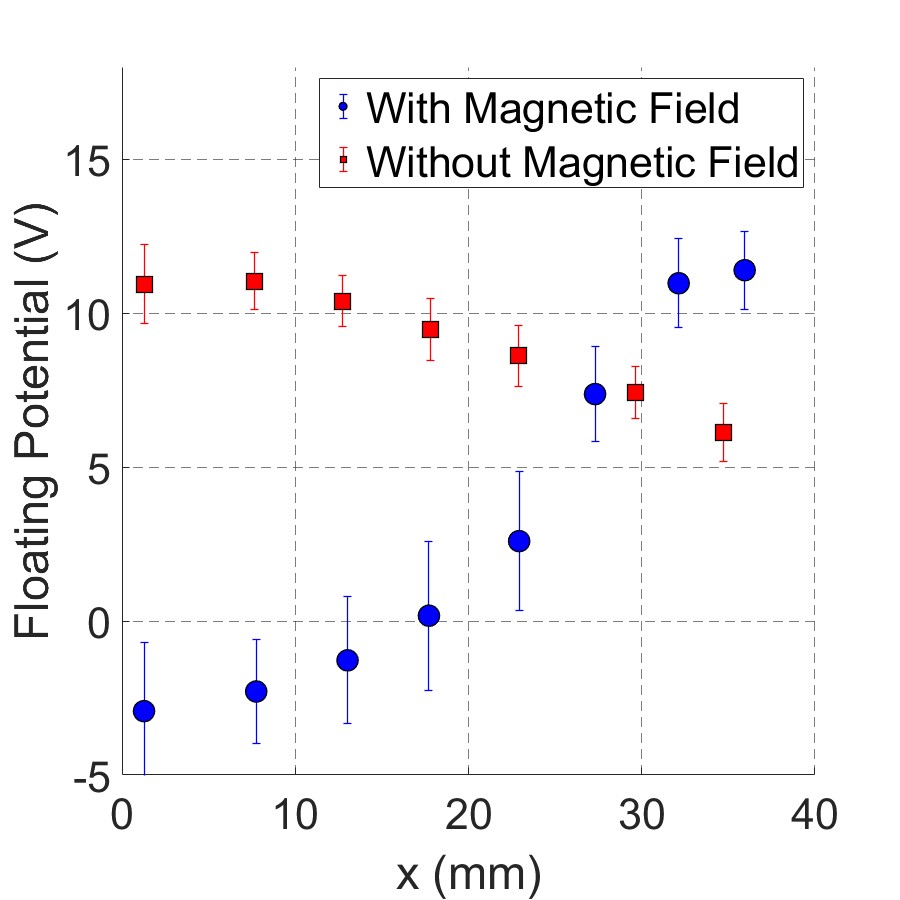}
    \caption{}
    \label{V_F_x}
  \end{subfigure}
   %\hfill
  \begin{subfigure}{0.45\textwidth}
    \includegraphics[width=\textwidth]{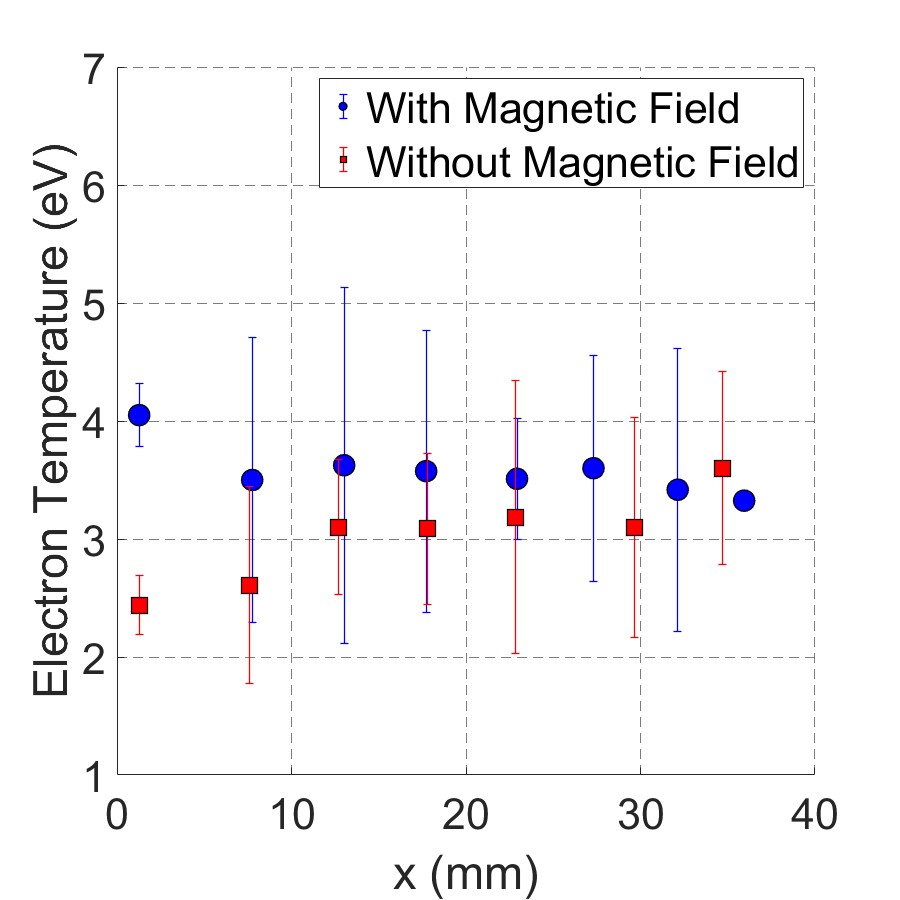} 
    \caption{}
    \label{T_e_x}
  \end{subfigure}

  \caption{Background Ar plasma parameters measured with a Langmuir probe. Radial variation of (a) floating potential  and (b) electron temperature at z = 12.5 mm, i.e., the mid plane between the two electrodes.}
  \label{Plasma_param}
\end{figure}

To understand why the dust cloud reacts differently to the presence of a magnetic field, it is important to explore changes in plasma conditions resulting from the presence of the magnets. Therefore, our investigation focuses on how the magnetic field has influenced plasma conditions, which can ultimately affect the spatial distribution of the dust cloud. Using a Langmuir probe in the background Ar plasma, we measured the plasma parameters of floating potential and electron temperature. A voltage sweep was done between - 40 V and 40 V, from which the current was measured. The floating potential is defined as the voltage measured when the current collected by the probe tip is zero. By approximating a linear fit to the logarithm of the electron current versus voltage beyond the floating potential, the electron temperature was calculated from the inverse of the slope \cite{hershkowitz1989langmuir}. The results of the measurements are shown in Fig. \ref{Plasma_param}. There is a radial variation in floating potential as shown in Fig. \ref{V_F_x}. However, there does not seem to be a radial variation of electron temperature. Even though these measurements were made in the absence of dust, we can suggest how the trend of the measurements will change. When the dust cloud has formed, the dust density is typically $\sim$ $10^{13}$ $m^{-3}$ \cite{tadsen2017amplitude}. This leads to an electron depletion in the background plasma due to the Havnes parameter being $> 1$ \cite{havnes1987dust}. We expect the overall floating potential of the plasma to slightly increase during the presence of the dusty nanoparticles\cite{ petersen2022decoupling}. At 500 Gauss, the changes in the floating potential, and consequently the radial electric field in the plasma, is possibly contributing to the faster cycle time. It is possible that the magnitude of several forces confining the particles in the plasma has decreased, thus leading to the faster cycle time. Additionally, it has already been shown that the electron temperature of particle growth as dusty plasma varies cyclically, in accordance with the dust growth cycles \cite{garofano2015}.

%For the two cases considered, with and without a magnetic field, Fig. \ref{V_F_x} clearly shows that the slope of the floating potential (in the absence of the dust), and by reasonable extension, the sign of the radial electric field change sign depending upon the presence of the magnet.  If this change of sign persists when the dust is present, this could be interpreted as a change in the radial, electrostatic confining force on the negatively charged, growing dust particles. 
Let us consider how this measurement of the background plasma parameter may be used to gain insights into the dynamics of the dust particle for the four cases discussed in this paper. Without a magnetic field, as indicated in Fig. \ref{cloud_int} (a) - (c), the apparent growth of both the carbonaceous and titania particles appear to fill the plasma volume in a very similar manner.  Although there is a left-right asymmetry (i.e., $x < 0$ mm vs. $x > 0$ mm) in the normalized light intensity, it is similar for both grown dust particle types without a magnetic field. By contrast, when evaluating the case with a magnetic field, as shown in Fig. \ref{cloud_int} (d) - (f), there is a difference in the spatial distribution of the grown dust of the two particle types in the presence of a magnetic field.  Under the assumption that both the carbonaceous and titania particles have the same negative charge, the change in the direction of the confining, radial electric field may contribute to the difference in the spatial distribution of the dust particles between the magnetized and unmagnetized cases.  However, the change in the direction of the electric field is insufficient to describe how a difference could arise the two particle types.

%%%%%%
%\textbf{Comment:  If the section above is rewritten, then this part also needs a rewrite.For all practical (plasma) purposes, magnetic fields in the 100-500 Gauss regime are effectively indistinguishable to the background plasma.  What may be important is the magnetic field gradient.  Looking at Fig. 5, there are several interesting features: 
%1.  Near the mid plane of the plasma (z ~ 12.5 mm +/- 5 mm), ∆B/∆z ~0.  I would interpret this to mean that where the particle cloud is growing, there is not a significant vertical (i.e., parallel to gravity), magnetic field gradient.
%2.  “Radially”, within x = 0 +/- 10 mm, there is no variation in the magnetic field; i.e, ∆B/∆x ~ 0.  However, once |x| ≥ 10 mm, there is a significant radial magnetic field gradient.  As a rough estimate, I used:  ∆B = 500- 200 = 280 G over ∆x = 10 mm to 30 mm
%So that:  |∆B/∆x| = 14 Gauss / mm = 1400 Gauss / m = 0.14 Tesla/m 
%** Here is where I need materials expertise.
%For the plasma components, it is the gradients that drive forces - would this be the same for the magnetic properties of a material - that it is the region where the gradient exists that you might see the largest impact on the material.
%Qualitatively, we see the difference in between the carbon and titanium in the region for |x| > 20 mm, which is where the magnetic field gradient exists.}
%%%%%%%

We now consider the effects of the magnetic field on the dust. First, according to the Hall parameter, a commonly employed metric for calculating the degree magnetisation of charged species in a plasma \cite{thomas2015, williams2022}, only the electrons were magnetised at 500 Gauss. Ions and dust particles were not magnetised, due to their heavier masses.  Therefore, there is no magnetisation due to charge accumulation on the dust that can explain the differences in spatial dust distribution of titania and carbonaceous nanoparticles. Second, there is a gradient in the radial direction of the magnetic field. According to Fig. \ref{radial_B}, the magnetic field, $B_z$ is constant in the region - 15 $\leq$ x(mm) $\leq$ 15. However, there is a change in $B_z$ with respect to x outside that range until the edge of the electrodes. It is possible that the gradient is contributing to the differences seen between the spatial distribution of the two dusty plasma. A different study of the magnetic properties of the materials is necessary in the future  in order to determine if they are contributing to the differences seen in the experiments due to the gradient.

\section{Conclusion} 
\label{sec:Conclusion}

We compared the growth of titania and carbonaceous nanoparticles as dusty plasmas, in the presence and absence of a weak magnetic field of 500 Gauss. Both particle growth displayed a cyclic behaviour, with a quicker cycle time at 500 Gauss. The cycle times were measured from OES of the plasma and NFT analysis of light emission intensity of the dust cloud. The particles reached their maximum size at the end of the cycle. The spatial distribution of the dust cloud appeared similar for both dusty plasma without magnetic field. However, during the presence of the magnetic field, the two dust cloud appeared differently. Measurement of background plasma parameters revealed changes in floating potential  during the presence of the magnetic field. However, these measurements were not enough to explain the differences between the two dust cloud during the presence of the magnetic field. Furthermore, the ions and dust were not magnetized according to the Hall parameter. Nevertheless, there was a gradient in the radial distribution of magnetic field. Future studies need to investigate the material properties and whether the gradient is responsible for the differences in the spatial distribution of the dust cloud between titania and carbonaceous dusty plasmas.

%%%%%%%%%%%%%%%%%%%%%%%%%%%%%%%%%%%%%%%%%%%
 \section*{Acknowledgement}
This work was supported with funding from the NSF EPSCoR program (OIA-2148653), and the U.S. Department of Energy – Plasma Science Facility (SC-0019176).  We thank Mr. Cameron Royer, Ms. Tamara Issac-Smith, Dr. Sarit Dhar, and Mr. Jeffrey Estep for technical assistance provided to implement this study.

%%%%%%%%%%%%%%%%%%%%%%%%%%%%%%%%%%%%%%%%%%%
\section*{Declaration of competing interest}

None.

%%%%%%%%%%%%%%%%%%%%%%%%%%%%%%%%%%%%%%%%%%%
\section*{Data Availability}
The data that support the findings of this study are available
from the corresponding author upon reasonable request.

%%%%%%%%%%%%%%%%%%%%%%%%%%%%%%%%%%%%%%%%%%%

\section*{References}
\providecommand{\newblock}{}

\end{document}